%% file: interplay_ms.tex
\documentclass[12pt,twocolumn]{iopart}
\usepackage{graphicx}
\usepackage{xspace}
\usepackage{hyperref} 
\usepackage[numbers,sort&compress]{natbib}
\usepackage{iopams}
\eqnobysec
\input{newcommands}

\begin{document}

\review{Interplay between magnetism and superconductivity in iron-chalcogenide superconductors: crystal growth and characterizations}

\author{Jinsheng Wen$^{1,2,3}$, Guangyong Xu$^{2}$, Genda Gu$^{2}$, J.~M.~Tranquada$^{2}$ and R.~J.~Birgeneau$^{1,3}$}

\address{$^{1}$Physics Department, University of California, Berkeley, California 94720, USA}
\address{$^{2}$Condensed Matter Physics and Materials Science Department, Brookhaven National Laboratory, Upton, New York 11973, USA}
\address{$^{3}$Materials Science Division, Lawrence Berkeley National Laboratory, Berkeley, California 94720, USA}
\eads{jinshengwen@berkeley.edu, jtran@bnl.gov}

\begin{abstract}
In this review, we present a summary of results on single crystal growth of two types of iron-chalcogenide superconductors, \fts (11), and $A_x$Fe$_{2-y}$Se$_2$ ($A=$ K, Rb, Cs, Tl, Tl/K, Tl/Rb), using Bridgman, zone-melting, vapor self-transport, and flux techniques. The superconducting and magnetic properties (the latter gained mainly from neutron scattering measurements) of these materials are reviewed to demonstrate the connection between magnetism and superconductivity. It will be shown that for the 11 system, while static magnetic order around the reciprocal lattice position (0.5,\,0) competes with superconductivity, spin excitations centered around (0.5,\,0.5) are closely coupled to the materials' superconductivity; this is made evident by the strong correlation between the spectral weight around (0.5,\,0.5) and the superconducting volume fraction.
The observation of a spin resonance below the superconducting temperature, \tc, and the magnetic-field dependence of the resonance, emphasize the close interplay between spin excitations and superconductivity, similar to cuprate superconductors. In $A_x$Fe$_{2-y}$Se$_2$, superconductivity with \tc $\sim30$~K borders an antiferromagnetic insulating phase; this is closer to the behavior observed in the cuprates but differs from that in other iron-based superconductors.

\end{abstract}

\tableofcontents
\maketitle
\section{Introduction}
A superconducting material can conduct electric current with zero-energy loss due to the absence of  electrical resistance below its superconducting transition temperature \tc; such behavior is clearly of great practical use~\cite{onnes1}. On the theoretical side, superconductivity was initially understood with the many-body theory developed by Bardeen, Cooper, and Schrieffer, (BCS theory)~\cite{bcstheory}; these authors explained the phenomenon of superconductivity on the basis of electron pairs whose interaction is mediated by  electron-phonon coupling.  In the 1980s, high-temperature superconductivity was discovered in lamellar copper-oxide materials whose \tc can be above liquid-nitrogen temperature~\cite{mullerlbco,ybcodis}. Here, the electron-phonon coupling as proposed in BCS theory is not sufficient to induce superconductivity at such high temperatures \cite{lee:17,carlsonbook1}. In these cuprate superconducting materials, superconductivity develops from electronically doping an antiferromagnetic Mott insulating phase with carriers~\cite{lee:17,carlsonbook1,birgeneau-2006,RevModPhys.70.897,orenstein}. It is thus hoped that one may eventually understand the high-\tc superconductivity in terms of the interplay between magnetism and superconductivity.

\begin{figure}[ht]
\begin{center}
  \includegraphics[width=\linewidth]{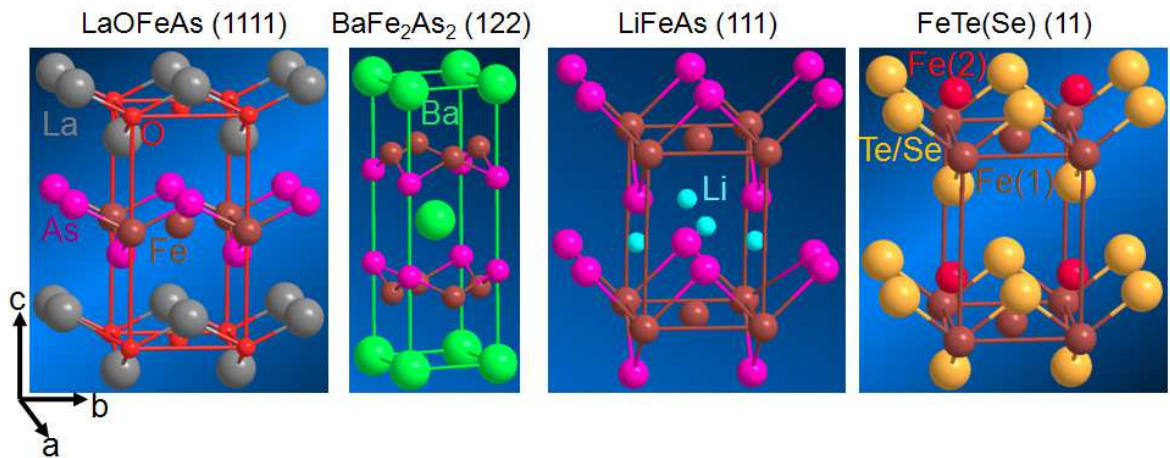}  \end{center}
  \caption{ Schematic crystal structures for the 1111, 122, 111 and 11 type iron-based superconductors.}\label{fig:ironbasedsc}
\end{figure}

Research on high-\tc superconductivity turned in a new direction in the year 2008 with the discovery of iron-pnictide superconductors~\cite{hosono_1,hosono_2}. The initial excitement came with the discovery by Hosono's group of superconductivity in LaFeAsO$_{1-x}$F$_x$ (labeled 1111, based on the elemental ratios in the chemical formula of the parent material) with $T_c=26$~K~\cite{hosono_2}, following an earlier report by the same group of superconductivity in LaFePO$_{1-x}$F$_x$ with \tc $\sim$~5~K~\cite{hosono_1}. The \tc was soon raised to 43~K, either by replacing La with Sm (SmFeAsO$_{1-x}$F$_x$)~\cite{chen-2008-453}, or by applying pressure~\cite{hosono4}. Several 1111 superconductors with $T_c>50$~K have been successively reported~\cite{JPSJtc54k,eplrentc51k,ren-2008-25,epltc56k}, and the current record is 56~K in Gd$_{1-x}$Th$_x$FeAsO~\cite{epltc56k}. Besides the 1111 system, four other families of iron-based superconductors have been discovered, typified by BaFe$_2$As$_2$ (122)~\cite{sefat:117004,rotter-2008-101,chen-2008-25,inosov:224503}, LiFeAs (111)~\cite{lifep,lifeas1,lifeas2,lifeas3,structure5,lifep2,nafeas1}, \fts (11)~\cite{hsu-2008,yeh-2008,sales:094521,chen:140509,fang-2008-78}, and Sr$_2$PO$_3$FePn (21311)~\cite{zhu:220512,sc21311}. Their crystal structures (\fref{fig:ironbasedsc}) are all tetragonal at room temperature, but the 122 family crystallizes in the $I4/mmm$ space group, while the space group is $P4/nmm$ for the others~\cite{structure5,zhao:132504,hosono_1,lester:144523,qiu:257002,kumar:144524}. One important common feature is that they are all layered structures, as in the cuprates~\cite{lee:17,birgeneau-2006,RevModPhys.70.897,orenstein,carlsonbook1}.

Extensive research has been carried out to study the magnetic correlations. It is now well established for these materials that long-range antiferromagnetic order is suppressed with
doping, while superconductivity appears above a certain doping
level~\cite{cruz,lester:144523,qiu:257002,kumar:144524,huang:257003,chen:064515,johannes-2009,wilson:184519,kofu-2009-11,zhao,luetkens-2008,drew-2009,rotter-2008-47,chen-2009-85,fang:140508,chu:014506,khasanov:140511,liupi0topp,0295-5075-90-2-27011,spinglass,JPSJ.79.102001,revisedfetesephase}. While there are a few cases where superconductivity
appears suddenly after magnetic order disappears~\cite{luetkens-2008}, it is more commonly found that magnetic order, either short- or long-ranged, coexists in some fashion with superconductivity over a finite range of doping~\cite{zhao,drew-2009,rotter-2008-47,chen-2009-85,fang:140508,chu:014506,spinglass,liupi0topp}.
The observations that superconductivity develops concomitantly with the suppression of  antiferromagnetic order in many iron-based superconductors~\cite{cruz,huang:054529,mcguire:094517,spinglass}, a behaviour similar to that in cuprates~\cite{lee:17,birgeneau-2006,RevModPhys.70.897,orenstein,carlsonbook1,kivelsonironnews},
suggest that there could also be a similar connection with the superconductivity. Conceptually, the simplest possibility would involve having the magnetic excitations replace phonons in the electron pairing interaction, and in the iron-based superconductors the finite momentum of the antiferromagnetic fluctuations is proposed to facilitate interband pairing~\cite{mazin:057003,kuroki-2008,ma:033111,dong-2008-83,cvetkovic-2009,graser-2009,JPSJ.77.113703}.  Of course, the role of the magnetic correlations could depend on the nature of the magnetism, which continues to be a subject of intense investigation \cite{Fang2008,Haule2009,Si2009,mazin-2009np,Kou2009epl,Medici2010,weiguounified,Arita2}.  Furthermore, four or five Fe $3d$ orbitals are involved in the multiple bands that cross the Fermi energy, and excitations among these orbitals have also been proposed as possible contributors to the pairing mechanism~\cite{Kontani2010}. In any case, the magnetic excitations respond strongly to the superconductivity: the low-energy fluctuations are gapped, and the spectral weight moves into a ``resonance'' peak above the gap; the resonance peak has been observed in a number of
iron-based superconductors~\cite{christianson-2008-456,lumsden:107005,chi:107006,shiliang-2009,
inosov-2009,resonance1111,qiu:067008,mook-2009-2,fieldeffectresonancewen}. Obviously, the iron-based superconductors provide new opportunities to address the long-standing challenge of high-\tc superconductivity, and this has made them some of the most heavily studied materials in condensed matter physics over the past three years. A number of reviews are already available on these superconductors~\cite{JPSJ.78.062001,JPSJS.77SC.1,lynn-2009-469,lumsdenreview1,Hosonoreviewphysicac,mkwreview1,canfieldreview,dcjohnstonreview,mazinnaturereview,JPSJ.79.102001,htcironbasereview}.

Among the five types of iron-based superconductors, the \tc for the 11 system is the lowest~\cite{lumsdenreview1}. However, the 11 materials are still of particular interest for a number of reasons: i) As seen from \fref{fig:ironbasedsc}, the crystal structure is the simplest, which makes it easier to study them; ii) 11 system does not contain As, thus making it safer to handle; iii) Most importantly, high-quality, large-size single crystals can be made available for this system (with $x\leq0.7$), as will be discussed in \sref{sec:crystalgrowth}. This is especially important for neutron scattering experiments because, although neutrons represent a powerful probe of magnetic correlations, the limitations of neutron source strength and scattering cross sections require the use of samples of considerable size ($\gtrsim 1$~cm$^3$) in order to obtain reliable data.

More recently, the discovery of superconductivity in a ternary iron-chalcogenide system with the approximate formula $A_{x}$Fe$_{2-y}$Se$_{2}$ ($A=$ K, Rb, Cs, Tl, Tl/K, Tl/Rb), whose highest \tc is $\sim$~33~K at ambient pressure, has stimulated considerable interest~\cite{scinkfese,kfese2,rbfese1,csfese1,tlrbfese1}. The structure at high temperature is equivalent to that of the 122 materials, as shown in \fref{fig:ironbasedsc} (122).

The remainder of this paper is organized as follows: in \sref{sec:crystalgrowth}, we present the efforts on crystal growth that make many further measurements possible; in \sref{sec:sc}, superconductivity in both the 11 and $A_{x}$Fe$_{2-y}$Se$_{2}$ systems is discussed; in \sref{sec:neutron}, the static and dynamic spin correlations obtained mainly by neutron scattering experiments are presented, followed by the conclusions in \sref{sec:conclusion}.

\section{Crystal growth}
\label{sec:crystalgrowth}
\begin{figure}[th]
\begin{center}
  \includegraphics[width=0.7\linewidth]{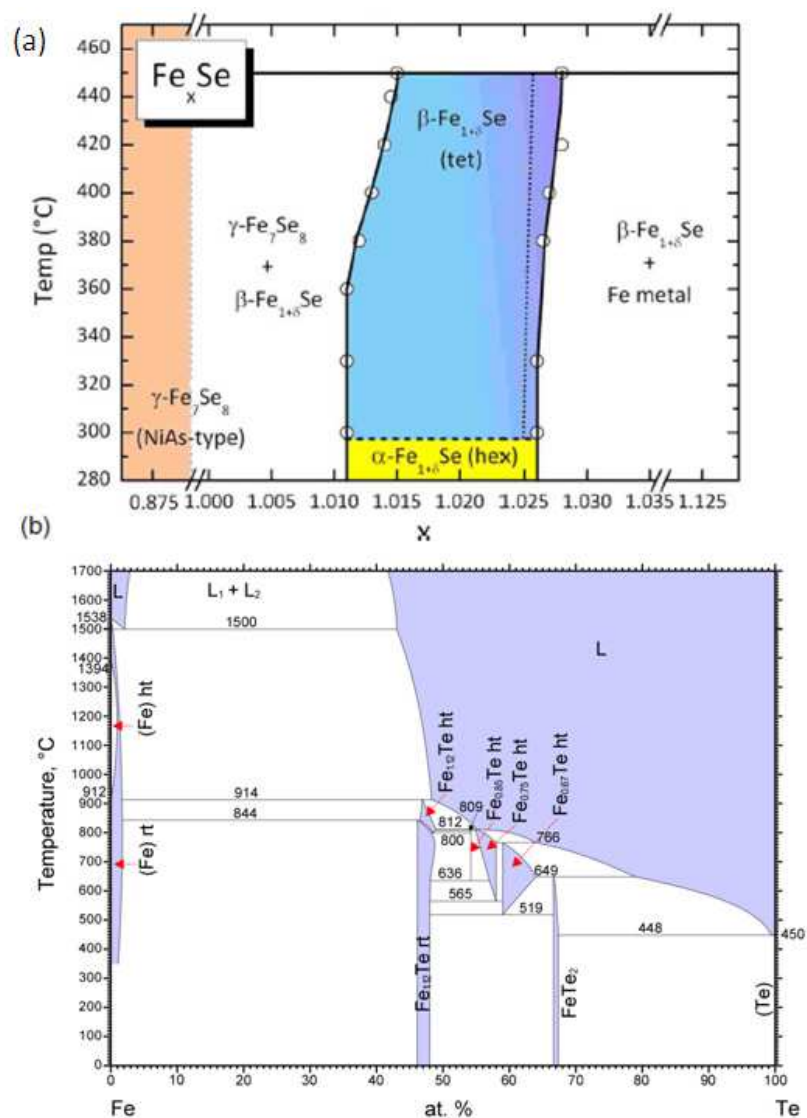}  \end{center}
  \caption{(a) Phase diagram for Fe-Se. Reprinted from \cite{mcqueen:014522}. $\copyright$ 2009 American Physical Society. (b) Phase diagram for Fe-Te. Reprinted from \cite{fetephasediagram_r}. $\copyright$ ASM International.}\label{fig:fetesebinaryphase}
\end{figure}

Fe-Se has a very complicated phase diagram~\cite{fesephasediagram_r}. The superconductivity in this system was found in the $\beta$-Fe$_{1+y}$Se, which has a tetragonal structure~\cite{hsu-2008,mcqueen:014522}. A modified phase diagram [\fref{fig:fetesebinaryphase}(a)] indicates that the $\beta$-phase only exists in a narrow temperature (300-440~$^\circ$C) and composition window (Fe:Se$=1.01$-$1.03$)~\cite{mcqueen:014522}.  Because there is no common phase boundary line between the liquid phase region and  the Fe$_{1+y}$Se solid phase region in the Fe-Se phase  diagram~\cite{fesephasediagram_r}, it is not possible to grow a single crystal of Fe$_{1+y}$Se directly from Fe-Se liquid. Therefore, methods that rely on growing a crystal directly from a melt can not be used in this case. Instead, other growth methods, such as vapor self-transport growth~\cite{patel:082508} and alkali-halide-flux growth~\cite{flux015020,Hu2011,kclfluxgrowth} have been reported.

In contrast, tetragonal Fe$_{1+y}$Te, though not superconducting, is stable in a larger temperature and composition space [\fref{fig:fetesebinaryphase}(b)], and large single crystals can be grown directly from the melt, with a certain amount of excess Fe~\cite{sales:094521,chen:140509}. It was quickly found that mixing Se with Te both optimizes \tc (at ambient pressure) and allows melt growth techniques. Indeed, to-date, large-size single crystals of \fts with $x\leq0.7$ have been successfully grown utilizing several standard melting techniques, including Bridgman (vertical and horizontal)~\cite{sales:094521,2011JAP93914M,liu:174509,chen:140509,JPSJ.79.084711,JPSJ.79.074704,wen:104506,spinglass,homesopticalprb,shcouplingprb,fieldeffectresonancewen,xudoping11,PhysRevB.80.214511,taen:092502,PhysRevB.81.020509,PhysRevB.80.214514}, and optical zone-melting~\cite{yeh-2009}. For the newly discovered $A_{x}$Fe$_{2-y}$Se$_{2}$ system, typically the Bridgman method has been employed~\cite{scinkfese,2011arXiv1101.0789W,2011arXiv1102.1010L,kfese2,rbfese1,kcsfese1,2010arXiv1012.5236F,2011arXiv1103.2904G}.

\subsection{Bridgman method}
\label{sec:bridgman}
\begin{figure}[th]
\begin{center}
  \includegraphics[width=0.8\linewidth]{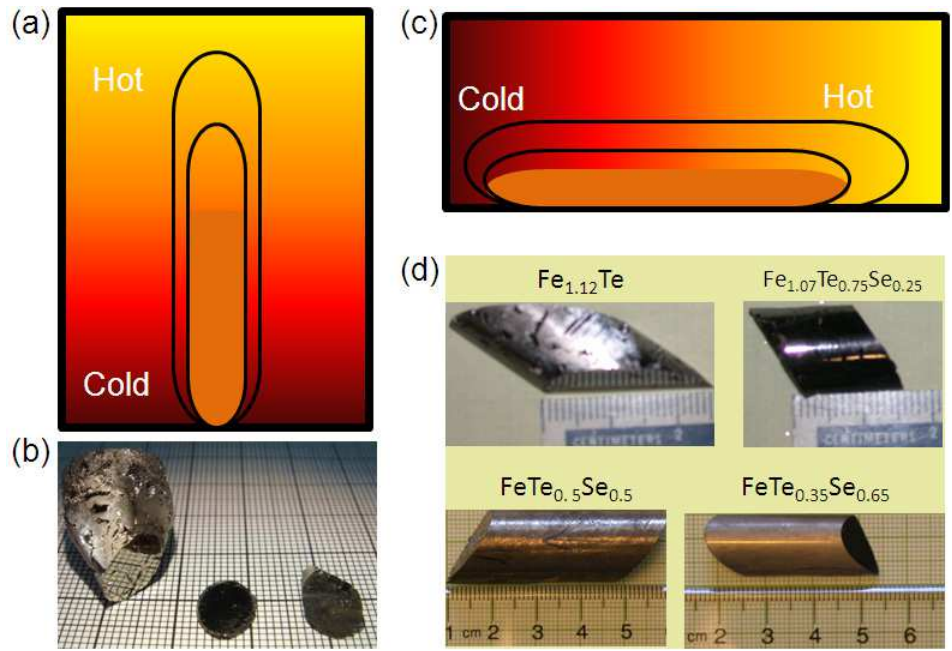}  \end{center}
  \caption{ (a) Schematic of the vertical Bridgman growth. (b) Crystals grown using the method shown in (a). Reprinted from \cite{sales:094521}. $\copyright$ 2009 American Physical Society. (c) Horizontal Bridgman growth. (d) Crystals grown using the method shown in (c)~\cite{wenphdthesis}.}\label{fig:bridgman}
\end{figure}

In \fref{fig:bridgman}(a) and (c) we show the schematics for the vertical and horizontal versions of the Bridgman technique.  An example of a synthesis scheme is as follows.  To start with, the raw materials (99.999\% Te, 99.999\% Se, and 99.98\% Fe) are weighed and mixed with the desired molar ratio, and then sealed into an  evacuated, high-purity (99.995\%) quartz tube. Since the tube often cracks during the cooling cycle, it is sealed inside a larger evacuated tube. The doubly-sealed materials are then put into the furnace vertically (or horizontally) for pre-melting, with the following sequence: ramp to 660~$^\circ$C in 2 hrs; hold for 1~hr; ramp to 900~$^\circ$C in 1~hr; hold for 1~hr; ramp to 1050~$^\circ$C in 1~hr; hold for 3~hrs; shut down the furnace and cool to room temperature. The reacted materials are then crushed and double-sealed into evacuated quartz tubes for the growth sequence: ramp to 660~$^\circ$C in 3 hrs; hold for 1~hr; ramp to 900~$^\circ$C in 2~hrs; hold for 1~hr; ramp to 1000~$^\circ$C in 1~hr; hold for 12~hrs; cool to 300~$^\circ$C with a cooling rate of $-0.5$ or $-1$~$^\circ$C/hr; shut down the furnace and cool to room temperature. In the furnace, there should be a small temperature gradient from one end to the other (\eg, at 850~$^\circ$C, $\Delta T/$distance~$\approx5$~$^\circ$C/cm) as shown in \fref{fig:bridgman}(a) and (c), which allows directional solidification of the melted liquid. (Some may choose to skip the premelting step.) By using this method, large-size ($>10$ g) high-quality single crystals can be obtained for \fts with $x\le0.7$. \Fref{fig:bridgman}(b) and (d) show some of the crystals grown using vertical and horizontal Bridgman methods. The crystals can have nice mirror-like cleavage surfaces, corresponding to the $a$-$b$ planes; however, in our experience, such crystals have excess Fe ($y>0$) and a reduced superconducting volume fraction.  Crystals with a large superconducting volume fraction (with $y\approx 0$) tend to have a textured cleavage surface associated with slight misorientation of grains, due to strain effects that develop on cooling in the constrained cross section of a quartz tube.

\begin{figure}[th]
\begin{center}
  \includegraphics[width=0.9\linewidth]{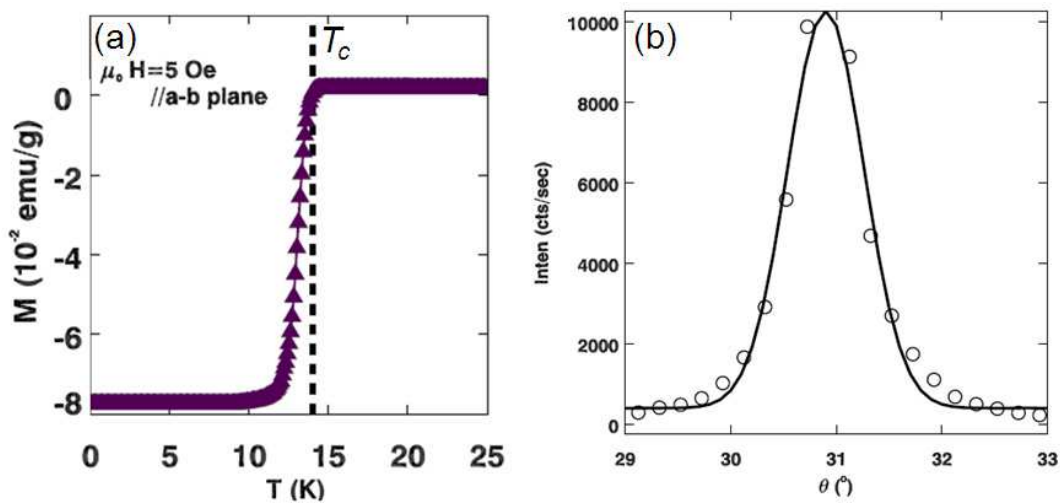}  \end{center}
  \caption{ (a) and (b), magnetization and (110) rocking curve of FeTe$_{0.5}$Se$_{0.5}$ samples.}\label{fig:samplequality}
\end{figure}

Many measurements such as resistivity, magnetization, x-ray diffraction (XRD) and neutron diffraction have been carried out to characterize such crystals~\cite{sales:094521,2011JAP93914M,liu:174509,chen:140509,JPSJ.79.084711,JPSJ.79.074704,wen:104506,spinglass,homesopticalprb,shcouplingprb,fieldeffectresonancewen,xudoping11,PhysRevB.80.214511,taen:092502,PhysRevB.81.020509,PhysRevB.80.214514}. In \fref{fig:samplequality} we show the bulk magnetization and rocking curve for FeTe$_{0.5}$Se$_{0.5}$ samples. The transition width, $\Delta T_c$, is $\sim1$~K [\fref{fig:samplequality}(a)], indicating that the sample is reasonably homogeneous. From neutron diffraction measurements, we obtained a mosaic spread of 0.85$^\circ$ for a 9-g sample [\fref{fig:samplequality}(b)]. These demonstrate that high-quality, large-size single crystals on \fts ($x\leq0.7$) can be successfully grown using the Bridgman method.

\begin{figure}[ht]
\begin{center}
  \includegraphics[width=0.8\linewidth]{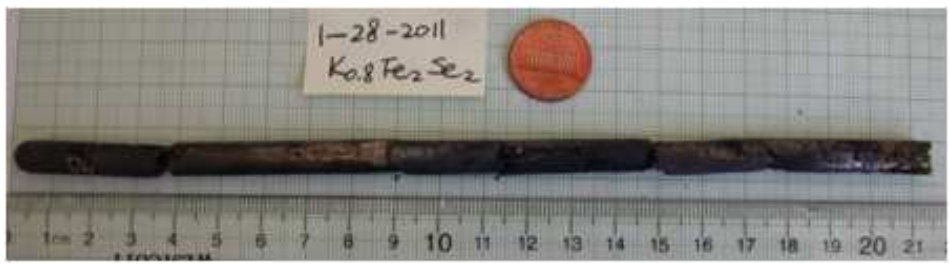}  \end{center}
  \caption{Single crystal rods of K$_{0.8}$Fe$_{2}$Se$_2$ grown using the vertical Bridgman method~\cite{kfesecrystalgrowth}.}\label{fig:bridgmankfese}
\end{figure}

To grow single crystals of $A_{x}$Fe$_{2-y}$Se$_{2}$, a similar method has been used. First, FeSe is prepared as the precursor by reacting Fe and Se in the appropriate ratio at 700~$^\circ$C for 4 hrs. The alkali and the precursor FeSe are loaded into an alumina crucible with the nominal composition, which was then sealed into a tube; Wang \et \cite{2011arXiv1101.0789W} have demonstrated the benefits of using an arc-welded Ta tube. The tube is then put into an evacuated quartz tube and sealed. A typical growth sequence is: ramp to 1050$^\circ$C in 15~hrs; hold for 4~hrs; cool to 750~$^\circ$C at a rate of $-1$ to $-3$~$^\circ$C/hr; shut down the furnace and cool to room temperature. Large-size single crystals can be obtained using this method, and single crystal rods so-obtained are shown in \fref{fig:bridgmankfese}. A one-step method using a similar growth procedure but without first reacting Fe and Se as the precursor has also been reported~\cite{2011arXiv1103.2904G}.

\subsection{Optical zone-melting technique}
\label{sec:zonemelting}
\begin{figure}[ht]
\begin{center}
\includegraphics[width=0.8\linewidth]{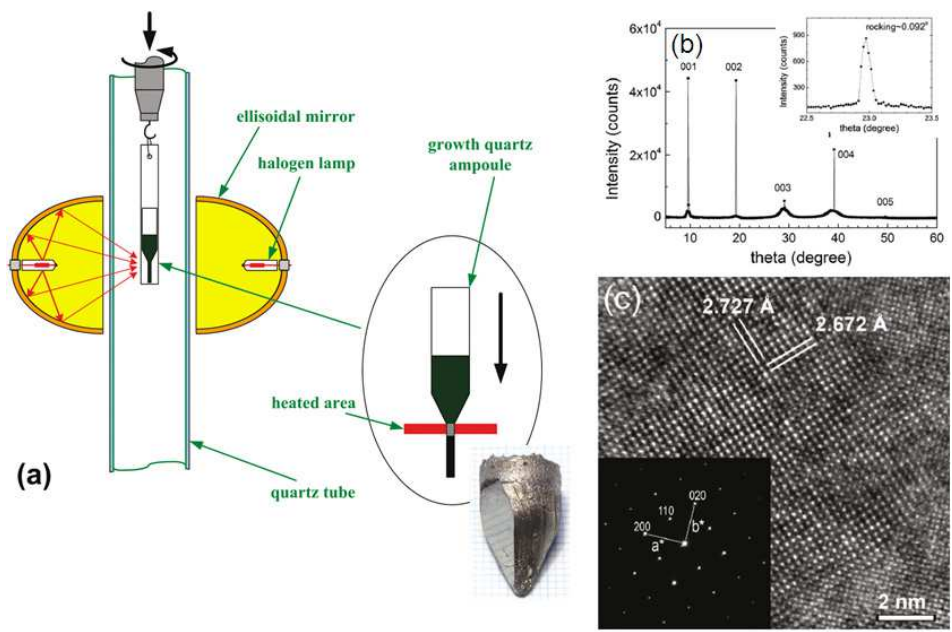}\end{center}
  \caption{ (a) Schematic diagram of apparatus setup of the optical zone-melting method. The inset shows an as-grown FeTe$_{0.7}$Se$_{0.3}$ single crystal on a 1-mm grid. The shiny surface is the $a$-$b$ plane. (b) XRD pattern for an FeSe$_{0.3}$Te$_{0.7}$ crystal. Miller indices for the tetragonal-PbO structure are shown. The inset shows the rocking curve for the (101) peak. (c) High-resolution TEM image of an FeSe$_{0.3}$Te$_{0.7}$ crystal and the electron diffraction indexed with a tetragonal lattice. Reprinted from \cite{yeh-2009}. $\copyright$ 2009 American Chemical Society.}\label{fig:zonemelting}
\end{figure}

Yeh \et~\cite{yeh-2009} have used the optical zone-melting method to grow \fts single crystals with $x$ ranging from 0 to 0.7. The advantage of this technique is that it offers the possibility of visually examining the locally melted zone, as well as providing convenient control of the growth rate using the image furnace. The schematic of the setup is shown in \fref{fig:zonemelting}(a). To grow single crystals, powders of 99.9\% Fe, 99.9\% Se, and 99.999\% Te were combined with the desired stoichiometry and mixed in a ball mill for 4~hrs. The mixed powders were cold pressed into discs under 400-kg/cm$^2$ uniaxial pressure, and then sealed into an evacuated quartz tube. The tube was heated at 600~$^\circ$C for 20~hrs. The reacted bulk material was reground to fine powder and then sealed into an evacuated quartz tube. The tube was sealed into a second evacuated tube, and then loaded into an optical floating-zone furnace with two 1500-W halogen lamps installed inside the mirrors as infrared radiation sources, as shown in \fref{fig:zonemelting}(a). The ampule was rotated at a rate of 20~rpm and moved downwards at a rate of 1-2 mm/hr. The as-grown crystals are annealed with the following sequence: ramp to 700-800~$^\circ$C in 7~hrs; hold for 48~hrs; cool to 420~$^\circ$C in 4~hrs; hold for 30~hrs; shut down the furnace and cool to room temperature.

The inset of \fref{fig:zonemelting}(a) shows an as-grown crystal for FeSe$_{0.3}$Te$_{0.7}$, which has an easily cleaved surface perpendicular to the $c$ axis~\cite{yeh-2009}. As shown in \fref{fig:zonemelting}(b), only (00$L$) peaks were found in the typical XRD pattern on a cleaved surface of the crystal, indicating that the cleaved surface corresponds to the $a$-$b$ plane. In the inset of \fref{fig:zonemelting}(b) it is shown that the crystal has a mosaic as small as 0.092$^\circ$. The phase purity was examined by x-ray powder diffraction with powder obtained by crushing the crystals. Yeh \et~\cite{yeh-2009} found that all diffraction peaks of crystals with $x<0.7$ belong to the tetragonal phase and no secondary phase was observed. High-resolution transmission electron microscopy (TEM) measurements showed that the diffraction patterns could be well indexed with a tetragonal structure, as shown in the inset of \fref{fig:zonemelting}(c)~\cite{yeh-2009}.

\subsection{Vapor self-transport and flux method}
\label{sec:vaporandflux}
The phase diagram of Fe-Se is such that one cannot grow large crystals of the $\beta$-FeSe directly from standard melting methods; Patel \et~\cite{patel:082508} used vapor self-transport as an alternative approach. They began by mixing 99.9\% Fe and 99.99\% Se powders with the desired molar ratio. The mixture was ground and sealed into an evacuated quartz tube. The material was then heated in a three-zone furnace with the following procedure: 1) heat the material with the temperatures for the three zones set to 700~$^\circ$C (where the Fe and Se mixture is located), 900~$^\circ$C, and 900~$^\circ$C, respectively, for 5 days; 2) crystal nucleation is initiated by setting the temperature of all zones to 700~$^\circ$C, and the temperature held constant for 5~hrs; 3) the crystals are left growing for 30 days with a temperature setting of 825~$^\circ$C, 700~$^\circ$C, and 825~$^\circ$C, respectively; 4) the system is cooled to 400~$^\circ$C rapidly, and the temperature is kept constant for 10~hrs; 5) cool to room temperature at a rate of $-3$~$^\circ$C/min. X-ray diffraction indicated that the dominant phase in the as-grown crystal was superconducting $\beta$-Fe$_{1+y}$Se, with a minority of non-superconducting $\alpha$-FeSe~\cite{patel:082508}.

Another approach is to use an alkali-halide flux. Zhang \et~\cite{flux015020} made use of a NaCl/KCl flux.  First, Fe and Se powders with the desired stoichiometry were reacted to obtain Fe$_{1+y}$Se polycrystalline samples. This material was mixed with NaCl/KCl flux (ratio 1:1), and the combination was ground and sealed in an evacuated quartz tube. The quartz tube was heated to 850~$^\circ$C. The temperature was maintained at 850~$^\circ$C for 2~hrs before cooling to 600~$^\circ$C at a rate of $-3$~$^\circ$C/hr. Finally, the furnace was cooled rapidly to room temperature. The crystals were separated from the flux by dissolving the NaCl/KCl flux in deionized water. There is also a report of using KCl(KBr) as the flux~\cite{kclfluxgrowth}.

Hu \et \cite{Hu2011} have recently demonstrated an improved method, utilizing a flux of LiCl/CsCl. Refinement of synchrotron XRD data yielded a stoichiometry of Fe$_{1.00(2)}$Se$_{1.00(3)}$.

\subsection{Remaining challenges}
As described above, it has been demonstrated that large crystals of \fts with $x\leq0.7$ can be grown from the melt; however, challenges remain in obtaining homogeneous samples~\cite{PhysRevB.82.020502,phasese11}. Hu \et~\cite{phasese11} performed scanning transmission
electron microscopy (STEM) and electron energy loss spectroscopy (EELS) measurements on a series of as-grown \fts crystals. They found that all these samples showed nanoscale
phase separation and chemical inhomogeneity of the Te/Se content. They attributed this to the presence of a miscibility gap in the phase diagram. The same conclusion was reached in an extended x-ray absorption fine-structure (EXAFS) study on \fts, where it was observed that the local structure differed from the average~\cite{PhysRevB.82.020502}.  Scanning tunneling microscopy/spectroscopy (STM/STS) measurements on a FeTe$_{0.55}$Se$_{0.45}$ sample also showed that there were Te- and Se-rich regions, whereas the averaged Te and Se contents were still 0.55 and 0.45, respectively~\cite{He2011}. The sample inhomogeneity may account for the broad transitions in these materials~\cite{mkwreview1,yeh-2009}. One way to improve crystals is to use a smaller cooling rate during the growth. For example, from our own experience, using a -1~$^\circ$C/hr cooling rate yields better crystals than using -3~$^\circ$C/hr. Annealing the as-grown crystals can also be useful~\cite{taen:092502,JPSJ.79.084711,yeh-2009,revisedfetesephase}.  Vacuum annealing can make the Se/Te distribution more homogeneous, thus improving the superconducting volume fraction \cite{taen:092502,JPSJ.79.084711}.  Alternatively, it has been reported that annealing in air reduces the amount of excess iron~\cite{revisedfetesephase}, which can be important for observing bulk superconductivity in samples with large Te concentrations.  Annealing in oxygen has been used to induce bulk superconductivity in the sulphide version, Fe$_{1+y}$Te$_{1-x}$S$_x$, at small $x$  \cite{2011JAP93914M}. It is more challenging to grow crystals of \fts with $x>0.7$~\cite{yeh-2009}. So far, large crystals (on the order of 1 g) have not yet been reported in this range. Even for small crystals or polycrystalline samples of Fe$_{1+y}$Se, impurity phases such as $\alpha$-FeSe, Fe$_7$Se$_8$, Fe$_3$O$_4$, and Fe are often found~\cite{patel:082508,Hu2011}, because of the complex phase diagram, as shown in \fref{fig:fetesebinaryphase}(a)~\cite{mcqueen:014522,fesephasediagram_r}.  Nevertheless, a recent study has reported the growth of phase pure and stoichiometric crystals of superconducting FeSe \cite{Hu2011}.

The major obstacle that limits the understanding of the intrinsic properties of $A_{x}$Fe$_{2-y}$Se$_{2}$ is the difficulty in preparing phase-pure superconducting samples. An initial TEM study on a superconducting sample of nominal composition KFe$_{2-y}$Se$_2$ ($0.2\le y\le0.3$) clearly demonstrates that there is nanoscale phase separation along the $c$ axis---in some regions, there is Fe-vacancy ordering, while in others the vacancies are disordered~\cite{PhysRevB.83.140505}. In a recent ARPES study, a variety of results were interpreted in terms of the coexistence of different phases separated mesoscopically~\cite{2011arXiv1106.3026C}. More effort on this system will be necessary in order to achieve predictable synthesis of single-phase samples.

\section{Superconductivity in iron chalcogenides}
\label{sec:sc}
\subsection{Superconductivity in the 11 system}
\begin{figure}[ht]
\begin{center}
 \includegraphics[width=0.8\columnwidth]{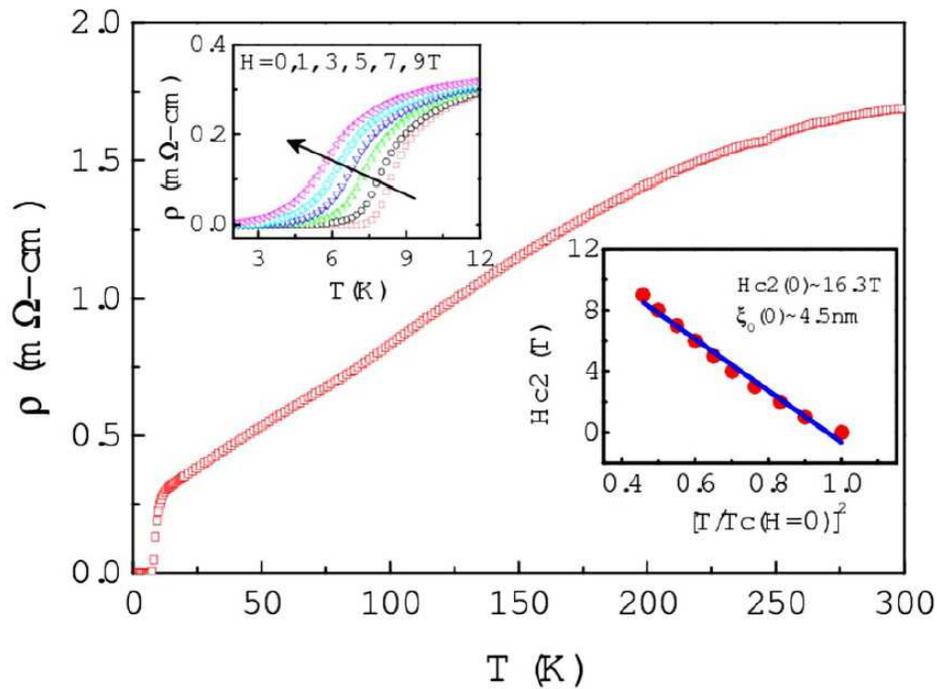}\end{center}
    \caption{Temperature dependence of electrical resistivity ($\rho$) of Fe$_{1.14}$Se. Left inset shows resistivity \vs temperature under different field strength below 12~K. Right inset shows the temperature dependence of the upper critical field. Reprinted from~\cite{hsu-2008}. $\copyright$ 2008 National Academy of Sciences of the U.S.A.}\label{fig:scinfese}
\end{figure}

Superconductivity in the iron chalcogenides was first discovered in Fe$_{1+y}$Se, with zero resistance at $T=8$~K, as shown in \fref{fig:scinfese}~\cite{hsu-2008}. The upper critical field at zero temperature, $\mu_0H_{c2}(0)$, was estimated to be 16.3 T~\cite{hsu-2008}.  The composition of the superconducting phase was initially reported to be FeSe$_{0.88}$ by Hsu {\it et al.} \cite{hsu-2008}, and a similar composition, FeSe$_{0.92}$, was identified by Margadonna {\it et al.} \cite{B813076K}; however, these powder samples showed a significant fraction of secondary phases in diffraction measurements.  McQueen {\it et al.} \cite{mcqueen:014522} followed this work with a careful study of synthesis conditions and composition.  They showed that synthesis from powders generally leads to problems with oxygen contamination, resulting in an overestimate of the Fe content in the sample.  By using large pieces of Fe and Se as starting materials, they found that the stable composition range of Fe$_{1+y}$Se is $0.01\le y\le 0.03$ (consistent with much earlier work on the phase diagram \cite{fesephasediagram_r}), with $T_c=8.5$~K at $y=0.01$ and dropping to zero at 0.03.  Synchrotron x-ray diffraction indicated single-phase samples.  As a further test, Pomjakushina {\it et al.} \cite{PhysRevB.80.024517} prepared a series of samples with different initial compositions, and analyzed the composition by Rietveld refinement of neutron powder diffraction data, obtaining FeSe$_{0.975}$ (or Fe$_{1.025}$Se) for the superconducting phase in these multiphase samples, close to stoichiometry. Finally, there is the impurity-free single crystal of Hu {\it et al.} \cite{Hu2011} that has been analyzed as stoichiometric.  It seems reasonable to conclude that the superconducting composition is close to stoichiometry. To establish convincingly the nature of any defects (Fe interstitials, Se vacancies, etc.), one would need to test for diffuse scattering from a single crystal.  Such measurements on Fe$_{1+y}$Te  have revealed a clear signature of atomic displacements associated with Fe interstitials \cite{Liu2011}.

At room temperature, Fe$_{1+y}$Se has a tetragonal PbO-type structure belonging to the $P4/nmm$ space group as shown in \fref{fig:ironbasedsc}(11). It has an iron-based planar sublattice equivalent to that of the iron pnictides~\cite{hosono_1,hosono_2}.  On cooling below $\sim90$~K, there is a structural transition to an orthorhombic phase (space group $Cmma$) \cite{fesephasetr,B813076K,PhysRevB.80.024517}, resulting in a subtle distortion of the FeSe$_4$ tetrahedra.  The transition temperature is observed to be sensitive to composition \cite{fesephasetr}.

\begin{figure}[ht]
\begin{center}
  \includegraphics[width=\linewidth]{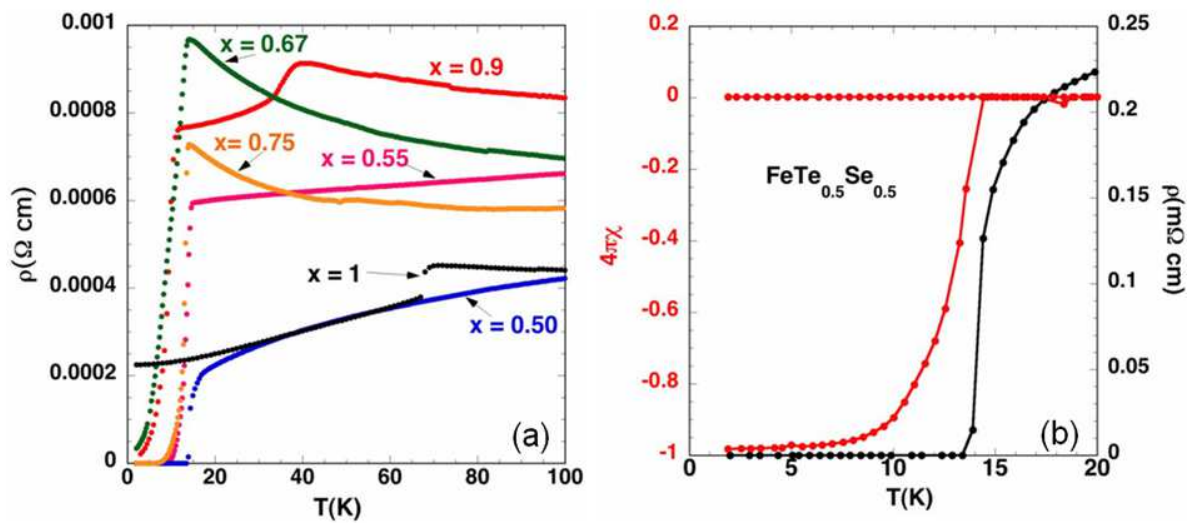}  \end{center}
  \caption{ (a) In-plane electrical resistivity $\rho$ as a function of temperature for FeTe$_x$Se$_{1-x}$. (b) Magnetic susceptibility of a FeTe$_{0.5}$Se$_{0.5}$ crystal measured with $\mu_0H=20$~Oe using zero-field cooling and field-cooling protocols. The resistivity data from the same sample are also shown on the right axis. Reprinted from~\cite{sales:094521}. $\copyright$ 2008 American Physical Society.}\label{fig:sc11}
\end{figure}

Systematic results for the in-plane electrical resistivity of FeTe$_x$Se$_{1-x}$ are shown in \fref{fig:sc11}(a)~\cite{sales:094521}. Fe$_{1+y}$Te is not superconducting; instead, it exhibits coincident magnetic and structural transitions at  $\sim65$~K~\cite{bao-2009,li-2009-79,liupi0topp,PhysRevB.81.094115}. This behaviour is similar to that of the undoped phases of the 1111 and 122 materials~\cite{cruz,qiu:257002,huang:257003,chen:064515,johannes-2009,wilson:184519,kofu-2009-11,zhao,luetkens-2008,drew-2009,rotter-2008-47,chen-2009-85,fang:140508,chu:014506,khasanov:140511,liupi0topp,0295-5075-90-2-27011,spinglass,JPSJ.79.102001,revisedfetesephase}, and it is therefore often referred to as the parent compound for the 11 system~\cite{bao-2009,li-2009-79,mizuguchi-2009-94,mizuguchi-2009-469,fang-2008-78}. By replacing Te with Se, the structural transition temperature is gradually suppressed \cite{PhysRevB.81.094115}. With 6\%\ Se doping, a trace of superconductivity starts to appear, and coexists with the antiferromagnetic order~\cite{liupi0topp}. With increasing Se content, the superconducting volume fraction improves, and $T_c$ also becomes optimized for $x\sim0.5$~\cite{yeh-2008,fang-2008-78,sales:094521}. Superconductivity extends all of the way to $x=1$ in Fe$_{1+y}$Se, as discussed above.

Other combinations of chalcogenides are also possible.  For example, one can substitute S for Se in Fe$_{1+y}$Se$_{1-z}$S$_z$.  For $z=0.2$,  \tc is increased slightly,  to 10~K~\cite{JPSJ.78.074712}. The \tc increase is accompanied by suppression of the tetragonal-to-orthorhombic transition temperature, with the transition already absent at $z=0.1$~\cite{JPSJ.78.074712}. For $z>0.2$, \tc gradually decreases, with no superconductivity observed for 0.4 S~\cite{JPSJ.78.074712}. Doping S into the parent compound Fe$_{1+y}$Te can also induce superconductivity~\cite{mizuguchi-2009-94,2011JAP93914M,PhysRevB.80.214514}, while suppressing the magnetic and structural transitions. The highest \tc in the Fe$_{1+y}$Te$_{1-z}$S$_z$ system is 10~K in FeTe$_{0.8}$S$_{0.2}$~\cite{mizuguchi-2009-94}.

\Fref{fig:sc11}(b) shows the magnetic susceptibility and resistivity of the optimally-doped sample, FeTe$_{0.5}$Se$_{0.5}$~\cite{sales:094521}. It has a \tc of 14~K, highest in the 11 system at ambient pressure, with a superconducting volume fraction close to 100\%~\cite{sales:094521,fieldeffectresonancewen}. One thing to note is that in some of the thin film samples, an enhancement of \tc was observed~\cite{si:052504,APEX.4.053101} (which may be related to strain effects, as pressure can enhance \tc; see \sref{sec:pressure}), while in other cases the opposite effect has been reported~\cite{imai-2009,mkwreview2}. For the optimally doped compound, the estimated $\mu_0H_{c2}(0)$ is $\gtrsim70$~T~\cite{fieldeffectresonancewen}, which makes it comparable with other iron-based and cuprate superconductors~\cite{si:052504,kida-2009,njphysuf1}. Similar to the iron-pnictide superconductors, the anisotropy for the 11 system is small. For instance, in Fe$_{1.03}$Te$_{0.7}$Se$_{0.3}$, the anisotropy coefficient $\gamma$ (ratio of $\mu_0H_{c2}$ measured in and out of plane) is less than 2~\cite{chen:140509}.

\begin{figure}[ht]
\begin{center}  \includegraphics[width=0.8\linewidth]{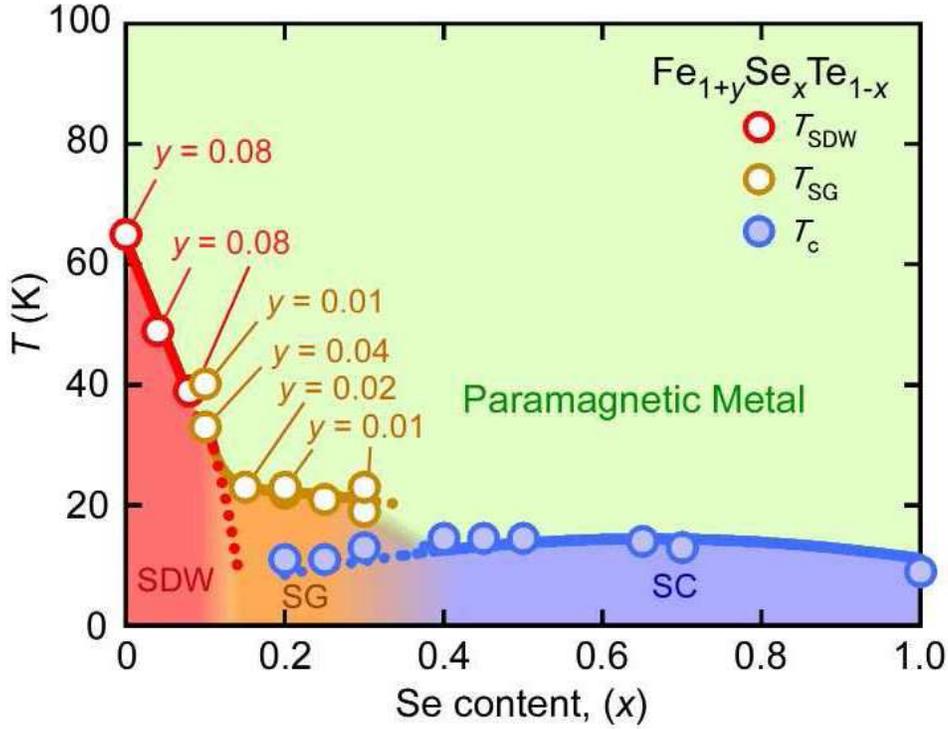}
\end{center}  \caption{Phase diagram of \fts with $y=0$ as a function of $x$ and $T$, constructed from single crystal bulk magnetization data. The nominal Fe content is $y=0$, unless it is specified. SDW, SG and SC stand for long-range antiferromagnetic, spin-glass and superconducting order respectively. Reprinted from \cite{spinglass}. $\copyright$ 2010 Physical Society of Japan.}\label{fig:fetesephasedgm}
\end{figure}

A number of groups have reported similar phase diagrams for \fts \cite{khasanov:140511,liupi0topp,0295-5075-90-2-27011,spinglass,JPSJ.79.102001,revisedfetesephase}; one version, obtained from magnetization measurements, is shown in \fref{fig:fetesephasedgm}. At first glance, this phase diagram is similar to those of the cuprate~\cite{lee:17,birgeneau-2006,RevModPhys.70.897,orenstein,carlsonbook1}
and iron-pnictide superconductors~\cite{cruz,qiu:257002,huang:257003,chen:064515,johannes-2009,wilson:184519,kofu-2009-11,zhao,luetkens-2008,drew-2009,rotter-2008-47,chen-2009-85,fang:140508,chu:014506,khasanov:140511,liupi0topp,0295-5075-90-2-27011,JPSJ.79.102001,revisedfetesephase}. The undoped compound is an antiferromagnet; with doping, the transition temperature is reduced, and superconductivity appears above 0.1 Se doping (see \cite{khasanov:140511,liupi0topp,0295-5075-90-2-27011}), with \tc \vs doping $x$ having a dome shape. However, there are several important differences which need to be pointed out: i) Unlike most other high-\tc superconductors, where ``doping" refers to substitution of elements with a different valence, here the replacement of Te with Se is isovalent. The positive hall coefficient in Fe$_{1.03}$Te$_{0.7}$Se$_{0.3}$ indicates that hole-carriers are dominant with Se doping~\cite{chen:140509}; ii) Superconductivity survives to $x=1$, unlike most other systems where superconductivity disappears with carrier doping above a certain value~\cite{lumsdenreview1}; iii) The material's properties can be tuned not only by doping with Se, but also by adjusting the amount of excess Fe.
For example, a number of studies of \fts with $x\lesssim0.5$ have shown that the presence of Fe interstitials correlates with a reduction in superconducting volume fraction \cite{liu:174509,Roessler2010,Bendele2010,2010arXiv1012.0590R,xudoping11}. Without Se, it appears difficult to reduce Fe content below $y=0.05$, but the minimum interstitial density decreases as Se is added \cite{PhysRevB.81.094115}.
The relationship between excess Fe and the magnetic correlations will be discussed in \sref{sec:neutron}.

\subsection{Pressure study on the 11 system}
\label{sec:pressure}
\begin{figure}[ht]
\begin{center}  \includegraphics[width=0.7\linewidth]{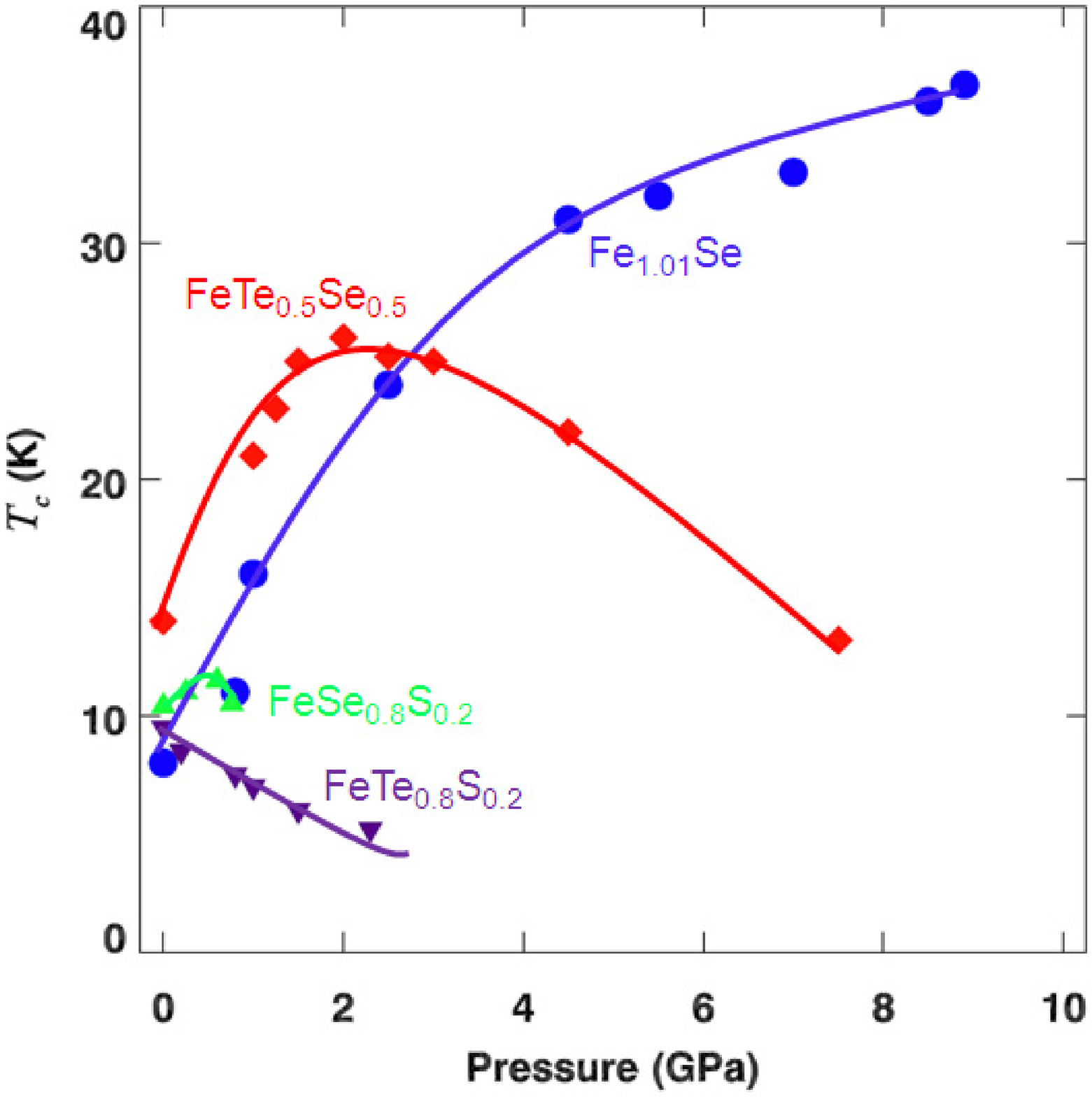}
\end{center}  \caption{Pressure dependence of the \tc for Fe$_{1.01}$Se,~\cite{medvedev-2009-8} FeTe$_{0.5}$Se$_{0.5}$,~\cite{JPSJ.78.063705} FeSe$_{0.8}$S$_{0.2}$,~\cite{Mizuguchi2010S353} and  FeTe$_{0.8}$S$_{0.2}$.~\cite{pressurefetes}}\label{fig:pressuresum}
\end{figure}

In \fref{fig:pressuresum} we plot the pressure dependence of \tc for four samples. For Fe$_{1+y}$Se, there is a strong pressure dependence~\cite{medvedev-2009-8,margadonna:064506,JPSJ.78.063704,tissen:092507}. Applying a pressure of 1.48~GPa raises \tc  to 27~K and $\mu_0H_{c2}(0)$ to 72~T~\cite{mizuguchi:152505}. With further increase of pressure, \tc appears to have a maximum of 37~K for pressures above 4~GPa~\cite{medvedev-2009-8,margadonna:064506,JPSJ.78.063704}. While \tc increases with pressure, the superconducting volume fraction is reported to decrease rapidly for pressure $(P)>1.5$~GPa, suggesting significant inhomogeneity~\cite{miyoshi-2009-78}. Margadonna \et~\cite{margadonna:064506} also found that the \tc-$P$ curve coincides with anomalies in the pressure evolution of the interlayer spacing, suggesting an intimate relationship between structure and superconductivity.

In the compound with mixed Se and Te, the pressure dependence of \tc is similar to that in Fe$_{1+y}$Se, though less dramatic~\cite{JPSJ.78.063705,pressurefetese,Mizuguchi2010S353}. In FeTe$_{0.5}$Se$_{0.5}$, the \tc onset increases rapidly from 13.5 to 26.2~K upon application of a pressure of 2 GPa~\cite{JPSJ.78.063705}. Above 2~GPa, \tc decreases linearly, and a metallic (non-superconducting) ground state is observed at 14~GPa~\cite{JPSJ.78.063705}. In other Se-doped samples, \eg, FeTe$_{0.43}$Se$_{0.57}$ and FeTe$_{0.75}$Se$_{0.25}$, the pressure effects are reported to be smaller~\cite{pressurefetese,Mizuguchi2010S353,JPSJ.79.102001}.

For FeSe$_{0.8}$S$_{0.2}$, a study with pressure up to 0.76~GPa shows only a weak effect on \tc, with an indication of a decline in \tc at higher pressure~\cite{Mizuguchi2010S353}. Contrary to the pressure enhancement of \tc discussed so far, application of pressure to Fe$_{1+y}$Te$_{1-x}$S$_x$ causes \tc to decrease monotonically, as shown in \fref{fig:pressuresum}~\cite{pressurefetes}. One might guess that applying pressure to the parent compound of the 11 system, Fe$_{1+y}$Te, could make it superconducting; however, superconductivity was not observed in Fe$_{1.09}$Te with hydrostatic pressure up to 2.5~GPa~\cite{JPSJ.78.083709}, although the structural transition temperature was reduced~\cite{JPSJ.78.083709}, and the lattice collapsed~\cite{PhysRevB.80.144519}.  In contrast, superconductivity with \tc up to 13~K has indeed been reported for thin film samples of FeTe under tensile stress~\cite{PhysRevLett.104.017003}.

\subsection{Doping with 3$d$ transition metals into \fts}
One can also change the properties of the 11 system by doping with 3$d$ transition metals (Mn, Co, Ni, Cu, etc). There are several initial studies on Fe$_{1+y}$Se~\cite{danielfeseco,williams-2009-21,PhysRevB.82.104502,mkwreview1}. Thomas \et~\cite{danielfeseco} have found that 2.5\%\ Co reduces \tc to below 4~K, and no superconductivity is observed for Co doping at and above 5\%, though the system is still metallic with 20\%\ Co ~\cite{JPSJ.78.074712}. Cu seems to have a larger effect in reducing \tc.  As little as 1.5\%\ Cu substitution results in the absence of a diamagnetic response~\cite{williams-2009-21}.  More interestingly, with increased Cu doping (4\%), the system evolves into an insulator~\cite{williams-2009-21,PhysRevB.82.104502}. Analysis of the temperature dependence of the resistivity suggests that the sample with 10\%\ Cu doping is a Mott insulator~\cite{PhysRevB.82.104502}. This result has interesting implications for the nature of electronic correlations in the iron chalcogenides. Ni doping is also found to suppress \tc~\cite{Zhang20091958,JPSJ.78.074712}, but does not drive the system to an insulating state for Ni concentations up to 20\%~\cite{JPSJ.78.074712}.  Mn substitution has little effect on \tc, and the system remains metallic and superconducting with Mn doping up to 5.5\%~\cite{PhysRevB.82.104502}.

Similar studies have been carried out in \fts with $x=0.35$~\cite{2010arXiv1010.4217G} and 0.5~\cite{fetesenico1}.  It is found that Co, Ni, and Cu all suppress \tc~\cite{2010arXiv1010.4217G}, and a metal-to-insulator transition is induced with both Co and Ni doping in FeTe$_{0.5}$Se$_{0.5}$ samples~\cite{fetesenico1}. There is also a report on Ni-doped Fe$_{1.1}$Te, which reveals that the lattice constants decrease with increasing Ni content~\cite{feteni1}.

\subsection{Superconductivity in $A_{x}$Fe$_{2-y}$Se$_{2}$ ($A=$ K, Rb, Cs, Tl, Tl/K, Tl/Rb)}
\begin{figure}[ht]
\begin{center}  \includegraphics[width=0.7\linewidth]{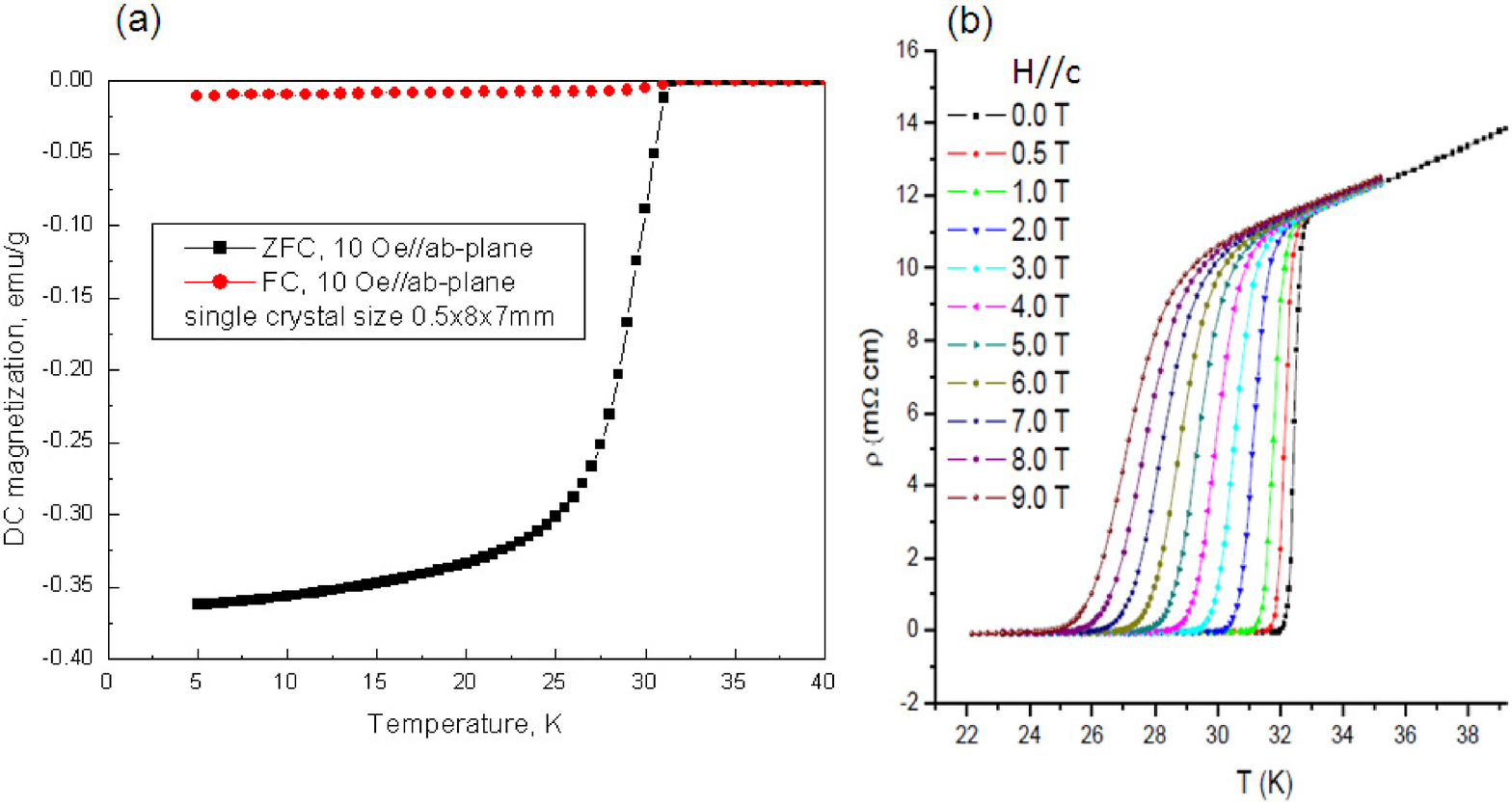}
\end{center}  \caption{(a) Temperature dependence of the magnetization for a K$_{0.8}$Fe$_2$Se$_2$ single crystal. (b) Resistivity measured in the $a$-$b$ plane with field applied along the $c$ axis~\cite{kfesecrystalgrowth}.}\label{fig:kfesemrho}
\end{figure}

At the end of the year 2010, Guo \et~\cite{scinkfese} reported that superconductivity with \tc above 30~K had been achieved in a ternary iron chalcogenide, K$_{0.8}$Fe$_2$Se$_2$ by intercalating the metal K in between FeSe layers. In \fref{fig:kfesemrho} we show results for the temperature dependence of the magnetization and in-plane resistivity for a K$_{0.8}$Fe$_2$Se$_2$ single crystal~\cite{kfesecrystalgrowth}. The \tc was $\sim$~32.5~K, the highest among all iron-chalcogenide superconductors at ambient pressure. This new superconductor with such a high \tc soon ignited another wave of intensive activity. To-date, besides K$_x$Fe$_{2-y}$Se$_{2}$~\cite{scinkfese,kfese2,2011arXiv1101.5117Z,2011arXiv1102.1931H,2011arXiv1103.0098L,2011arXiv1102.1010L,2011arXiv1101.4967T,2011arXiv1102.1381Y,2011arXiv1103.0507M,2011arXiv1101.4572K}, superconductivity has been found in a series of compounds with the formula $A_{x}$Fe$_{2-y}$Se$_{2}$ where $A=$ Rb, Cs, Tl, Tl/K, or Tl/Rb~\cite{rbfese1,2011arXiv1102.3380P,2011arXiv1101.5670L,2011arXiv1102.2464S,2011arXiv1102.1919P,2011arXiv1102.2783L,2011arXiv1101.1873S,kcsfese1,csfese1,2011arXiv1102.2783L,tlrbfese1,2010arXiv1012.5236F}, and the highest \tc achieved so far is $\sim33$~K. By substituting S (K$_x$Fe$_{2-y}$Se$_{2-z}$S$_z$), \tc does not decrease for $z$ up to 0.4~\cite{2011arXiv1102.2434L,2011arXiv1101.5327L}, but with $z=1.6$, superconductivity is completely suppressed~\cite{2011arXiv1102.2434L}. In contrast, Co doping in K$_{0.8}$Fe$_2$Se$_2$ has a very strong effect. With 0.5\% Co, \tc is reduced to 5~K~\cite{2011arXiv1102.3506Z}.  A number of the newly discovered $A_{x}$Fe$_{2-y}$Se$_{2}$ superconductors and their $T_c$'s are summarized in \tref{tbl:scikfese}, from which one can see that the \tc is not very sensitive to the stoichiometry, except for the Co-doped sample.

\begin{table}
\caption{Summary of \tc for the newly discovered $A_{x}$Fe$_{2-y}$Se$_{2}$ superconductors.}
\label{tbl:scikfese}
\begin{tabular}{llll}
\br
    Material & \tc (K) & Material & \tc (K)\\
  \mr
  K$_{0.8}$Fe$_2$Se$_2$~\cite{scinkfese,kfese2,2011arXiv1102.1010L,2011arXiv1101.4967T,2011arXiv1101.5670L,2011arXiv1101.4572K} &30-33&  K$_{0.8}$Fe$_{2}$Se$_{0.8}$S$_{1.2}$~\cite{2011arXiv1102.2434L} &18.2\\
  K$_{0.8}$Fe$_{1.6}$Se$_2$~\cite{2011arXiv1102.0830B} &32&  K$_{0.8}$Fe$_{1.995}$Co$_{0.005}$Se$_{2}$~\cite{2011arXiv1102.3506Z} &5\\
  K$_{0.8}$Fe$_{1.76}$Se$_2$~\cite{2011arXiv1103.0507M} &32&  Rb$_{0.83(2)}$Fe$_{1.70(1)}$Se$_{2}$~\cite{2011arXiv1102.3380P} &31.5\\
  K$_{0.65}$Fe$_{1.41}$Se$_2$~\cite{2011arXiv1102.2217W} &32&Rb$_{0.8}$Fe$_{2}$Se$_{2}$~\cite{2010arXiv1012.5637L,2011arXiv1101.5670L} &32\\
  K$_{0.85}$Fe$_{1.83}$Se$_{2.09}$~\cite{2011arXiv1103.0059R} &30&  Rb$_{0.88}$Fe$_{1.81}$Se$_{2}$~\cite{rbfese1} &31.5\\
  K$_{0.85}$Fe$_{2}$Se$_{1.8}$~\cite{2011arXiv1101.1234Y} &32.6&  Cs$_{0.86}$Fe$_{1.66}$Se$_{2}$~\cite{2011arXiv1101.1234Y,kcsfese1,2011arXiv1101.5670L} &30\\
  K$_{0.86}$Fe$_{2}$Se$_{1.82}$~\cite{2011arXiv1101.1234Y,kcsfese1} &31.1&  Cs$_{0.83(2)}$Fe$_{1.71(1)}$Se$_{1.96}$~\cite{2011arXiv1102.1919P} &28.5\\
  K$_{0.7}$Fe$_{1.8}$Se$_2$~\cite{2011arXiv1102.1381Y} &28& Cs$_{0.8}$Fe$_{2}$Se$_{2}$~\cite{2011arXiv1102.2464S} &32 \\
  K$_{0.86}$Fe$_{1.73}$Se$_2$~\cite{2011arXiv1103.0098L} &25&  TlFe$_{1.7}$Se$_2$~\cite{2010arXiv1012.5236F} & 22.8\\
  K$_{0.86}$Fe$_{1.84}$Se$_{2}$~\cite{2011arXiv1102.1931H} &30& Tl$_{0.58}$K$_{0.42}$Fe$_{1.72}$Se$_{2}$~\cite{tlrbfese1,2011arXiv1101.4556M} &32\\
  K$_{0.83(1)}$Fe$_{1.66(1)}$Se$_{2}$~\cite{2011arXiv1102.3380P} &29.5&  Tl$_{0.64}$K$_{0.36}$Fe$_{1.83}$Se$_{2}$~\cite{2010arXiv1012.5236F} &31\\
  K$_{0.8}$Fe$_{1.7}$SeS~\cite{2011arXiv1102.3505G} &25&  Tl$_{0.63}$K$_{0.37}$Fe$_{1.78}$Se$_{2}$~\cite{hongdingarpeskfese} &29\\
  K$_{0.8}$Fe$_{2}$Se$_{1.4}$S$_{0.4}$~\cite{2011arXiv1101.5327L} &32.8&Tl$_{0.47}$Rb$_{0.34}$Fe$_{1.63}$Se$_2$~\cite{2011arXiv1102.3888M} & 33\\
  K$_{0.8}$Fe$_{2}$Se$_{1.6}$S$_{0.4}$~\cite{2011arXiv1102.2434L} &33.2&Tl$_{0.58}$Rb$_{0.42}$Fe$_{1.72}$Se$_2$~\cite{tlrbfese1}&32\\
  K$_{0.8}$Fe$_{2}$Se$_{1.2}$S$_{0.8}$~\cite{2011arXiv1102.2434L} &24.6 &Tl$_{0.4}$Rb$_{0.4}$Fe$_{2-y}$Se$_2$~\cite{2011arXiv1102.2783L} & 31.8\\
  \br
\end{tabular}
\end{table}

There have been some pressure studies on these systems. In a K$_x$Fe$_2$Se$_2$ sample, which had a \tc of 32.6~K, \tc decreased with increasing pressure, while in a sample with a \tc of 31.1~K, a dome-shaped \tc-$P$ curve was observed, with a maximum \tc of 32.7~K at 0.48~GPa~\cite{2011arXiv1101.1234Y}. A similar dome shape was observed for Cs$_x$Fe$_2$Se$_2$,   with \tc maximized to 31.1~K at $P=0.82$~GPa, compared to 30~K at ambient pressure~\cite{2011arXiv1101.1234Y}. In a sample with nominal composition of K$_{0.8}$Fe$_2$Se$_2$, which had a \tc of 33~K, it was found that, although the onset \tc increased with pressure, the temperature where the resistivity reached zero decreased~\cite{2011arXiv1101.0896K}. In all of the samples studied, the large normal-state resistance was greatly reduced by pressure~\cite{2011arXiv1101.1234Y,2011arXiv1101.0896K,2011arXiv1102.2464S,2011arXiv1101.0092G}. In Cs$_{0.8}$Fe$_2$Se$_2$, the resistance decreased by two orders of magnitude, concomitant with a sudden \tc reduction at $\sim$~8~GPa~\cite{2011arXiv1102.2464S}.

In the normal state of these materials, the resistivity-temperature curve exhibits a very pronounced hump~\cite{scinkfese}, which is believed to be irrelevant to superconductivity~\cite{2011arXiv1101.5670L,2011arXiv1101.0789W,2011arXiv1101.1234Y}. However, in K$_{0.8}$Fe$_{1.7}$Se$_2$, Guo \et~\cite{2011arXiv1101.0092G} found an interesting correlation between superconductivity and the hump: with increasing pressure, the magnitude of the hump went down as \tc did, and the hump disappeared at $P=9.2$~GPa, in coincidence with the complete suppression of the superconductivity. By normalizing the resistance to the room-temperature value, Seyfarth \et~\cite{2011arXiv1102.2464S} found that the hump still existed at $P=9$~GPa, at which pressure the sample was not superconducting; based on this, they concluded that the hump was not related to the superconductivity.

These newly discovered superconductors are particularly interesting and fundamentally important in part due to their electronic structures. First-principle calculations have shown that the band structure in this new system is quite different from that of other iron-based superconductors~\cite{PhysRevLett.106.087005,2011arXiv1102.2215Y,2010arXiv1012.5164S,2011arXiv1101.0051N,2010arXiv1012.5621C,2011arXiv1101.0533C}. In particular, the band near the Brillouin zone center $\Gamma$ point sinks well below the Fermi level~\cite{PhysRevLett.106.087005,2011arXiv1102.2215Y,2010arXiv1012.5164S,2011arXiv1101.0051N,2010arXiv1012.5621C,2011arXiv1101.0533C}.
Experimentally, this prediction has been verified by several initial angle-resolved photoemission spectroscopy (ARPES) studies, which show that there is no hole pocket near the $\Gamma$ point~\cite{kfesearpes1,2010arXiv1012.6017Q}. Instead, there are electron pockets near both the $\Gamma$ and M (Brillouin zone corner) points~\cite{2011arXiv1101.4556M,hongdingarpeskfese,2011arXiv1102.1057Z}.  Although this result does not rule out the possibility that interband scattering between the electron pockets at $\Gamma$ and M could be important to the superconductivity~\cite{2011arXiv1101.4988M}, the sign-changed $s_\pm$ pairing symmetry which has been suggested in other iron-based superconductors is not likely to apply here~\cite{mazin-2009,mazin:057003,kuroki-2008}. Several alternatives, including $d$-wave~\cite{epl9357003,2011arXiv1101.4988M}, $s$-wave~\cite{2011arXiv1101.4462Z,2011arXiv1103.1902}, and $s_{++}$-wave~\cite{2011arXiv1102.4049S} superconductivity have been proposed theoretically. A number of ARPES~\cite{2011arXiv1101.4556M,hongdingarpeskfese,2011arXiv1102.1057Z,kfesearpes1,2010arXiv1012.6017Q}, nuclear magnetic resonance (NMR)~\cite{2011arXiv1101.4967T,2011arXiv1101.3687M,2011arXiv1101.4572K}, and specific heat measurements~\cite{2011arXiv1101.5117Z} have been carried out and all of them indicate that the gap is nodeless.

Another interesting feature which distinguishes the $A_x$Fe$_{2-y}$Se$_2$ superconductors from other iron-based superconductors is that their superconductivity is in close vicinity to an insulating state, similar to the case of cuprate superconductors~\cite{lee:17,birgeneau-2006,RevModPhys.70.897,orenstein,carlsonbook1}, but not to that of other iron-based superconductors~\cite{dcjohnstonreview}. Interestingly, it is reported that the insulating K$_x$Fe$_{2-y}$Se sample becomes superconducting by annealing and quenching, and after a few days at room temperature, it is insulating again, showing that the insulating and superconducting states are reversible~\cite{2011arXiv1103.1347}. Based on these observations, it is suggested that disordering of Fe vacancies plays a key role in rendering the superconductivity~\cite{2011arXiv1103.1347}. The experimentally observed insulating ground state~\cite{2011arXiv1102.3674B,2010arXiv1012.5236F,2011arXiv1102.2434L,2011arXiv1101.5670L,2011arXiv1102.3505G} has stimulated several theoretical calculations~\cite{PhysRevLett.106.087005,2011arXiv1102.2215Y,2011arXiv1101.3307Y}. Two possible origins of the insulating behaviour have been proposed which are to be verified experimentally: the system might be a Mott insulator~\cite{2011arXiv1101.3307Y,2011arXiv1102.3505G,2011arXiv1101.0533C,2011arXiv1103.4599F} (the Mott behaviour may have a complex origin where both Fe vacancies and 3$d$ electron-electron correlations play roles as suggested in \cite{2011arXiv1101.0533C}), or it might be a band insulator resulting from the electronic structure reconstruction due to ordered Fe vacancies~\cite{PhysRevLett.106.087005,2011arXiv1102.2215Y}.

\subsection{Summary}
To summarize this section, it is found that the parent compound for the 11 system, Fe$_{1+y}$Te, is non-superconducting, and superconductivity is achieved by replacing Te with Se or S. The optimal superconductivity with \tc of 14~K is found to be in samples with $y=0$ and $x\approx0.5$. Fe$_{1+y}$Se is where superconductivity was initially discovered for the 11 system; it has a \tc of 8~K. The \tc increases to $\sim$~37~K upon application of hydrostatic pressure. The superconductivity seems to be vulnerable to doping with 3$d$ transition metals, which can tune the system to a (possibly Mott) insulating regime. More recently, a series of ternary iron-chalcogenide compounds with the chemical formula $A_x$Fe$_{2-y}$Se were found to be superconducting with \tc up to 33~K, the highest among all Fe-Se-based superconductors at ambient pressure.

\section{Magnetic correlations}
\label{sec:neutron}
\subsection{Magnetic order}
\label{subsec:magneticorder}
\subsubsection{\fts system}

\begin{figure}[ht]
\begin{center}
  \includegraphics[width=0.9\linewidth]{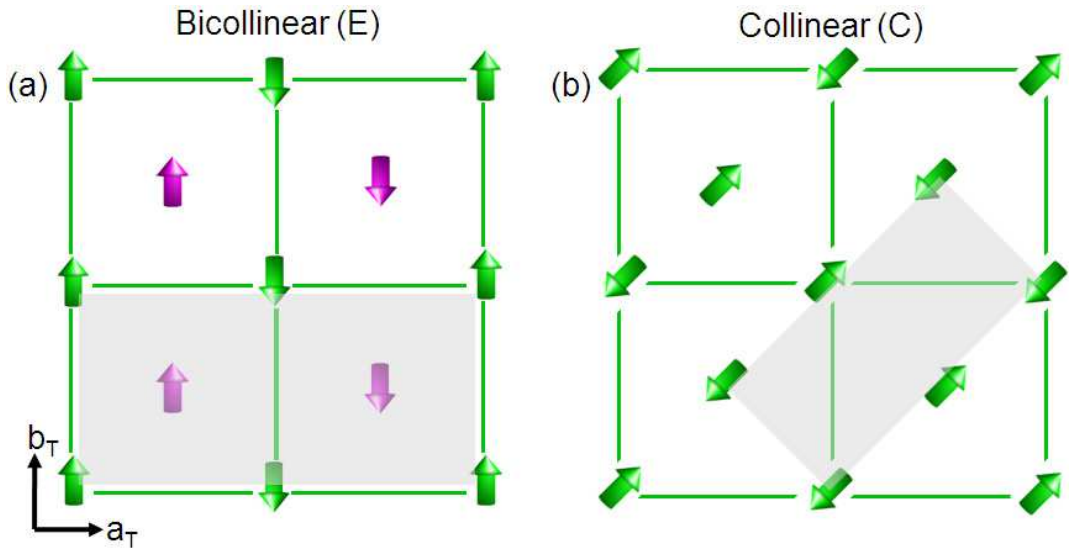}\end{center}
  \caption{(a) Schematic for the bicollinear (E-type) in-plane spin structure in tetragonal notation of the 11 compound. (b) Collinear (C-type) magnetic structure for iron pnictides. Shadow represents the magnetic unit cell.}\label{fig:magneticstructure}
\end{figure}

The magnetic order in \fts has attracted considerable attention because of its rich physics.
An initial band-structure calculation predicted that the Fermi-surface topology in this system should be similar to that of the iron pnictides~\cite{subedi-2008-78}; this has been confirmed by ARPES measurements~\cite{xia037002,PhysRevLett.105.197001}. From the Fermi-surface-nesting picture~\cite{subedi-2008-78}, one would expect that this system would have collinear (C-type) spin-density-wave (SDW) order with an in-plane wave vector (0.5,\,0.5), assuming a unit cell containing two irons as shown in \fref{fig:magneticstructure}(b). However, several decades ago, Fruchart \et \cite{structure6} determined that the magnetic ordering vector is (0.5,\,0) in Fe$_{1.125}$Te. This result has been confirmed by Bao \et~\cite{bao-2009} in Fe$_{1.075}$Te, and by Li \et~\cite{li-2009-79} in Fe$_{1.068}$Te; each of these has a bicollinear (E-type) spin structure as shown in \fref{fig:magneticstructure}(a). Even more surprising is the fact that ARPES measurements have observed no SDW nesting instability along (0.5,\,0), although there is a weak hole pocket around the X point~\cite{xia037002,Physics.2.59}. Clearly, a simple nesting mechanism cannot account for these experimental results.  More recent first-principle calculations have identified the role of local-moments and the importance of Hund's exchange coupling, and have provided better agreement with the experimental observations~\cite{han:067001,ma-2009,loalm1,johannes-2009,fang-2009,moon057003,weiguounified}.

The magnetic order in Fe$_{1+y}$Te is long-ranged, with a maximum moment size of $\sim2.5\ \mu_{\rm B}$/Fe for $y=0.05$ \cite{PhysRevB.81.094115}; the moment size decreases for larger $y$~\cite{structure6,li-2009-79,bao-2009}.  The moment size is significant compared to that in iron-pnictide antiferromagnets, but it is small compared to the effective moment estimated from the magnetic susceptibility in the paramagnetic phase \cite{Zaliznyak2011}.  The moment is found to be aligned mostly along the $b$ axis as shown in \fref{fig:magneticstructure}(a)~\cite{li-2009-79,bao-2009,PhysRevB.81.094115}.  Bao \et~\cite{bao-2009} have shown that the order can be commensurate or incommensurate depending upon the amount of excess Fe---with more Fe, the incommensurability is larger. Upon doping with Se, the order is suppressed, with a reduced ordering temperature (from 65 to $\lesssim30$~K with 0.1 Se)~\cite{spinglass}, reduced size of the ordered magnetic moment (from 2.1 to 0.27 $\mu_{\rm B}$/Fe)~\cite{liupi0topp}, and shorter correlation length (the magnetic peak is resolution limited in the parent compound, while with 0.25 Se doping, the order is short-ranged with a correlation length of $\sim$~4~\AA~\cite{wen:104506,bao-2009}). For Se content above 0.15, there is a phase where spin-glass order and superconductivity coexist~\cite{wen:104506,spinglass,khasanov:140511,liupi0topp,PhysRevB.81.094115}. In two single crystal samples, Fe$_{1.07}$Te$_{0.75}$Se$_{0.25}$ and FeTe$_{0.7}$Se$_{0.3}$, both exhibiting short-range incommensurate order below 40~K, it is found that with increasing excess Fe, the incommensurability is larger, similar to the case in the parent compound~\cite{wen:104506,bao-2009}. In FeTe$_{0.7}$Se$_{0.3}$, which has more Se and less Fe than the Fe$_{1.07}$Te$_{0.75}$Se$_{0.25}$ sample, the spin-glass order is depressed, with weaker peak intensity and shorter correlation length, while superconductivity is enhanced, with higher \tc and superconducting volume fraction~\cite{wen:104506}. Interestingly, a magnetic peak is only observed on one side of the commensurate wave vector (0.5,\,0), [\ie (0.5-$\delta$,\,0) and not (0.5$\pm\delta$,\,0) with $\delta$ being the incommensurability]; this is likely a result of an imbalance of ferromagnetic/antiferromagnetic correlations between neighbouring spins~\cite{wen:104506}.  (A closely related picture of spin clusters is indicated by a recent study of Fe$_{1.1}$Te \cite{Zaliznyak2011}.) With further increase of the Se concentration, the superconductivity optimizes with $x\approx0.5$, and no static magnetic order is observed (for $y\sim0$)~\cite{fieldeffectresonancewen,khasanov:140511,liupi0topp}.

Next, we turn to the intriguing case of superconducting Fe$_{1+y}$Se.  Although this system exhibits a symmetry-lowering structural transition on cooling through $\sim90$~K \cite{fesephasetr,B813076K,PhysRevB.80.024517}, measurements with local probes such as M\"ossbauer spectroscopy \cite{fesephasetr,mcqueen:014522} and  $^{77}$Se NMR \cite{imai-2009-102} indicate an absence of static magnetic order.  This is in sharp contrast with the case of Fe$_{1+y}$Te; nevertheless, NMR measurements indicate that spin fluctuations increase on cooling towards \tc \cite{imai-2009-102}.   In studies of a sample that showed an increase of $T_c$ to 37~K under pressure, M\"ossbauer measurements indicated an absence of magnetic order up to $\sim30$~GPa~\cite{medvedev-2009-8}.   NMR measurements on the same material found that spin fluctuations are enhanced under pressure, along with the superconductivity~\cite{imai-2009-102}.  In contrast,  Bendele \et~\cite{PhysRevLett.104.087003} reported evidence for short-range magnetic order for $P\gtrsim0.8$~GPa based on muon-spin rotation ($\mu$SR) measurements; however, this sample showed a much more modest impact of pressure on the superconductivity, with a maximum \tc of 13~K at 0.7~GPa.  M\"{o}ssbauer studies have shown that doping Cu into Fe$_{1+y}$Se induces a local magnetic moment, with the size of the moment increasing with Cu doping while superconductivity is suppressed~\cite{williams-2009-21}. Spin-glass behavior has been inferred from magnetization measurements on these Cu-doped samples~\cite{williams-2009-21}.

The magnetic and superconducting phases in \fts are summarized in \fref{fig:fetesephasedgm}. Overall, the behaviour is consistent with the common belief that static magnetic order competes with superconductivity, while spin fluctuations promote it.

\subsubsection{$A_x$Fe$_{2-y}$Se$_2$ materials}
\label{subsec:afesem}
Antiferromagnetic order was found in TlFe$_{2-y}$Se$_2$ by neutron and M\"{o}ssbauer measurements ~\cite{afmafese1,afmafese2} long before the recent discovery of superconductivity in the related $A_x$Fe$_{2-y}$Se$_2$ compounds. Bao \et~\cite{2011arXiv1102.0830B} have recently used neutron powder diffraction to determine the crystalline and magnetic structure for  K$_{x}$Fe$_{2-x}$Se$_2$.  At high temperature, the crystal structure is equivalent to that of the 122 materials. At lower temperature, the Fe vacancy order drives the system to an $I4/m$ phase with an enlarged unit cell of $\sqrt5\times\sqrt5\times1$~\cite{C1SC00070E}, which contains a pair of the Fe-Se layers related by inversion symmetry~\cite{2011arXiv1102.0830B}. It is found that $T_N$ is as high as 559~K and that the magnetic moment is as large as 3.31~$\mu_{\rm B}$/Fe, forming a collinear antiferromagnetic structure with (101) being the propagation wave vector~\cite{2011arXiv1102.3674B}. The moment points mainly along the $c$ axis~\cite{2011arXiv1102.0830B}. Such a collinear magnetic structure has been reproduced by several theoretical works~\cite{PhysRevLett.106.087005,2011arXiv1102.2215Y,2011arXiv1102.1344C,2011arXiv1103.4599F}. It is intriguing that the magnetic order occurs in a tetragonal phase without breaking the four-fold symmetry, although the presence of Fe vacancies may reduce any magnetic frustration \cite{2011arXiv1103.4599F}. More surprisingly, it has been reported that superconductivity coexists with the strong antiferromagnetic order, both in the neutron study of K$_{0.8}$Fe$_2$Se$_2$~\cite{2011arXiv1102.0830B}  and in the $\mu$SR study of Cs$_{0.8}$Fe$_2$Se$_{1.96}$~\cite{2011arXiv1101.1873S}. There have been many further observations of the coexistence of superconductivity and antiferromagnetic order~\cite{2011arXiv1102.2882Y,2011arXiv1102.3380P,2011arXiv1103.0059R,2011arXiv1102.2783L,2011arXiv1102.1919P}.
However, due to the difficulty in obtaining samples with high superconducting volume fractions, as well as evidence pointing to phase separations~\cite{PhysRevB.83.140505,2011arXiv1104.2008S,2011arXiv1106.3026C,2011arXiv1103.1347,2011arXiv1101.5670L,2011arXiv1107.0412R,phasesep122_1}, it is not yet clear whether the superconductivity and antiferromagnetic order coexist locally or in different regions of a sample.

\subsection{Spin excitations in \fts}
\label{subsec:spinexcitation}
\subsubsection{Spin dynamics near (0.5,\,0.5)}
\label{subsec:resonance}
\begin{figure}
\centering\includegraphics[width=.9\linewidth]{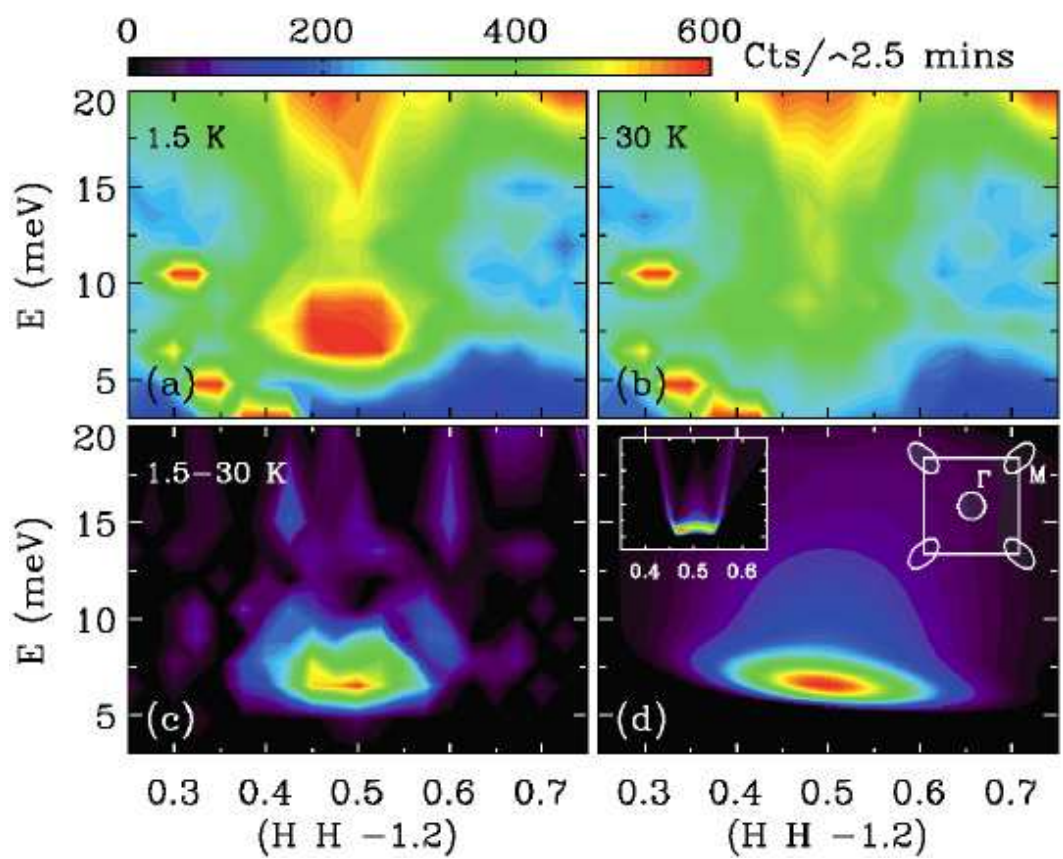}
\caption{ Spin resonance in FeTe$_{0.6}$Se$_{0.4}$. (a) and (b) shows the spin excitation spectrum as a function of $\bf Q$ and energy at 1.5 and 30 K respectively.  (c) shows the resonance intensity as determined by the difference between the 1.5 and 30 K spectra.  (d) Calculated intensity difference from a simplified two-band model with resolution convolved. The left inset shows the resolution-free calculated intensity and the
right inset is the normal-state Fermi surface employed in the calculation. Reprinted from \cite{qiu:067008}. $\copyright$ 2009 American Physical Society.}\label{feteseresonance}
\end{figure}

The ``resonance" mode in the magnetic excitations, which is defined as the energy at which there is a significant increase in spectral weight when the system enters the superconducting phase, has been the subject of extensive measurements. The resonance is predicted to occur at a particular wave vector if it connects portions of the Fermi surface that have opposite signs of the superconducting gap function. Therefore, observations of the resonance may provide important information relevant to the pairing symmetry~\cite{maier:020514,maier:134520,PhysRevLett.75.4126,PhysRevB.64.172508}. (Note that these analyses make the assumption that the magnetism is due to the same electrons that participate in the superconductivity, while the validation of the assumption is still under debate.~\cite{Physics.2.59,2011arXiv1103.5073Z,shcouplingprb}) Resonance excitations have been observed in a number of iron-based superconductors~\cite{christianson-2008-456,resonance1111}, consistent with the presumed gap-sign change between the hole and electron pockets~\cite{mazin-2009}. In \fts, despite the fact that the magnetic ordering wave vector is different from that of iron pnictides by 45$^\circ$, the resonance excitation was observed to be near the same (0.5,\,0.5) wave vector by several groups~\cite{qiu:067008,mook-2009-2,incomfetese1,shcouplingprb,fieldeffectresonancewen,PhysRevLett.105.157002,fetesecomp142202}.
Compared to the magnetic excitation spectrum above \tc, the low-temperature spectral weight is greatly enhanced around (0.5,\,0.5) and the resonance energy of $\sim$~6.5~meV, as shown in \fref{feteseresonance}(c)~\cite{qiu:067008}. The resonance energy corresonds to $\sim 5k_{\rm B}T_c$, similar to the situation in other high-\tc superconductors~\cite{lumsdenreview1}.  Accompanying the resonance, there is a spin gap with an energy of $\sim$~4~meV; the intensity below the spin gap is shifted to the resonance in the superconducting state~\cite{qiu:067008,fieldeffectresonancewen}.

\begin{figure}[ht]
\begin{center}\includegraphics[width=0.8\linewidth]{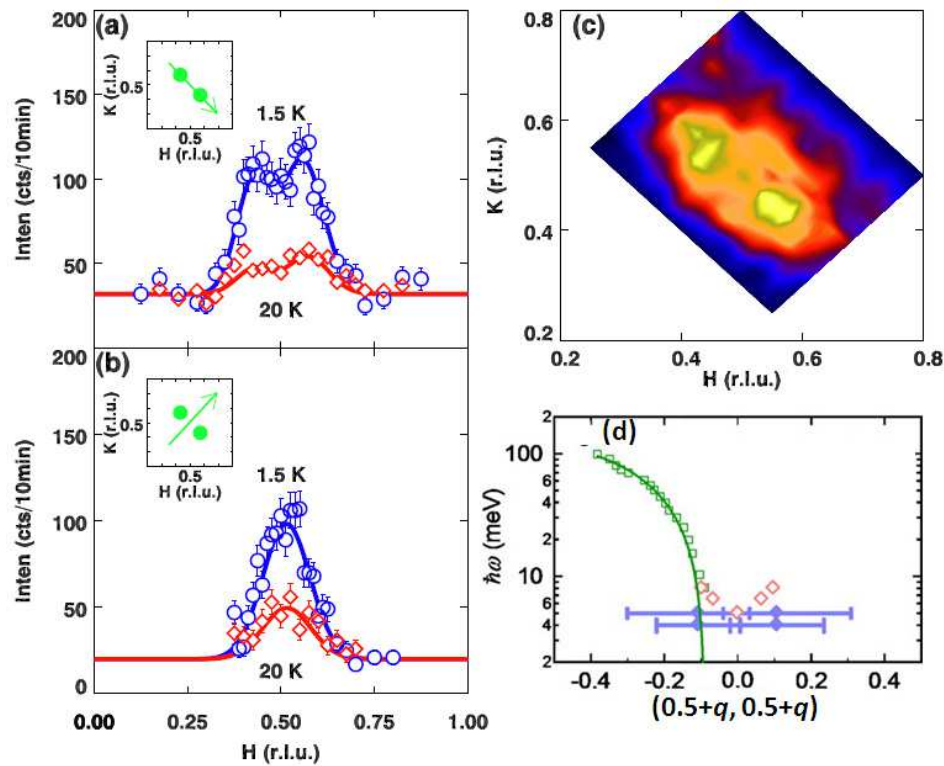}
\end{center}\caption{Incommensurate resonance in {\bf Q}, peaking at ($0.5\pm\delta$,$0.5\mp\delta$), transversely to (0.5,\,0.5), at $\hbar\omega=6.5$~meV. (a) and (b) show the {\bf Q} scans at 1.5 and 20~K, below and above \tc, with scan directions shown in the left insets. Upper right inset is obtained by subtracting 20-K data from 1.5-K data. (c) is the plot of the {\bf Q} dependence of the intensity at 6.5~meV at 1.5~K. (d) shows the dispersion from (0.5,\,0.5) in semi-log scale. (a)-(c) reprinted from \cite{wenphdthesis}. (d) reprinted from \cite{shcouplingprb}; $\copyright$ 2010 American Physical Society.} \label{fig:incomresonance}
\end{figure}

Interestingly, it is found that the resonance is incommensurate in {\bf Q}, peaking at ($0.5\pm\delta$,$0.5\mp\delta$), in a direction transverse to (0.5,\,0.5) \cite{shcouplingprb}. Results on the {\bf Q} dependence of the magnetic response at 6.5~meV are plotted in \fref{fig:incomresonance}. Transverse scans along [$1\bar10$] exhibit a pair of peaks as shown in \fref{fig:incomresonance}(a), while longitudinal scans show a single broad peak centered at (0.5,\,0.5), as shown in \fref{fig:incomresonance}(b). In both cases, the intensity is enhanced when the sample is cooled below \tc. The color-coded plot of intensity \vs ${\bf Q}$ at 6.5~meV and $T=1.5$~K, \fref{fig:incomresonance}(c), demonstrates an intriguing anisotropy: the transverse peaks are not reproduced along the longitudinal direction.

For other energies, the anisotropy still persists~\cite{lumsden-2009,mook-2009-2,incomfetese1,PhysRevLett.105.157002,fetesecomp142202}, showing that the magnetic excitations are anisotropic, dispersing only along the direction transverse to (0.5,\,0.5), as shown in \fref{fig:incomresonance}(d). This is certainly not a spin-wave like excitation, as in CaFe$_2$As$_2$~\cite{diallo:187206,zhao-2009-5}, since in that case one would expect to see a cone-shaped dispersion. ARPES measurements on the sample found that the Fermi surface near (0.5,\,0.5) appears to consist of four incommensurate pockets~\cite{shcouplingprb}. While Fermi-surface nesting is in principle compatible with the observation of the incommensurate resonance, the dispersion of isolated intensity peaks along a single direction is quite unusual and requires consideration of coupling of spin and orbital effects~\cite{shcouplingprb}.

Since superconductivity, and hence the pairing, is sensitive to magnetic field, one would
naturally expect that an external magnetic field could impact the resonance, as seen in YBa$_2$Cu$_3$O$_{6.6}$~\cite{nature965} and in
La$_{1.82}$Sr$_{0.18}$CuO$_4$~\cite{tranquada-2004}. Qiu \et~\cite{qiu:067008} found no significant change in the resonance in FeTe$_{0.6}$Se$_{0.4}$ in the presence of a 7-T field on a cold-neutron spectrometer with low flux. Another experiment on FeTe$_{0.5}$Se$_{0.5}$ using a spectrometer with higher flux concluded that the resonance starts to appear at a lowered $T_c$,
12~K, with reduced intensity, due to the suppression of the superconductivity; however, there was no detectable change in either the resonance energy or the width of the resonance peak~\cite{fieldeffectresonancewen}. With a field of 14.5~T, Zhao \et~\cite{zhao-2010field122} have shown that in BaFe$_{1.9}$Ni$_{0.1}$As$_2$, both the resonance energy and intensity are reduced, and the resonance peak is slightly broadened.

One commonly adopted view is that the spin resonance is a singlet-to-triplet excitation~\cite{PhysRevB.64.172508,PhysRevLett.75.4126,Bourges200545}. In principle, this hypothesis can be tested by experiments in magnetic field which should induce a Zeeman splitting  and lift the degeneracy of the triplet excited state~\cite{nature965}. Bao \et~\cite{wbao2} tried to address this problem by applying a 14-T field on FeSe$_{0.4}$Te$_{0.6}$, and it appears that the field induces a peak splitting. However, a more recent experiment shows that the field only reduces the spectral weight around the resonance mode~\cite{2011arXiv1105.4923L}. We applied a 16-T magnetic field and examined the field effect on the resonance. No splitting was observed in this measurement either. The nature of the resonance mode is still an open question.

\subsubsection{Doping dependence}
\label{subsec:weighttransfer}
\begin{figure}[ht]
\begin{center}\includegraphics[width=0.9\linewidth]{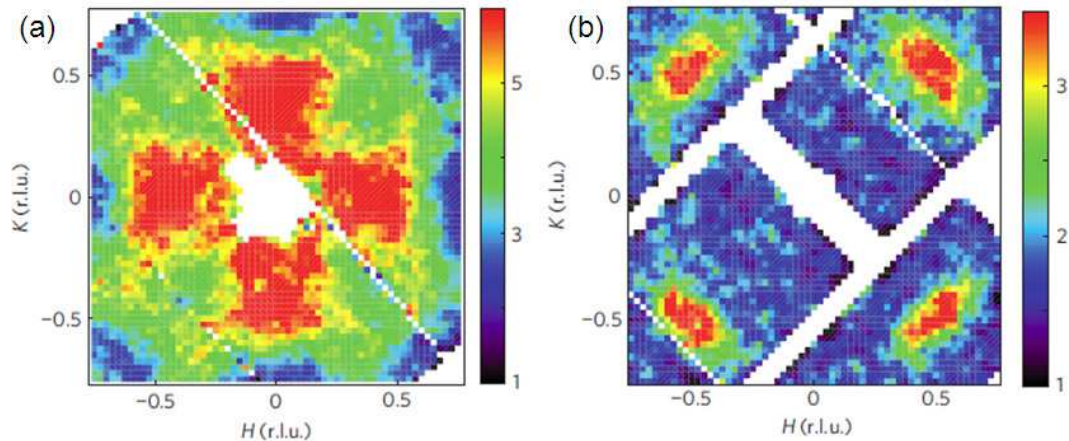}
\end{center}\caption{(a) Constant-energy cut of the magnetic excitation spectrum at an energy transfer of $6\pm1$~meV for the non-superconducting sample Fe$_{1.04}$Te$_{0.73}$Se$_{0.27}$ at 5~K. (b) Plot for the superconducting sample FeTe$_{0.51}$Se$_{0.49}$ at 3.5~K. Reprinted from \cite{lumsden-2009}. $\copyright$ 2010 Macmillan Publishers Ltd.}
\label{fig:weighttransfer}
\end{figure}

As discussed above, the non-superconducting parent compound of the 11 system, Fe$_{1+y}$Te, has static magnetic order with a bicollinear spin configuration. In samples with robust superconducting properties, there is strong magnetic scattering around (0.5,\,0.5) with a spin resonance below \tc, corresponding to spin correlations of the collinear type.  A number of theoretical~\cite{weiguounified,fang-2009} and experimental~\cite{lumsden-2009,liupi0topp,xudoping11,fetesecomp142202} studies have been carried out on the doping evolution of the magnetic correlations (static and dynamic), and their correlation with superconductivity. Lumsden \et~\cite{lumsden-2009} have performed measurements on time-of-flight spectrometers which cover a large momentum-energy space on two samples, a non-superconducting Fe$_{1.04}$Te$_{0.73}$Se$_{0.27}$, and a superconducting FeTe$_{0.51}$Se$_{0.49}$. It is clearly shown in \fref{fig:weighttransfer}(a) that at 5-7~meV, for the non-superconducting sample, the spectral weight is mostly concentrated around (0.5,\,0), where static magnetic order is observed. For the superconducting sample, magnetic excitations with a spin resonance near (0.5,\,0.5) are dominant, as shown in \fref{fig:weighttransfer}(b). On the other hand, the high-energy ($>120$~meV) spectrum looks qualitatively similar for these two samples~\cite{lumsden-2009}.  In the non-superconducting parent compound, Fe$_{1.05}$Te, spin waves dispersing up to $\sim250$~meV have been measured by Lipscombe \et ~\cite{PhysRevLett.106.057004}. These have been modelled in terms of spin waves calculated from a Heisenberg Hamiltonian with nearest- and next-nearest-neighbour couplings~\cite{PhysRevLett.106.057004}; however, a recent study of Fe$_{1.1}$Te, identifying distinct patterns of diffuse scattering and anomalous thermal enhancement of the effective moment, raises questions about such an approach \cite{Zaliznyak2011}.

As mentioned previously, the properties of \fts are sensitive not only to the Se concentration, but also to the amount of excess Fe. Xu \et~\cite{xudoping11}  demonstrated this by measuring the magnetic spectrum around (0.5,\,0) and (0.5,\,0.5) in four samples: FeTe$_{0.7}$Se$_{0.3}$, Fe$_{1.05}$Te$_{0.7}$Se$_{0.3}$; FeTe$_{0.5}$Se$_{0.5}$, and Fe$_{1.05}$Te$_{0.55}$Se$_{0.45}$. Both samples with $y=0$ are superconducting with the same \tc of 14~K; a resonance in the spin excitation spectrum is observed below \tc near (0.5,\,0.5), although the one with $x=0.3$ has a smaller superconducting volume fraction, and less spectral weight around (0.5,\,0.5). Also, there is short-range static magnetic order near (0.5,\,0) in FeTe$_{0.7}$Se$_{0.3}$, while the sample with $x=0.5$ does not exhibit magnetic order, short- or long-ranged, and the low-energy excitations close to (0.5,\,0) also disappear. With 0.05 extra Fe, superconductivity in both samples is fully suppressed, leading to the absence of the resonance, and the spectral-weight transfers from (0.5,\,0.5) to (0.5,\,0). In both samples, there is short-range static order and strong spectral weight around (0.5,\,0). Recently, Stock \et~\cite{2011arXiv1103.1811} have shown that by changing $y$ in Fe$_{1+y}$Te, the low-energy magnetic excitation spectrum can be dramatically modified, thus demonstrating the important role of excess Fe.

\subsection{Summary}
\label{subsec:neutronsum}
In the parent compound of the 11 system, there is long-range antiferromagnetic order with an ordering temperature of 65~K, and an in-plane wave vector (0.5,\,0); importantly, no Fermi-surface nesting is found along this wave vector. The direction of the wave vector is different from that of the iron pnicitides by 45$^\circ$. The magnetic ordering wave vector can become incommensurate with larger amounts of excess Fe.

Upon Se doping, the antiferromagnetic order is suppressed and becomes short-ranged, followed by the appearance of superconductivity. For samples with $x\approx0.5$ and robust superconductivity, there is no static order, and the low-energy spin excitations around (0.5,\,0) also disappear. The spectral weight is shifted to (0.5,\,0.5), where an incommensurate spin resonance is observed; the spin resonance is demonstrated to be intimately tied to the superconductivity from both the temperature and magnetic-field dependence. The magnetic excitation spectrum around (0.5,\,0.5) shows an interesting anisotropy, which apparently cannot be explained by either a Fermi-surface-nesting or a local-moment model. In Fe$_{1+y}$Se, there is no static order, while there are spin fluctuations which are enhanced near \tc. The superconducting and magnetic properties of the system can also be modified by adjusting the amount of excess Fe with fixed Se content. The extra Fe is found to enhance the magnetic correlations around (0.5,\,0) and suppress superconductivity as well as the spin excitations around (0.5,\,0.5).

The results on the evolution of the magnetic excitation spectrum with the tuning parameters (Se/Fe), clearly indicate that static bicollinear magnetic order in the 11 system competes with superconductivity. It appears that only when the system evolves towards fluctuating collinear  magnetic correlations, does superconductivity appear; this seems to be universal across all known iron-based superconductor families. Despite these agreements, the origin of the magnetism in the 11 compound is quite controversial. Some believe that the magnetism arises from itinerant electrons~\cite{mook-2009-2,incomfetese1}. Recently there has been work suggesting a large local moment associated with the low-energy magnetic excitations in a superconducting FeTe$_{0.35}$Se$_{0.65}$ sample; this observation is incompatible with predictions from a weakly coupled itinerant picture~\cite{xulocal1}. The fact that a simple Fermi-surface-nesting picture~\cite{subedi-2008-78} cannot explain many of the experimental observations~\cite{xia037002,bao-2009,shcouplingprb,xulocal1} leads to arguments for a significant local-moment character to the magnetism~\cite{Physics.2.59,xia037002,han:067001,ma-2009,loalm1,johannes-2009,fang-2009,moon057003}.

For the $A_x$Fe$_{2-y}$Se$_2$ system, it is found that superconductivity is in close proximity to an insulating phase. This feature is different from that in other iron-based superconductors, but mirrors that in the high-\tc cuprates. Whether the microscopic coexistence of strong antiferromagnetic order with high-\tc superconductivity is true or not requires further investigation. If they do coexist, it requires further studies to understand whether or not the superconducting pairing mechanism in this system is the same as that in other high-\tc superconductors.

\section{Conclusions}
\label{sec:conclusion}
Thanks to the sustained efforts on sample synthesis for the \fts system, some high-quality samples have been made available. In particular, by using the Bridgman technique, many large-size, high-quality single crystals have been grown. Measurements performed on these samples yield a plethora of fascinating results. It has been found that the static antiferromagnetic order in the parent compound is centered around the wave vector (0.5,\,0) with a bicollinear spin structure that competes with superconductivity, while superconducting samples are characterized by collinear magnetic correlations with magnetic excitations centered around the wave vector (0.5,\,0.5). The argument that there is an intimate relationship between the spin excitations around (0.5,\,0.5) and superconductivity has been reinforced by the temperature and magnetic-field dependence of the resonance. Generally, this system shows strong similarities to other high-\tc superconductors, where it is believed that spin excitations play a progenitive role in the superconductivity.

For the newly discovered $A_x$Fe$_{2-y}$Se$_2$ superconductors, which have superconducting transition temperatures, \tc, up to 33~K, crystals can be obtained using the Bridgman technique. This system is unique: i) It has a very different Fermi-surface topology with two electron pockets at the Brillouin zone center; ii) The superconducting phase borders an insulating parent phase, as in the case of the cuprates, but different from all previously investigated iron-based superconductors.

While $A_x$Fe$_{2-y}$Se$_2$ has attracted significant attention, there are still many basic questions to be answered. For instance, reports on the spin dynamics in this system have been very limited~\cite{2011arXiv1105.4675W}. This will provide important clues on whether or not this system shares the same basic physics underlying the superconductivity as other high-temperature superconductors. Also, further efforts on obtaining single crystals with improved sample quality and larger size will certainly be needed in order to elucidate many unresolved issues. For the 11 system, sample inhomogeneity is a less significant issue. Future work to control the stoichiometry more precisely will be helpful.

\section{Acknowledgements}
The work at Brookhaven National Laboratory (J.W., G.X., G.G., and J.M.T.) was supported by the Office of Basic Energy Sciences, Division of Materials Science and Engineering, U.S. Department of Energy, under Contract No. DE-AC02-98CH10886.  J.M.T. is also supported in part by the Center for Emergent Superconductivity, an Energy Frontier Research Center. Work at Lawrence Berkeley National Laboratory (J.W. and R.J.B.) was supported by the same Office under Contract No. DE-AC02-05CH11231. The authors thank all of their collaborators listed in the references. We are also grateful to our colleagues and collaborators for allowing us to reproduce their work here.

\bibliography{references}
\bibliographystyle{unsrtjw}

\end{document}

%% file: newcommands.tex
\providecommand*{\eg}{\emph{e.\,g.}\xspace}%
\providecommand*{\ie}{\emph{i.\,e.}\xspace}%
\providecommand*{\et}{\emph{et\,al.}\xspace}%
\providecommand*{\vs}{\emph{vs}\xspace}%
\providecommand*{\tc}{$T_c$\xspace}%
\providecommand*{\fts}{Fe$_{1+y}$Te$_{1-x}$Se$_x$\xspace}%

%% file: interplay_ms.bbl
\begin{thebibliography}{100}

\bibitem{onnes1}
H.~K. Onnes.
\newblock The resistance of pure mercury at helium temperatures.
\newblock {\em Comm. Leiden.}, 120b, 1911.

\bibitem{bcstheory}
J.~Bardeen, L.~N. Cooper, and J.~R. Schrieffer.
\newblock Theory of Superconductivity.
\newblock {\em Phys. Rev.}, 108:1175, 1957.

\bibitem{mullerlbco}
J.~G. Bednorz and K.~A. M\"uller.
\newblock Possible High $T_c$ Superconductivity in the Ba-La-Cu-O system.
\newblock {\em Z. Phys. B}, 64:189, 1986.

\bibitem{ybcodis}
M.~K. Wu, J.~R. Ashburn, C.~J. Torng, P.~H. Hor, R.~L. Meng, L.~Gao, Z.~J.
  Huang, Y.~Q. Wang, and C.~W. Chu.
\newblock Superconductivity at 93 K in a new mixed-phase Y-Ba-Cu-O compound
  system at ambient pressure.
\newblock {\em Phys. Rev. Lett.}, 58:908, 1987.

\bibitem{lee:17}
Patrick~A. Lee, Naoto Nagaosa, and Xiao-Gang Wen.
\newblock Doping a Mott insulator: Physics of high-temperature
  superconductivity.
\newblock {\em Rev. Mod. Phys.}, 78:17, 2006.

\bibitem{carlsonbook1}
E.~W. {Carlson}, V.~J. {Emery}, S.~A. {Kivelson}, and D.~{Orgad}.
\newblock {\em The Physics of Conventional and Unconventional Superconductors},
  chapter Concepts in High Temperature Superconductivity.
\newblock Springer-Verlag, 2002.

\bibitem{birgeneau-2006}
R.~J. Birgeneau, C.~Stock, J.~M. Tranquada, and K.~Yamada.
\newblock Magnetic neutron scattering in hole doped cuprate superconductors.
\newblock {\em J. Phys. Soc. Jpn.}, 75:111003, 2006.

\bibitem{RevModPhys.70.897}
M.~A. Kastner, R.~J. Birgeneau, G.~Shirane, and Y.~Endoh.
\newblock Magnetic, transport, and optical properties of monolayer copper
  oxides.
\newblock {\em Rev. Mod. Phys.}, 70:897, 1998.

\bibitem{orenstein}
J.~Orenstein and A.~J. Millis.
\newblock Advances in the Physics of High-Temperature Superconductivity.
\newblock {\em Science}, 288:468, 2000.

\bibitem{hosono_1}
Yoichi Kamihara, Hidenori Hiramatsu, Masahiro Hirano, Ryuto Kawamura, Hiroshi
  Yanagi, Toshio Kamiya, and Hideo Hosono.
\newblock Iron-Based Layered Superconductor: LaOFeP.
\newblock {\em J. Am. Chem. Soc.}, 128:10012, 2006.

\bibitem{hosono_2}
Yoichi Kamihara, Takumi Watanabe, Masahiro Hirano, and Hideo Hosono.
\newblock Iron-Based Layered Superconductor La[O$_{1-x}$F$_x$]FeAs
  (x=0.05-0.12) with $T_c$=26 K.
\newblock {\em J. Am. Chem. Soc.}, 130:3296, 2008.

\bibitem{chen-2008-453}
X.~H. Chen, T.~Wu, G.~Wu, R.~H. Liu, H.~Chen, and D.~F. Fang.
\newblock Superconductivity at 43 K in Samarium-arsenide Oxides
  SmFeAsO$_{1-x}$F$_x$.
\newblock {\em Nature}, 453:761, 2008.

\bibitem{hosono4}
Hiroki Takahashi, Kazumi Igawa, Kazunobu Arii, Yoichi Kamihara, Masahiro
  Hirano, and Hideo Hosono.
\newblock Superconductivity at 43 K in an iron-based layered compound
  LaO$_{1-x}$F$_x$FeAs.
\newblock {\em Nature}, 453:376, 2008.

\bibitem{JPSJtc54k}
Hijiri Kito, Hiroshi Eisaki, and Akira Iyo.
\newblock Superconductivity at 54\,K in F-Free NdFeAsO$_{1-y}$.
\newblock {\em J. Phys. Soc. Jpn.}, 77:063707, 2008.

\bibitem{eplrentc51k}
Zhi-An Ren, Jie Yang, Wei Lu, Wei Yi, Xiao-Li Shen, Zheng-Cai Li, Guang-Can
  Che, Xiao-Li Dong, Li-Ling Sun, Fang Zhou, and Zhong-Xian Zhao.
\newblock Superconductivity in the iron-based F-doped layered quaternary
  compound NdO$_{1-x}$F$_x$FeAs.
\newblock {\em Euro. Phys. Lett.}, 82:57002, 2008.

\bibitem{ren-2008-25}
Zhi-An Ren, Wei Lu, Jie Yang, Wei Yi, Xiao-Li Shen, Zheng-Cai Li, Guang-Can
  Che, Xiao-Li Dong, Li-Ling Sun, Fang Zhou, and Zhong-Xian Zhao.
\newblock Superconductivity at 55 K in iron-based F-doped layered quaternary
  compound SmO$_{1-x}$F$_x$FeAs.
\newblock {\em Chin. Phys. Lett.}, 25:2215, 2008.

\bibitem{epltc56k}
Cao Wang, Linjun Li, Shun Chi, Zengwei Zhu, Zhi Ren, Yuke Li, Yuetao Wang, Xiao
  Lin, Yongkang Luo, Shuai Jiang, Xiangfan Xu, Guanghan Cao, and Zhu'an Xu.
\newblock Thorium-doping-induced superconductivity up to 56 K in
  Gd$_{1-x}$Th$_x$FeAsO.
\newblock {\em Euro. Phys. Lett.}, 83:67006, 2008.

\bibitem{sefat:117004}
Athena~S. Sefat, Rongying Jin, Michael~A. McGuire, Brian~C. Sales, David~J.
  Singh, and David Mandrus.
\newblock Superconductivity at 22 K in Co-Doped BaFe$_2$As$_2$ Crystals.
\newblock {\em Phys. Rev. Lett.}, 101:117004, 2008.

\bibitem{rotter-2008-101}
Marianne Rotter, Marcus Tegel, and Dirk Johrendt.
\newblock Superconductivity at 38 K in the iron arsenide
  Ba$_{1-x}$K$_x$Fe$_2$As$_2$.
\newblock {\em Phys. Rev. Lett.}, 101:107006, 2008.

\bibitem{chen-2008-25}
G.~F. Chen, Z.~Li, G.~Li, W.~Z. Hu, J.~Dong, X.~D. Zhang, P.~Zheng, N.~L. Wang,
  and J.~L. Luo.
\newblock Superconductivity in hole-doped (Sr$_{1-x}$K$_x$)Fe$_2$As$_2$.
\newblock {\em Chin. Phys. Lett.}, 25:3403, 2008.

\bibitem{inosov:224503}
D.~S. Inosov, A.~Leineweber, Xiaoping Yang, J.~T. Park, N.~B. Christensen,
  R.~Dinnebier, G.~L. Sun, Ch. Niedermayer, D.~Haug, P.~W. Stephens, J.~Stahn,
  O.~Khvostikova, C.~T. Lin, O.~K. Andersen, B.~Keimer, and V.~Hinkov.
\newblock Suppression of the structural phase transition and lattice softening
  in slightly underdoped Ba$_{1-x}$K$_x$Fe$_2$As$_2$ with electronic phase
  separation.
\newblock {\em Phys. Rev. B}, 79:224503, 2009.

\bibitem{lifep}
Z.~Deng, X.~C. Wang, Q.Q. Liu, S.~J. Zhang, Y.~X. Lv, J.~L. Zhu, R.C. Yu, and
  C.Q. Jin.
\newblock A new 111 type iron pnictide superconductor LiFeP.
\newblock {\em Euro. Phys. Lett.}, 87:3704, 2009.

\bibitem{lifeas1}
X.~C. {Wang}, Q.~Q. {Liu}, Y.~X. {Lv}, W.~B. {Gao}, L.~X. {Yang}, R.~C. {Yu},
  F.~Y. {Li}, and C.~Q. {Jin}.
\newblock {The superconductivity at 18 K in LiFeAs system}.
\newblock {\em Solid State Commun.}, 148:538, 2008.

\bibitem{lifeas2}
M.~J. {Pitcher}, D.~R. {Parker}, P.~{Adamson}, S.~J.~C. {Herkelrath}, A.~T.
  {Boothroyd}, and S.~J. {Clarke}.
\newblock {Structure and superconductivity of LiFeAs}.
\newblock {\em Chem. Commun.}, 45:5918, 2008.

\bibitem{lifeas3}
C.~W. {Chu}, F.~{Chen}, M.~{Gooch}, A.~M. {Guloy}, B.~{Lorenz}, B.~{Lv},
  K.~{Sasmal}, Z.~J. {Tang}, J.~H. {Tapp}, and Y.~Y. {Xue}.
\newblock {The synthesis and characterization of LiFeAs and NaFeAs}.
\newblock {\em Physica C}, 469:326, 2009.

\bibitem{structure5}
Joshua~H. Tapp, Zhongjia Tang, Bing Lv, Kalyan Sasmal, Bernd Lorenz, Paul C.~W.
  Chu, and Arnold~M. Guloy.
\newblock LiFeAs: An intrinsic FeAs-based superconductor with $ T_c =18$ K.
\newblock {\em Phys. Rev. B}, 78:060505, 2008.

\bibitem{lifep2}
Z.~{Deng}, X.~C. {Wang}, Q.~Q. {Liu}, S.~J. {Zhang}, Y.~X. {Lv}, J.~L. {Zhu},
  R.~C. {Yu}, and C.~Q. {Jin}.
\newblock {A new ``111'' type iron pnictide superconductor LiFeP}.
\newblock {\em Euro. Phys. Lett.}, 87:37004, 2009.

\bibitem{nafeas1}
S.~J. {Zhang}, X.~C. {Wang}, Q.~Q. {Liu}, Y.~X. {Lv}, X.~H. {Yu}, Z.~J. {Lin},
  Y.~S. {Zhao}, L.~{Wang}, Y.~{Ding}, H.~K. {Mao}, and C.~Q. {Jin}.
\newblock {Superconductivity at 31 K in the ``111''-type iron arsenide
  superconductor Na$_{1-x}$FeAs induced by pressure}.
\newblock {\em Euro. Phys. Lett.}, 88:47008, 2009.

\bibitem{hsu-2008}
Fong-Chi Hsu, Jiu-Yong Luo, Kuo-Wei Yeh, Ta-Kun Chen, Tzu-Wen Huang, Phillip~M.
  Wu, Yong-Chi Lee, Yi-Lin Huang, Yan-Yi Chu, Der-Chung Yan, and Maw-Kuen Wu.
\newblock Superconductivity in the PbO-type Structure alpha-FeSe.
\newblock {\em Proc. Natl. Acad. Sci. USA}, 105:14262, 2008.

\bibitem{yeh-2008}
Kuo-Wei Yeh, Tzu-Wen Huang, Yi-Lin Huang, Ta-Kun Chen, Fong-Chi Hsu, Phillip~M.
  Wu, Yong-Chi Lee, Yan-Yi Chu, Chi-Liang Chen, Jiu-Yong Luo, Der-Chung Yan,
  and Maw kuen Wu.
\newblock Tellurium substitution effect on superconductivity of the
  $\alpha$-phase Iron Selenide.
\newblock {\em Euro. Phys. Lett.}, 84:37002, 2008.

\bibitem{sales:094521}
B.~C. Sales, A.~S. Sefat, M.~A. McGuire, R.~Y. Jin, D.~Mandrus, and
  Y.~Mozharivskyj.
\newblock Bulk superconductivity at 14 K in single crystals of
  Fe$_{1+y}$Te$_x$Se$_{1-x}$.
\newblock {\em Phys. Rev. B}, 79:094521, 2009.

\bibitem{chen:140509}
G.~F. Chen, Z.~G. Chen, J.~Dong, W.~Z. Hu, G.~Li, X.~D. Zhang, P.~Zheng, J.~L.
  Luo, and N.~L. Wang.
\newblock Electronic properties of single-crystalline Fe$_{1.05}$Te and
  Fe$_{1.03}$Se$_{0.30}$Te$_{0.70}$.
\newblock {\em Phys. Rev. B}, 79:140509(R), 2009.

\bibitem{fang-2008-78}
M.~H. Fang, H.~M. Pham, B.~Qian, T.~J. Liu, E.~K. Vehstedt, Y.~Liu, L.~Spinu,
  and Z.~Q. Mao.
\newblock Superconductivity close to magnetic instability in
  Fe(Se$_{1-x}$Te$_x$)$_{0.82}$.
\newblock {\em Phys. Rev. B}, 78:224503, 2008.

\bibitem{zhu:220512}
Xiyu Zhu, Fei Han, Gang Mu, Peng Cheng, Bing Shen, Bin Zeng, and Hai-Hu Wen.
\newblock Transition of stoichiometric Sr$_2$VO$_3$FeAs to a superconducting
  state at 37.2 K.
\newblock {\em Phys. Rev. B}, 79:220512, 2009.

\bibitem{sc21311}
Hiraku Ogino, Yutaka Matsumura, Yukari Katsura, Koichi Ushiyama, Shigeru Horii,
  Kohji Kishio, and Jun ichi Shimoyama.
\newblock Superconductivity at 17 K in (Fe$_2$P$_2$)(Sr$_4$Sc$_2$O$_6$): a new
  superconducting layered pnictide oxide with a thick perovskite oxide layer.
\newblock {\em Supercon. Sci. Tech.}, 22:075008, 2009.

\bibitem{zhao:132504}
Jun Zhao, Q.~Huang, Clarina {de la Cruz}, J.~W. Lynn, M.~D. Lumsden, Z.~A. Ren,
  Jie Yang, Xiaolin Shen, Xiaoli Dong, Zhongxian Zhao, and Pengcheng Dai.
\newblock Lattice and magnetic structures of PrFeAsO,
  PrFeAsO$_{0.85}$F$_{0.15}$, and PrFeAsO$_{0.85}$.
\newblock {\em Phys. Rev. B}, 78:132504, 2008.

\bibitem{lester:144523}
C.~Lester, Jiun-Haw Chu, J.~G. Analytis, S.~C. Capelli, A.~S. Erickson, C.~L.
  Condron, M.~F. Toney, I.~R. Fisher, and S.~M. Hayden.
\newblock Neutron scattering study of the interplay between structure and
  magnetism in Ba(Fe$_{1-x}$Co$_x$)$_2$As$_2$.
\newblock {\em Phys. Rev. B}, 79:144523, 2009.

\bibitem{qiu:257002}
Y.~Qiu, Wei Bao, Q.~Huang, T.~Yildirim, J.~M. Simmons, M.~A. Green, J.~W. Lynn,
  Y.~C. Gasparovic, J.~Li, T.~Wu, G.~Wu, and X.~H. Chen.
\newblock Crystal Structure and Antiferromagnetic Order in NdFeAsO$_{1-x}$F$_x$
  (x = 0.0 and 0.2) Superconducting Compounds from Neutron Diffraction
  Measurements.
\newblock {\em Phys. Rev. Lett.}, 101:257002, 2008.

\bibitem{kumar:144524}
Neeraj Kumar, Songxue Chi, Ying Chen, Kumari~Gaurav Rana, A.~K. Nigam,
  A.~Thamizhavel, William Ratcliff, II, S.~K. Dhar, and Jeffrey~W. Lynn.
\newblock Evolution of the bulk properties, structure, magnetic order, and
  superconductivity with Ni doping in CaFe$_{2-x}$Ni$_x$As$_2$.
\newblock {\em Phys. Rev. B}, 80:144524, 2009.

\bibitem{cruz}
C.~{de la Cruz}, Q.~Huang, J.~W. Lynn, J.~Li, W.~Ratcliff, J.~L. Zarestzky,
  H.~A. Mook, G.~F. Chen, J.~L. Luo, N.~L. Wang, and P.~Dai.
\newblock Magnetic order close to superconductivity in the iron-based layered
  LaO$_{1-x}$F$_x$FeAs systems.
\newblock {\em Nature}, 453:899, 2008.

\bibitem{huang:257003}
Q.~Huang, Y.~Qiu, Wei Bao, M.~A. Green, J.~W. Lynn, Y.~C. Gasparovic, T.~Wu,
  G.~Wu, and X.~H. Chen.
\newblock Neutron-Diffraction Measurements of Magnetic Order and a Structural
  Transition in the Parent BaFe$_2$As$_2$ Compound of FeAs-Based
  High-Temperature Superconductors.
\newblock {\em Phys. Rev. Lett.}, 101:257003, 2008.

\bibitem{chen:064515}
Ying Chen, J.~W. Lynn, J.~Li, G.~Li, G.~F. Chen, J.~L. Luo, N.~L. Wang,
  Pengcheng Dai, C.~{de la Cruz}, and H.~A. Mook.
\newblock Magnetic order of the iron spins in NdFeAsO.
\newblock {\em Phys. Rev. B}, 78:064515, 2008.

\bibitem{johannes-2009}
M.~D. Johannes and I.~I. Mazin.
\newblock Microscopic origin of magnetism and magnetic interactions in
  ferropnictides.
\newblock {\em Phys. Rev. B}, 79:220510(R), 2009.

\bibitem{wilson:184519}
Stephen~D. Wilson, Z.~Yamani, C.~R. Rotundu, B.~Freelon, E.~Bourret-Courchesne,
  and R.~J. Birgeneau.
\newblock Neutron diffraction study of the magnetic and structural phase
  transitions in BaFe$_2$As$_2$.
\newblock {\em Phys. Rev. B}, 79:184519, 2009.

\bibitem{kofu-2009-11}
M.~Kofu, Y.~Qiu, Wei bao, S.~H. Lee, S.~Chang, T.~Wu, G.~Wu, and X.~H. Chen.
\newblock Neutron scattering investigation of the magnetic order in single
  crystalline BaFe$_2$As$_2$.
\newblock {\em New J. Phys.}, 11:055001, 2009.

\bibitem{zhao}
Jun Zhao, Q.~Huang, C.~{de la Cruz}, Shiliang Li, J.~W. Lynn, Y.~Chen, M.~A.
  Green, G.~F. Chen, G.~Li, Z.~Li, J.~L. Luo, N.~L. WANG, , and Pengcheng Dai.
\newblock Structural and magnetic phase diagram of CeFeAsO$_{1-x}$F$_x$ and its
  relation to high-temperature superconductivity.
\newblock {\em Nature Mater.}, 7:953, 2008.

\bibitem{luetkens-2008}
H.~Luetkens, H.~H. Klauss, M.~Kraken, F.~J. Litterst, T.~Dellmann,
  R.~Klingeler, C.~Hess, R.~Khasanov, A.~Amato, C.~Baines, J.~Hamann-Borrero,
  N.~Leps, A.~Kondrat, G.~Behr, J.~Werner, and B.~Buechner.
\newblock The electronic phase diagram of the LaO$_{1-x}$F$_x$FeAs
  superconductor.
\newblock {\em Nature Mater.}, 8:305, 2009.

\bibitem{drew-2009}
A.~J. Drew, Ch~Niedermayer, P.~J. Baker, F.~L. Pratt, S.~J. Blundell,
  T.~Lancaster, R.~H. Liu, G.~Wu, X.~H. Chen, I.~Watanabe, V.~K. Malik,
  A.~Dubroka, M.~Roessle, K.~W. Kim, C.~Baines, and C.~Bernhard.
\newblock Coexistence of static magnetism and superconductivity in
  SmFeAsO$_{1-x}$F$_x$ as revealed by muon spin rotation.
\newblock {\em Nature Mater.}, 8:310, 2009.

\bibitem{rotter-2008-47}
Marianne Rotter, Michael Pangerl, Marcus Tegel, and Dirk Johrendt.
\newblock Superconductivity and Crystal Structures of
  (Ba$_{1-x}$K$_x$)Fe$_2$As$_2$ (x = 0 - 1).
\newblock {\em Angew. Chem. Int.. Ed.}, 47:7949, 2008.

\bibitem{chen-2009-85}
H.~Chen, Y.~Ren, Y.~Qiu, Wei bao, R.~H. Liu, G.~Wu, T.~Wu, Y.~L. Xie, X.~F.
  Wang, Q.~Huang, and X.~H. Chen.
\newblock Coexistence of the spin-density-wave and superconductivity in the
  (Ba,K)Fe$_2$As$_2$.
\newblock {\em Euro. Phys. Lett.}, 85:17006, 2009.

\bibitem{fang:140508}
Lei Fang, Huiqian Luo, Peng Cheng, Zhaosheng Wang, Ying Jia, Gang Mu, Bing
  Shen, I.~I. Mazin, Lei Shan, Cong Ren, and Hai-Hu Wen.
\newblock Roles of multiband effects and electron-hole asymmetry in the
  superconductivity and normal-state properties of
  Ba(Fe$_{1-x}$Co$_x$)$_2$As$_2$.
\newblock {\em Phys. Rev. B}, 80:140508(R), 2009.

\bibitem{chu:014506}
Jiun-Haw Chu, James~G. Analytis, Chris Kucharczyk, and Ian~R. Fisher.
\newblock Determination of the phase diagram of the electron-doped
  superconductor Ba(Fe$_{1-x}$Co$_x$)$_2$As$_2$.
\newblock {\em Phys. Rev. B}, 79:014506, 2009.

\bibitem{khasanov:140511}
R.~Khasanov, M.~Bendele, A.~Amato, P.~Babkevich, A.~T. Boothroyd,
  A.~Cervellino, K.~Conder, S.~N. Gvasaliya, H.~Keller, H.-H. Klauss,
  H.~Luetkens, V.~Pomjakushin, E.~Pomjakushina, and B.~Roessli.
\newblock Coexistence of incommensurate magnetism and superconductivity in
  Fe$_{1+y}$Se$_x$Te$_{1-x}$.
\newblock {\em Phys. Rev. B}, 80:140511, 2009.

\bibitem{liupi0topp}
T.~J. {Liu}, J.~{Hu}, B.~{Qian}, D.~{Fobes}, Z.~Q. {Mao}, W.~{Bao},
  M.~{Reehuis}, S.~A.~J. {Kimber}, K.~{Prokes}, S.~{Matas}, D.~N. {Argyriou},
  A.~{Hiess}, A.~{Rotaru}, H.~{Pham}, L.~{Spinu}, Y.~{Qiu}, V.~{Thampy}, A.~T.
  {Savici}, J.~A. {Rodriguez}, and C.~{Broholm}.
\newblock {From ($\pi$, 0) magnetic order to superconductivity with ($\pi$,
  $\pi$) magnetic resonance in Fe$_{1.02}$(Te$_{1-x}$Se$_x$)}.
\newblock {\em Nature Mater.}, 9:716, 2010.

\bibitem{0295-5075-90-2-27011}
P.~L. Paulose, C.~S. Yadav, and K.~M. Subhedar.
\newblock Magnetic phase diagram of Fe$_{1.1}$Te$_{1-x}$Se$_x$ : A comparative
  study with the stoichiometric superconducting FeTe$_{1-x}$Se$_x$ system.
\newblock {\em Euro. Phys. Lett.}, 90:27011, 2010.

\bibitem{spinglass}
N.~Katayama, S.~Ji, D.~Louca, S.-H. Lee, M.~Fujita, T.~J. Sato, J.~S. Wen,
  Z.~J. Xu, G.~D. Gu, G.~Xu, Z.~W. Lin, M.~Enoki, S.~Chang, K.~Yamada, and
  J.~M. Tranquada.
\newblock Investigation of the spin-glass regime between the antiferromagnetic
  and superconducting phases in Fe$_{1+y}$Se$_x$Te$_{1-x}$.
\newblock {\em J. Phys. Soc. Jpn.}, 79:113702, 2010.

\bibitem{JPSJ.79.102001}
Yoshikazu Mizuguchi and Yoshihiko Takano.
\newblock Review of Fe Chalcogenides as the Simplest Fe-Based Superconductor.
\newblock {\em J. Phys. Soc. Jpn.}, 79:102001, 2010.

\bibitem{revisedfetesephase}
C.~{Dong}, H.~{Wang}, Z.~{Li}, J.~{Chen}, H.~Q. {Yuan}, and M.~{Fang}.
\newblock {Revised Phase Diagram for FeTe$_{1-x}$Se$_x$ system with less excess
  Fe atoms}.
\newblock {\em arXiv:1012.5188}, 2010.

\bibitem{huang:054529}
Q.~Huang, Jun Zhao, J.~W. Lynn, G.~F. Chen, J.~L. Luo, N.~L. Wang, and
  Pengcheng Dai.
\newblock Doping evolution of antiferromagnetic order and structural distortion
  in LaFeAsO$_{1-x}$F$_x$.
\newblock {\em Phys. Rev. B}, 78:054529, 2008.

\bibitem{mcguire:094517}
Michael~A. McGuire, Andrew~D. Christianson, Athena~S. Sefat, Brian~C. Sales,
  Mark~D. Lumsden, Rongying Jin, E.~Andrew Payzant, David Mandrus, Yanbing
  Luan, Veerle Keppens, Vijayalaksmi Varadarajan, Joseph~W. Brill,
  Rapha\"{e}l~P. Hermann, Moulay~T. Sougrati, Fernande Grandjean, and Gary~J.
  Long.
\newblock Phase transitions in LaFeAsO: Structural, magnetic, elastic, and
  transport properties, heat capacity and M\"{o}ssbauer spectra.
\newblock {\em Phys. Rev. B}, 78:094517, 2008.

\bibitem{kivelsonironnews}
S.~A. {Kivelson} and H.~{Yao}.
\newblock {Iron-based superconductors: Unity or diversity?}
\newblock {\em Nature Mater.}, 7:927, 2008.

\bibitem{mazin:057003}
I.~I. Mazin, D.~J. Singh, M.~D. Johannes, and M.~H. Du.
\newblock Unconventional Superconductivity with a Sign Reversal in the Order
  Parameter of LaFeAsO$_{1-x}$F$_x$.
\newblock {\em Phys. Rev. Lett.}, 101:057003, 2008.

\bibitem{kuroki-2008}
Kazuhiko Kuroki, Seiichiro Onari, Ryotaro Arita, Hidetomo Usui, Yukio Tanaka,
  Hiroshi Kontani, and Hideo Aoki.
\newblock Unconventional Pairing Originating from the Disconnected Fermi
  Surfaces of Superconducting LaFeAsO$_{1-x}$F$_x$.
\newblock {\em Phys. Rev. Lett.}, 101:087004, 2008.

\bibitem{ma:033111}
Fengjie Ma and Zhong-Yi Lu.
\newblock Iron-based layered compound LaFeAsO is an antiferromagnetic
  semimetal.
\newblock {\em Phys. Rev. B}, 78:033111, 2008.

\bibitem{dong-2008-83}
J.~Dong, H.~J. Zhang, G.~Xu, Z.~Li, G.~Li, W.~Z. Hu, D.~Wu, G.~F. Chen, X.~Dai,
  J.~L. Luo, Z.~Fang, and N.~L. Wang.
\newblock Competing Orders and Spin-Density-Wave Instability in
  La(O$_{1-x}$F$_x$)FeAs.
\newblock {\em Euro. Phys. Lett.}, 83:27006, 2008.

\bibitem{cvetkovic-2009}
V.~Cvetkovic and Z.~Tesanovic.
\newblock Multiband magnetism and superconductivity in Fe-based compounds.
\newblock {\em Euro. Phys. Lett.}, 85:37002, 2009.

\bibitem{graser-2009}
S~Graser, T~A Maier, P~J Hirschfeld, and D~J Scalapino.
\newblock Near-degeneracy of several pairing channels in multiorbital models
  for the Fe pnictides.
\newblock {\em New J. Phys.}, 11:025016, 2009.

\bibitem{JPSJ.77.113703}
Hisashi Kotegawa, Satoru Masaki, Yoshiki Awai, Hideki Tou, Yoshikazu Mizuguchi,
  and Yoshihiko Takano.
\newblock Evidence for Unconventional Superconductivity in Arsenic-Free
  Iron-Based Superconductor FeSe: A $^{77}$Se-NMR Study.
\newblock {\em J. Phys. Soc. Jpn.}, 77:113703, 2008.

\bibitem{Fang2008}
C.~Fang, H.~Yao, W.~F. Tsai, J.~P. Hu, and S.~A. Kivelson.
\newblock Theory of electron nematic order in LaOFeAs.
\newblock {\em Phys. Rev. B}, 77:224509, 2008.

\bibitem{Haule2009}
K~Haule and G~Kotliar.
\newblock {Coherence--incoherence crossover in the normal state of iron
  oxypnictides and importance of Hund's rule coupling}.
\newblock {\em New J. Phys.}, 11:025021, 2009.

\bibitem{Si2009}
Qimiao Si, Elihu Abrahams, Jianhui Dai, and Jian-Xin Zhu.
\newblock {Correlation effects in the iron pnictides}.
\newblock {\em New J. Phys.}, 11:045001, 2009.

\bibitem{mazin-2009np}
I.~I. Mazin and M.~D. Johannes.
\newblock A key role for unusual spin dynamics in ferropnictides.
\newblock {\em Nature Phys.}, 5:141, 2009.

\bibitem{Kou2009epl}
Su-Peng Kou, Tao Li, and Zheng-yu Weng.
\newblock Coexistence of itinerant electrons and local moments in iron-based
  superconductors.
\newblock {\em Euro. Phys. Lett.}, 88:17010, 2009.

\bibitem{Medici2010}
Luca {de' Medici}, Syed~R. Hassan, and Massimo Capone.
\newblock {Genesis of coexisting itinerant and localized electrons in Iron
  Pnictides}.
\newblock {\em J. Supercond. Nov. Magn.}, 22:535, 2009.

\bibitem{weiguounified}
Wei-Guo Yin, Chi-Cheng Lee, and Wei Ku.
\newblock Unified Picture for Magnetic Correlations in Iron-Based
  Superconductors.
\newblock {\em Phys. Rev. Lett.}, 105:107004, 2010.

\bibitem{Arita2}
Ryotaro Arita and Hiroaki Ikeda.
\newblock Is Fermi-Surface Nesting the Origin of Superconductivity in Iron
  Pnictides?: A Fluctuation-Exchange-Approximation Study.
\newblock {\em J. Phys. Soc. Jpn.}, 78(11):113707, 2009.

\bibitem{Kontani2010}
Hiroshi Kontani and Seiichiro Onari.
\newblock {Orbital-Fluctuation-Mediated Superconductivity in Iron Pnictides:
  Analysis of the Five-Orbital Hubbard-Holstein Model}.
\newblock {\em Phys. Rev. Lett.}, 104:157001, 2010.

\bibitem{christianson-2008-456}
A.~D. Christianson, E.~A. Goremychkin, R.~Osborn, S.~Rosenkranz, M.~D. Lumsden,
  C.~D. Malliakas, l~S.~Todorov, H.~Claus, D.~Y. Chung, M.~G. Kanatzidis, R.~I.
  Bewley, and T.~Guidi.
\newblock Unconventional superconductivity in Ba$_{0.6}$K$_0.4$Fe$_2$As$_2$
  from inelastic neutron scattering.
\newblock {\em Nature}, 456:930, 2008.

\bibitem{lumsden:107005}
M.~D. Lumsden, A.~D. Christianson, D.~Parshall, M.~B. Stone, S.~E. Nagler,
  G.~J. MacDougall, H.~A. Mook, K.~Lokshin, T.~Egami, D.~L. Abernathy, E.~A.
  Goremychkin, R.~Osborn, M.~A. McGuire, A.~S. Sefat, R.~Jin, B.~C. Sales, and
  D.~Mandrus.
\newblock Two-dimensional resonant magnetic excitation in
  BaFe$_{1.84}$Co$_{0.16}$As$_2$.
\newblock {\em Phys. Rev. Lett.}, 102:107005, 2009.

\bibitem{chi:107006}
Songxue Chi, Astrid Schneidewind, Jun Zhao, Leland~W. Harriger, Linjun Li,
  Yongkang Luo, Guanghan Cao, Zhu'an Xu, Micheal Loewenhaupt, Jiangping Hu, and
  Pengcheng Dai.
\newblock Inelastic Neutron-Scattering Measurements of a Three-Dimensional Spin
  Resonance in the FeAs-Based BaFe$_{1.9}$Ni$_{0.1}$As$_2$ Superconductor.
\newblock {\em Phys. Rev. Lett.}, 102:107006, 2009.

\bibitem{shiliang-2009}
Shiliang Li, Ying Chen, Sung Chang, Jeffrey~W. Lynn, Linjun Li, Yongkang Luo,
  Guanghan Cao, Zhu'an Xu, and Pengcheng Dai.
\newblock Spin gap and magnetic resonance in superconducting
  BaFe$_{1.9}$Ni$_{0.1}$As$_2$.
\newblock {\em Phys. Rev. B}, 79:174527, 2009.

\bibitem{inosov-2009}
D.~S. Inosov, J.~T. Park, P.~Bourges, D.~L. Sun, Y.~Sidis, A.~Schneidewind,
  K.~Hradil, D.~Haug, C.~T. Lin, B.~Keimer, and V.~Hinkov.
\newblock Normal-State Spin Dynamics and Temperature-Dependent Spin Resonance
  Energy in an Optimally Doped Iron Arsenide Superconductor.
\newblock {\em Nature Phys.}, 6:178, 2010.

\bibitem{resonance1111}
Shin-ichi Shamoto, Motoyuki Ishikado, Andrew~D. Christianson, Mark~D. Lumsden,
  Shuichi Wakimoto, Katsuaki Kodama, Akira Iyo, and Masatoshi Arai.
\newblock Inelastic neutron scattering study of the resonance mode in the
  optimally doped pnictide superconductor LaFeAsO$_{0.92}$F$_{0.08}$.
\newblock {\em Phys. Rev. B}, 82:172508, 2010.

\bibitem{qiu:067008}
Yiming Qiu, Wei Bao, Y.~Zhao, Collin Broholm, V.~Stanev, Z.~Tesanovic, Y.~C.
  Gasparovic, S.~Chang, Jin Hu, Bin Qian, Minghu Fang, and Zhiqiang Mao.
\newblock Spin Gap and Resonance at the Nesting Wave Vector in Superconducting
  FeSe$_{0.4}$Te$_{0.6}$.
\newblock {\em Phys. Rev. Lett.}, 103:067008, 2009.

\bibitem{mook-2009-2}
H.~A. Mook, M.~D. Lumsden, A.~D. Christianson, S.~E. Nagler, Brian~C. Sales,
  Rongying Jin, Michael~A. McGuire, Athena~S. Sefat, D.~Mandrus, T.~Egami, and
  Clarina dela Cruz.
\newblock Unusual Relationship between Magnetism and Superconductivity in
  FeTe$_{0.5}$Se$_{0.5}$.
\newblock {\em Phys. Rev. Lett.}, 104:187002, 2010.

\bibitem{fieldeffectresonancewen}
Jinsheng Wen, Guangyong Xu, Zhijun Xu, Zhi~Wei Lin, Qiang Li, Ying Chen,
  Songxue Chi, Genda Gu, and J.~M. Tranquada.
\newblock Effect of magnetic field on the spin resonance in
  FeTe$_{0.5}$Se$_{0.5}$ as seen via inelastic neutron scattering.
\newblock {\em Phys. Rev. B}, 81:100513(R), 2010.

\bibitem{JPSJ.78.062001}
Kenji Ishida, Yusuke Nakai, and Hideo Hosono.
\newblock To What Extent Iron-Pnictide New Superconductors Have Been Clarified:
  A Progress Report.
\newblock {\em J. Phys. Soc. Jpn.}, 78:062001, 2009.

\bibitem{JPSJS.77SC.1}
Hideo Hosono.
\newblock Layered Iron Pnictide Superconductors: Discovery and Current Status.
\newblock {\em J. Phys. Soc. Jpn. Supplement C}, 77:1, 2008.

\bibitem{lynn-2009-469}
Jeffrey~W. Lynn and Pengcheng Dai.
\newblock Neutron Studies of the Iron-based Family of High $T_C$ Magnetic
  Superconductors.
\newblock {\em Physcia C}, 469:469, 2009.

\bibitem{lumsdenreview1}
M.~D. {Lumsden} and A.~D. {Christianson}.
\newblock {Magnetism in Fe-based superconductors}.
\newblock {\em J. Phys.: Conden. Matter}, 22:203203, 2010.

\bibitem{Hosonoreviewphysicac}
Hideo Hosono.
\newblock Two classes of superconductors discovered in our material research:
  Iron-based high temperature superconductor and electride superconductor.
\newblock {\em Physica C}, 469:314, 2009.

\bibitem{mkwreview1}
M.K. Wu, F.C. Hsu, K.W. Yeh, T.W. Huang, J.Y. Luo, M.J. Wang, H.H. Chang, T.K.
  Chen, S.M. Rao, B.H. Mok, C.L. Chen, Y.L. Huang, C.T. Ke, P.M. Wu, A.M.
  Chang, C.T. Wu, and T.P. Perng.
\newblock The development of the superconducting PbO-type $\beta$-FeSe and
  related compounds.
\newblock {\em Physica C}, 469:340, 2009.

\bibitem{canfieldreview}
Paul~C. Canfield and Sergey~L. Bud'ko.
\newblock FeAs-Based Superconductivity: A Case Study of the Effects of
  Transition Metal Doping on BaFe$_2$As$_2$.
\newblock {\em Annual Rev. Conden. Matter Phys.}, 1:27, 2010.

\bibitem{dcjohnstonreview}
D.~C. Johnston.
\newblock The puzzle of high temperature superconductivity in layered iron
  pnictides and chalcogenides.
\newblock {\em Adv. Phys.}, 59:803, 2010.

\bibitem{mazinnaturereview}
I.~Mazin.
\newblock {Superconductivity gets an iron boost}.
\newblock {\em Nature}, 464:183, 2010.

\bibitem{htcironbasereview}
Johnpierre Paglione and Richard~L. Greene.
\newblock High-temperature superconductivity in iron-based materials.
\newblock {\em Nature Phys.}, 6:645, 2010.

\bibitem{scinkfese}
Jiangang Guo, Shifeng Jin, Gang Wang, Shunchong Wang, Kaixing Zhu, Tingting
  Zhou, Meng He, and Xiaolong Chen.
\newblock Superconductivity in the iron selenide K$_{x}$Fe$_{2}$Se$_{2}$
  $(0\le{}x\le{}1.0)$.
\newblock {\em Phys. Rev. B}, 82:180520, 2010.

\bibitem{kfese2}
Y.~{Mizuguchi}, H.~{Takeya}, Y.~{Kawasaki}, T.~{Ozaki}, S.~{Tsuda},
  T.~{Yamaguchi}, and Y.~{Takano}.
\newblock {Transport properties of the new Fe-based superconductor
  K$_{x}$Fe$_{2}$Se$_{2}$ ($T_c=33$ K)}.
\newblock {\em Appl. Phys. Lett.}, 98:042511, 2011.

\bibitem{rbfese1}
A.~F. Wang, J.~J. Ying, Y.~J. Yan, R.~H. Liu, X.~G. Luo, Z.~Y. Li, X.~F. Wang,
  M.~Zhang, G.~J. Ye, P.~Cheng, Z.~J. Xiang, and X.~H. Chen.
\newblock Superconductivity at 32 K in single-crystalline
  Rb$_{x}$Fe$_{2-y}$Se$_{2}$.
\newblock {\em Phys. Rev. B}, 83:060512, 2011.

\bibitem{csfese1}
A~Krzton-Maziopa, Z~Shermadini, E~Pomjakushina, V~Pomjakushin, M~Bendele,
  A~Amato, R~Khasanov, H~Luetkens, and K~Conder.
\newblock Synthesis and crystal growth of Cs$_{0.8}$(FeSe$_{0.98}$)$_2$: a new
  iron-based superconductor with $T_c=27$ K.
\newblock {\em J. Phys. Conden. Matter}, 23:052203, 2011.

\bibitem{tlrbfese1}
Hang-Dong Wang, Chi-Heng Dong, Zu-Juan Li, Qian-Hui Mao, Sha-Sha Zhu, Chun-Mu
  Feng, H.~Q. Yuan, and Ming-Hu Fang.
\newblock Superconductivity at 32 K and anisotropy in
  Tl$_{0.58}$Rb$_{0.42}$Fe$_{1.72}$Se$_2$ crystals.
\newblock {\em Euro. Phys. Lett.}, 93:47004, 2011.

\bibitem{fesephasediagram_r}
H.~Okamoto.
\newblock The Fe-Se (Iron-Selenium) system.
\newblock {\em J. Phase Equilib.}, 12:383, 1991.

\bibitem{mcqueen:014522}
T.~M. McQueen, Q.~Huang, V.~Ksenofontov, C.~Felser, Q.~Xu, H.~Zandbergen, Y.~S.
  Hor, J.~Allred, A.~J. Williams, D.~Qu, J.~Checkelsky, N.~P. Ong, and R.~J.
  Cava.
\newblock Extreme sensitivity of superconductivity to stoichiometry in
  Fe$_{1+\delta}$Se.
\newblock {\em Phys. Rev. B}, 79:014522, 2009.

\bibitem{patel:082508}
U.~Patel, J.~Hua, S.~H. Yu, S.~Avci, Z.~L. Xiao, H.~Claus, J.~Schlueter, V.~V.
  Vlasko-Vlasov, U.~Welp, and W.~K. Kwok.
\newblock Growth and superconductivity of FeSe$_x$ crystals.
\newblock {\em Appl. Phys. Lett.}, 94:082508, 2009.

\bibitem{flux015020}
S~B Zhang, Y~P Sun, X~D Zhu, X~B Zhu, B~S Wang, G~Li, H~C Lei, X~Luo, Z~R Yang,
  W~H Song, and J~M Dai.
\newblock Crystal growth and superconductivity of FeSe$_x$.
\newblock {\em Supercond. Sci. Technol.}, 22:015020, 2009.

\bibitem{Hu2011}
Rongwei Hu, Hechang Lei, Milinda Abeykoon, Emil~S. Bozin, Simon J.~L. Billinge,
  J.~B. Warren, Theo Siegrist, and C.~Petrovic.
\newblock Synthesis, crystal structure, and magnetism of
  $\beta{}$-Fe$_{1.00(2)}$Se$_{1.00(3)}$ single crystals.
\newblock {\em Phys. Rev. B}, 83:224502, 2011.

\bibitem{kclfluxgrowth}
B.~H. Mok, S.~M. Rao, M.~C. Ling, K.~J. Wang, C.~T. Ke, P.~M. Wu, C.~L. Chen,
  F.~C. Hsu, T.~W. Huang, J.~Y. Luo, D.~C. Yan, K.~W. Ye, T.~B. Wu, A.~M.
  Chang, and M.~K. Wu.
\newblock Growth and Investigation of Crystals of the New Superconductor
  $\alpha$-FeSe from KCl Solutions.
\newblock {\em Crystal Growth and Design}, 9(7):3260, 2009.

\bibitem{fetephasediagram_r}
H.~Okamoto and L.~E. Tanner.
\newblock Fe-Te (Iron-Tellurium).
\newblock {\em Binary Alloy Phase Diagrams}, 2:1781, 1990.

\bibitem{2011JAP93914M}
Y.~{Mizuguchi}, K.~{Deguchi}, Y.~{Kawasaki}, T.~{Ozaki}, M.~{Nagao},
  S.~{Tsuda}, T.~{Yamaguchi}, and Y.~{Takano}.
\newblock {Superconductivity in oxygen-annealed FeTe$_{1-x}$S$_{x}$ single
  crystal}.
\newblock {\em J. Appl. Phys.}, 109:013914, 2011.

\bibitem{liu:174509}
T.~J. Liu, X.~Ke, B.~Qian, J.~Hu, D.~Fobes, E.~K. Vehstedt, H.~Pham, J.~H.
  Yang, M.~H. Fang, L.~Spinu, P.~Schiffer, Y.~Liu, and Z.~Q. Mao.
\newblock Charge-carrier localization induced by excess Fe in the
  superconductor Fe$_{1+y}$Te$_{1-x}$Se$_x$.
\newblock {\em Phys. Rev. B}, 80:174509, 2009.

\bibitem{JPSJ.79.084711}
Takashi Noji, Takumi Suzuki, Haruki Abe, Tadashi Adachi, Masatsune Kato, and
  Yoji Koike.
\newblock Growth, Annealing Effects on Superconducting and Magnetic Properties,
  and Anisotropy of FeSe$_{1-x}$Te$_{x}$ ($0.5\leq x\leq 1$) Single Crystals.
\newblock {\em J. Phys. Soc. Jpn.}, 79:084711, 2010.

\bibitem{JPSJ.79.074704}
Jinhu Yang, Mami Matsui, Masatomo Kawa, Hiroto Ohta, Chishiro Michioka, Chiheng
  Dong, Hangdong Wang, Huiqiu Yuan, Minghu Fang, and Kazuyoshi Yoshimura.
\newblock Magnetic and Superconducting Properties in Single Crystalline
  Fe$_{1+\delta}$Te$_{1-x}$Se$_{x}$ ($x<0.50$) System.
\newblock {\em J. Phys. Soc. Jpn.}, 79:074704, 2010.

\bibitem{wen:104506}
{Jinsheng Wen}, {Guangyong Xu}, {Zhijun Xu}, {Zhi Wei Lin}, {Qiang Li},
  W.~Ratcliff, Genda Gu, and J.~M. Tranquada.
\newblock Short-range incommensurate magnetic order near the superconducting
  phase boundary in Fe$_{1+\delta}$Te$_{1-x}$Se$_x$.
\newblock {\em Phys. Rev. B}, 80:104506, 2009.

\bibitem{homesopticalprb}
C.~C. Homes, A.~Akrap, { J. S. Wen}, Z.~J. Xu, Z.~W. Lin, Q.~Li, and G.~D. Gu.
\newblock {Electronic correlations and unusual superconducting response in the
  optical properties of the iron-chalcogenide FeTe$_{0.55}$Se$_{0.45}$}.
\newblock {\em Phys. Rev. B}, 81:180508(R), 2010.

\bibitem{shcouplingprb}
S.-H. Lee, Guangyong Xu, W.~Ku, J.~S. Wen, C.~C. Lee, N.~Katayama, Z.~J. Xu,
  S.~Ji, Z.~W. Lin, G.~D. Gu, H.-B. Yang, P.~D. Johnson, Z.-H. Pan, T.~Valla,
  M.~Fujita, T.~J. Sato, S.~Chang, K.~Yamada, and J.~M. Tranquada.
\newblock Coupling of spin and orbital excitations in the iron-based
  superconductor FeTe$_{0.5}$Se$_{0.5}$.
\newblock {\em Phys. Rev. B}, 81:220502(R), 2010.

\bibitem{xudoping11}
Zhijun Xu, Jinsheng Wen, Guangyong Xu, Qing Jie, Zhiwei Lin, Qiang Li, Songxue
  Chi, D.~K. Singh, Genda Gu, and J.~M. Tranquada.
\newblock Disappearance of static magnetic order and evolution of spin
  fluctuations in Fe$_{1+\delta}$Se$_{x}$Te$_{1-x}$.
\newblock {\em Phys. Rev. B}, 82:104525, 2010.

\bibitem{PhysRevB.80.214511}
I.~Pallecchi, G.~Lamura, M.~Tropeano, M.~Putti, R.~Viennois, E.~Giannini, and
  D.~Van~der Marel.
\newblock Seebeck effect in Fe$_{1+x}$Te$_{1-y}$Se$_{y}$ single crystals.
\newblock {\em Phys. Rev. B}, 80:214511, 2009.

\bibitem{taen:092502}
T.~Taen, Y.~Tsuchiya, Y.~Nakajima, and T.~Tamegai.
\newblock Superconductivity at $T_c\sim14$ K in single-crystalline
  FeTe$_{0.61}$Se$_{0.39}$.
\newblock {\em Phys. Rev. B}, 80:092502, 2009.

\bibitem{PhysRevB.81.020509}
Minghu Fang, Jinhu Yang, F.~F. Balakirev, Y.~Kohama, J.~Singleton, B.~Qian,
  Z.~Q. Mao, Hangdong Wang, and H.~Q. Yuan.
\newblock Weak anisotropy of the superconducting upper critical field in
  Fe$_{1.11}$Te$_{0.6}$Se$_{0.4}$ single crystals.
\newblock {\em Phys. Rev. B}, 81:020509, 2010.

\bibitem{PhysRevB.80.214514}
Rongwei Hu, Emil~S. Bozin, J.~B. Warren, and C.~Petrovic.
\newblock Superconductivity, magnetism, and stoichiometry of single crystals of
  Fe$_{1+y}$(Te$_{1-x}$S$_{x}$)$_{z}$.
\newblock {\em Phys. Rev. B}, 80:214514, 2009.

\bibitem{yeh-2009}
K.~W. {Yeh}, C.~T. {Ke}, T.~W. {Huang}, T.~K. {Chen}, Y.~L. {Huang}, P.~M.
  {Wu}, and M.~K. {Wu}.
\newblock {Superconducting FeSe$_{1-x}$Te$_x$ Single Crystals Grown by Optical
  Zone-Melting Technique}.
\newblock {\em Cryst. Growth Des.}, 9:4847, 2009.

\bibitem{2011arXiv1101.0789W}
D.~M. Wang, J.~B. He, T.-L. Xia, and G.~F. Chen.
\newblock Effect of varying iron content on the transport properties of the
  potassium-intercalated iron selenide K$_x$Fe$_{2-y}$Se$_2$.
\newblock {\em Phys. Rev. B}, 83:132502, 2011.

\bibitem{2011arXiv1102.1010L}
Hechang Lei and C.~Petrovic.
\newblock Anisotropy in transport and magnetic properties of
  K$_{0.64}$Fe$_{1.44}$Se$_{2}$.
\newblock {\em Phys. Rev. B}, 83:184504, 2011.

\bibitem{kcsfese1}
J.~J. Ying, X.~F. Wang, X.~G. Luo, A.~F. Wang, M.~Zhang, Y.~J. Yan, Z.~J.
  Xiang, R.~H. Liu, P.~Cheng, G.~J. Ye, and X.~H. Chen.
\newblock Superconductivity and magnetic properties of single crystals of
  K$_{0.75}$Fe$_{1.66}$Se$_2$ and Cs$_{0.81}$Fe$_{1.61}$Se$_2$.
\newblock {\em Phys. Rev. B}, 83:212502, 2011.

\bibitem{2010arXiv1012.5236F}
Ming-Hu Fang, Hang-Dong Wang, Chi-Heng Dong, Zu-Juan Li, Chun-Mu Feng, Jian
  Chen, and H.~Q. Yuan.
\newblock Fe-based superconductivity with $T_c=31$ K bordering an
  antiferromagnetic insulator in (Tl,K)Fe$_x$Se$_2$.
\newblock {\em Euro. Phys. Lett.}, 94:27009, 2011.

\bibitem{2011arXiv1103.2904G}
Z.~{Gao}, Y.~{Qi}, L.~{Wang}, C.~{Yao}, D.~{Wang}, X.~{Zhang}, and Y.~{Ma}.
\newblock {Upper fields and critical current density of
  K$_{0.58}$Fe$_{1.56}$Se$_2$ single crystals grown by one step technique}.
\newblock {\em arXiv:1103.2904}, 2011.

\bibitem{wenphdthesis}
Jinsheng Wen.
\newblock {\em Interplay between magnetism and superconductivity in
  high-temperature superconductors La$_{2-x}$Ba$_x$CuO$_4$ and
  Fe$_{1+y}$Te$_{1-x}$Se$_x$: crystal growth and neutron scattering studies}.
\newblock PhD thesis, Stony Brook University, 2010.

\bibitem{kfesecrystalgrowth}
Zhijun Xu, Jinsheng Wen, Guangyong Xu, Sujung Han, Qiang Li, J.~M. Tranquada,
  and Genda Gu.
\newblock unpublished.

\bibitem{PhysRevB.82.020502}
B.~Joseph, A.~Iadecola, A.~Puri, L.~Simonelli, Y.~Mizuguchi, Y.~Takano, and
  N.~L. Saini.
\newblock Evidence of local structural inhomogeneity in FeSe$_{1-x}$Te$_{x}$
  from extended x-ray absorption fine structure.
\newblock {\em Phys. Rev. B}, 82:020502(R), 2010.

\bibitem{phasese11}
Hefei Hu, J.~M. Zuo, J.~S. Wen, Z.~J. Xu, Z.~W. Lin, Q.~Li, Genda Gu, W.~K.
  Park, and L.~H. Greene.
\newblock Phase Separation and Chemical Inhomogeneity in the Iron Chalcogenide
  Superconductor Fe$_{1+y}$Te$_x$Se$_{1-x}$.
\newblock {\em arXiv:1009.6010}, 2010.

\bibitem{He2011}
Xiaobo He, Guorong Li, Jiandi Zhang, A.~B. Karki, Rongying Jin, B.~C. Sales,
  A.~S. Sefat, M.~A. McGuire, D.~Mandrus, and E.~W. Plummer.
\newblock {Nanoscale chemical phase separation in FeTe$_{0.55}$Se$_{0.45}$ as
  seen via scanning tunneling spectroscopy}.
\newblock {\em Phys. Rev. B}, 83:220502, 2011.

\bibitem{PhysRevB.83.140505}
Z.~Wang, Y.~J. Song, H.~L. Shi, Z.~W. Wang, Z.~Chen, H.~F. Tian, G.~F. Chen,
  J.~G. Guo, H.~X. Yang, and J.~Q. Li.
\newblock Microstructure and ordering of iron vacancies in the superconductor
  system K$_{y}$Fe$_{x}$Se$_{2}$ as seen via transmission electron microscopy.
\newblock {\em Phys. Rev. B}, 83:140505, 2011.

\bibitem{2011arXiv1106.3026C}
F.~{Chen}, M.~{Xu}, Q.~Q. {Ge}, Y.~{Zhang}, Z.~R. {Ye}, L.~X. {Yang},
  J.~{Jiang}, B.~P. {Xie}, R.~C. {Che}, M.~{Zhang}, A.~F. {Wang}, X.~H. {Chen},
  D.~W. {Shen}, X.~M. {Xie}, M.~H. {Jiang}, J.~P. {Hu}, and D.~L. {Feng}.
\newblock {Electronic identification of the actual parental phase of
  K$_x$Fe$_{2-y}$Se$_2$ superconductor and its intrinsic mesoscopic phase
  separation}.
\newblock {\em arXiv:1106.3026}, 2011.

\bibitem{B813076K}
Serena Margadonna, Yasuhiro Takabayashi, Martin~T. McDonald, Karolina
  Kasperkiewicz, Yoshikazu Mizuguchi, Yoshihiko Takano, Andrew~N. Fitch,
  Emmanuelle Suard, and Kosmas Prassides.
\newblock Crystal structure of the new FeSe$_{1-x}$ superconductor.
\newblock {\em Chem. Commun.}, page 5607, 2008.

\bibitem{PhysRevB.80.024517}
E.~Pomjakushina, K.~Conder, V.~Pomjakushin, M.~Bendele, and R.~Khasanov.
\newblock Synthesis, crystal structure, and chemical stability of the
  superconductor FeSe$_{1-x}$.
\newblock {\em Phys. Rev. B}, 80:024517, 2009.

\bibitem{Liu2011}
X.~Liu, C.-C. Lee, Z.~J. Xu, J.~S. Wen, G.~Gu, W.~Ku, J.~M. Tranquada, and
  J.~P. Hill.
\newblock {X-ray diffuse scattering study of local distortions in Fe$_{1+x}$Te
  induced by excess Fe}.
\newblock {\em Phys. Rev. B}, 83:184523, 2011.

\bibitem{fesephasetr}
T.~M. McQueen, A.~J. Williams, P.~W. Stephens, J.~Tao, Y.~Zhu, V.~Ksenofontov,
  F.~Casper, C.~Felser, and R.~J. Cava.
\newblock Tetragonal-to-Orthorhombic Structural Phase Transition at 90 K in the
  Superconductor Fe$_{1.01}$Se.
\newblock {\em Phys. Rev. Lett.}, 103:057002, 2009.

\bibitem{bao-2009}
Wei Bao, Y.~Qiu, Q.~Huang, M.~A. Green, P.~Zajdel, M.~R. Fitzsimmons,
  M.~Zhernenkov, M.~Fang, B.~Qian, E.~K. Vehstedt, J.~Yang, H.~M. Pham,
  L.~Spinu, and Z.~Q. Mao.
\newblock Tunable ($\delta\pi, \delta\pi$)-Type Antiferromagnetic Order in
  $\alpha$-Fe(Te,Se) Superconductors.
\newblock {\em Phys. Rev. Lett.}, 102:247001, 2009.

\bibitem{li-2009-79}
Shiliang Li, Clarina {de la Cruz}, Q.~Huang, Y.~Chen, J.~W. Lynn, Jiangping Hu,
  Yi-Lin Huang, Fong-Chi Hsu, Kuo-Wei Yeh, Maw kuen Wu, and Pengcheng Dai.
\newblock First-order magnetic and structural phase transitions in
  Fe$_{1+y}$Se$_x$Te$_{1-x}$.
\newblock {\em Phys. Rev. B}, 79:054503, 2009.

\bibitem{PhysRevB.81.094115}
A.~Martinelli, A.~Palenzona, M.~Tropeano, C.~Ferdeghini, M.~Putti, M.~R.
  Cimberle, T.~D. Nguyen, M.~Affronte, and C.~Ritter.
\newblock From antiferromagnetism to superconductivity in
  Fe$_{1+y}$Te$_{1-x}$Se$_{x}$ $(0\le{}x\le{}0.20)$ : Neutron powder
  diffraction analysis.
\newblock {\em Phys. Rev. B}, 81:094115, 2010.

\bibitem{mizuguchi-2009-94}
Y.~Mizuguchi, F.~Tomioka, S.~Tsuda, T.~Yamaguchi, and Y.~Takano.
\newblock Superconductivity in S-substituted FeTe.
\newblock {\em Appl. Phys. Lett.}, 94:012503, 2009.

\bibitem{mizuguchi-2009-469}
Y.~Mizuguchi, F.~Tomioka, S.~Tsuda, T.~Yamaguchi, and Y.~Takano.
\newblock FeTe as a candidate material for new iron-based superconductor.
\newblock {\em Physica C}, 469:1027, 2009.

\bibitem{JPSJ.78.074712}
Yoshikazu Mizuguchi, Fumiaki Tomioka, Shunsuke Tsuda, Takahide Yamaguchi, and
  Yoshihiko Takano.
\newblock Substitution Effects on FeSe Superconductor.
\newblock {\em J. Phys. Soc. Jpn.}, 78:074712, 2009.

\bibitem{si:052504}
Weidong Si, Zhi-Wei Lin, Qing Jie, Wei-Guo Yin, Juan Zhou, Genda Gu, P.~D.
  Johnson, and Qiang Li.
\newblock Enhanced superconducting transition temperature in
  FeSe$_{0.5}$Te$_{0.5}$ thin films.
\newblock {\em Appl. Phys. Lett.}, 95:052504, 2009.

\bibitem{APEX.4.053101}
Ichiro Tsukada, Masafumi Hanawa, Takanori Akiike, Fuyuki Nabeshima, Yoshinori
  Imai, Ataru Ichinose, Seiki Komiya, Tatsuo Hikage, Takahiko Kawaguchi,
  Hiroshi Ikuta, and Atsutaka Maeda.
\newblock Epitaxial Growth of FeSe$_{0.5}$Te$_{0.5}$ Thin Films on CaF$_{2}$
  Substrates with High Critical Current Density.
\newblock {\em Applied Physics Express}, 4:053101, 2011.

\bibitem{imai-2009}
Yoshinori Imai, Ryo Tanaka, Takanori Akiike, Masafumi Hanawa, Ichiro Tsukada,
  and Atsutaka Maeda.
\newblock Superconductivity of FeSe$_{0.5}$Te$_{0.5}$ Thin Films Grown by
  Pulsed Laser Deposition.
\newblock {\em Jpn. J. Appl. Phys.}, 49:023101, 2010.

\bibitem{mkwreview2}
M.~K. Wu, K.~W. Yeh, H.~C. Hsu, T.~W. Huang, T.~K. Chen, J.~Y. Luo, M.~J. Wang,
  H.~H. Chang, C.~Ke, M.~H. Moh, and S.~M.~D Rao.
\newblock The development of the superconducting tetragonal PbO-type FeSe and
  related compounds.
\newblock {\em Physica Status Solidi (B)}, 247:500, 2010.

\bibitem{kida-2009}
Takanori Kida, Takahiro Matsunaga, Masayuki Hagiwara, Yoshikazu Mizuguchi,
  Yoshihiko Takano, and Koichi Kindo.
\newblock Upper critical fields of the 11-system iron-chalcogenide
  superconductor FeSe$_{0.25}$Te$_{0.75}$.
\newblock {\em J. Phys. Soc. Jpn.}, 78:113701, 2009.

\bibitem{njphysuf1}
C~S Yadav and P~L Paulose.
\newblock Upper critical field, lower critical field and critical current
  density of FeTe$_{0.60}$ Se$_{0.40}$ single crystals.
\newblock {\em New J. Phys.}, 11:103046, 2009.

\bibitem{Roessler2010}
S.~{R\"o{\ss}ler}, Dona Cherian, S.~Harikrishnan, H.~L. Bhat, Suja Elizabeth,
  J.~A. Mydosh, L.~H. Tjeng, F.~Steglich, and S.~Wirth.
\newblock {Disorder-driven electronic localization and phase separation in
  superconducting Fe$_{1+y}$Te$_{0.5}$Se$_{0.5}$ single crystals}.
\newblock {\em Phys. Rev. B}, 82:144523, 2010.

\bibitem{Bendele2010}
M.~Bendele, P.~Babkevich, S.~Katrych, S.~N. Gvasaliya, E.~Pomjakushina,
  K.~Conder, B.~Roessli, A.~T. Boothroyd, R.~Khasanov, and H.~Keller.
\newblock {Tuning the superconducting and magnetic properties of
  Fe$_{y}$Se$_{0.25}$Te$_{0.75}$ by varying the iron content}.
\newblock {\em Phys. Rev. B}, 82:212504, 2010.

\bibitem{2010arXiv1012.0590R}
E.~E. {Rodriguez}, C.~{Stock}, P.~{Hsieh}, N.~{Butch}, J.~{Paglione}, and M.~A.
  {Green}.
\newblock {Interstitial Iron Controlled Superconductivity in
  Fe$_{1+x}$Te$_{0.7}$Se$_{0.3}$}.
\newblock {\em arXiv:1012.0590}, 2010.

\bibitem{medvedev-2009-8}
S.~Medvedev, T.~M. McQueen, I.~Trojan, T.~Palasyuk, M.~I. Eremets, R.~J. Cava,
  S.~Naghavi, F.~Casper, V.~Ksenofontov, G.~Wortmann, and C.~Felser.
\newblock Electronic and magnetic phase diagram of $\alpha$-Fe$_{1.01}$Se with
  superconductivity at 36.7 K under pressure.
\newblock {\em Nature Mater.}, 8:630, 2009.

\bibitem{margadonna:064506}
S.~Margadonna, Y.~Takabayashi, Y.~Ohishi, Y.~Mizuguchi, Y.~Takano, T.~Kagayama,
  T.~Nakagawa, M.~Takata, and K.~Prassides.
\newblock Pressure evolution of the low-temperature crystal structure and
  bonding of the superconductor FeSe ($T_c$ = 37 K).
\newblock {\em Phys. Rev. B}, 80:064506, 2009.

\bibitem{JPSJ.78.063704}
Satoru Masaki, Hisashi Kotegawa, Yudai Hara, Hideki Tou, Keizo Murata,
  Yoshikazu Mizuguchi, and Yoshihiko Takano.
\newblock Precise Pressure Dependence of the Superconducting Transition
  Temperature of FeSe: Resistivity and $^{77}$Se-NMR Study.
\newblock {\em J. Phys. Soc. Jpn.}, 78:063704, 2009.

\bibitem{tissen:092507}
V.~G. Tissen, E.~G. Ponyatovsky, M.~V. Nefedova, A.~N. Titov, and V.~V.
  Fedorenko.
\newblock Effects of pressure-induced phase transitions on superconductivity in
  single-crystal Fe$_{1.02}$Se.
\newblock {\em Phys. Rev. B}, 80:092507, 2009.

\bibitem{mizuguchi:152505}
Yoshikazu Mizuguchi, Fumiaki Tomioka, Shunsuke Tsuda, Takahide Yamaguchi, and
  Yoshihiko Takano.
\newblock Superconductivity at 27 K in tetragonal FeSe under high pressure.
\newblock {\em Appl. Phys. Lett.}, 93:152505, 2008.

\bibitem{miyoshi-2009-78}
K.~Miyoshi, Y.~Takaichi, E.~Mutou, K.~Fujiwara, and J.~Takeuchi.
\newblock Anomalous Pressure Dependence of the Superconducting Transition
  Temperature in FeSe$_{1-x}$ Studied by DC Magnetic Measurements.
\newblock {\em J. Phys. Soc. Jpn}, 78:093703, 2009.

\bibitem{JPSJ.78.063705}
Kazumasa Horigane, Nao Takeshita, Chul-Ho Lee, Haruhiro Hiraka, and Kazuyoshi
  Yamada.
\newblock First Investigation of Pressure Effects on Transition from
  Superconductive to Metallic Phase in FeSe$_{0.5}$Te$_{0.5}$.
\newblock {\em J. Phys. Soc. Japn.}, 78:063705, 2009.

\bibitem{Mizuguchi2010S353}
Yoshikazu Mizuguchi, Fumiaki Tomioka, Keita Deguchi, Shunsuke Tsuda, Takahide
  Yamaguchi, and Yoshihiko Takano.
\newblock Pressure effects on FeSe family superconductors.
\newblock {\em Physica C}, 470:S353, 2010.

\bibitem{pressurefetes}
Y~Mizuguchi, Y~Hara, K~Deguchi, S~Tsuda, T~Yamaguchi, K~Takeda, H~Kotegawa,
  H~Tou, and Y~Takano.
\newblock Anion height dependence of $T_c$ for the Fe-based superconductor.
\newblock {\em Supercond. Sci. Technol.}, 23:054013, 2010.

\bibitem{pressurefetese}
Nathalie~C. Gresty, Yasuhiro Takabayashi, Alexey~Y. Ganin, Martin~T. McDonald,
  John~B. Claridge, Duong Giap, Yoshikazu Mizuguchi, Yoshihiko Takano, Tomoko
  Kagayama, Yasuo Ohishi, Masaki Takata, Matthew~J. Rosseinsky, Serena
  Margadonna, and Kosmas Prassides.
\newblock Structural Phase Transitions and Superconductivity in
  Fe$_{1+\delta}$Se$_{0.57}$Te$_{0.43}$ at Ambient and Elevated Pressures.
\newblock {\em J. Am. Chem. Soc.}, 131:16944, 2009.

\bibitem{JPSJ.78.083709}
Hironari Okada, Hiroyuki Takahashi, Yoshikazu Mizuguchi, Yoshihiko Takano, and
  Hiroki Takahashi.
\newblock Successive Phase Transitions under High Pressure in FeTe$_{0.92}$.
\newblock {\em J. Phys. Soc. Jpn.}, 78:083709, 2009.

\bibitem{PhysRevB.80.144519}
Chao Zhang, Wei Yi, Liling Sun, Xiao-Jia Chen, Russell~J. Hemley, Ho-kwang Mao,
  Wei Lu, Xiaoli Dong, Ligang Bai, Jing Liu, Antonio~F. Moreira Dos~Santos,
  Jamie~J. Molaison, Christopher~A. Tulk, Genfu Chen, Nanlin Wang, and
  Zhongxian Zhao.
\newblock Pressure-induced lattice collapse in the tetragonal phase of
  single-crystalline $Fe_{1.05}Te$.
\newblock {\em Phys. Rev. B}, 80:144519, 2009.

\bibitem{PhysRevLett.104.017003}
Y.~Han, W.~Y. Li, L.~X. Cao, X.~Y. Wang, B.~Xu, B.~R. Zhao, Y.~Q. Guo, and
  J.~L. Yang.
\newblock Superconductivity in Iron Telluride Thin Films under Tensile Stress.
\newblock {\em Phys. Rev. Lett.}, 104:017003, 2010.

\bibitem{danielfeseco}
Evan~Lyle Thomas, Winnie Wong-Ng, Daniel Phelan, and Jasmine~N. Millican.
\newblock Thermopower of Co-doped FeSe.
\newblock {\em J. Appl. Phys.}, 105:073906, 2009.

\bibitem{williams-2009-21}
A.~J. Williams, T.~M. McQueen, V.~Ksenofontov, C.~Felser, and R.~J. Cava.
\newblock The metal-insulator transition in Fe$_{1.01-x}$Cu$_x$Se.
\newblock {\em J. Phys. Condens. Matter}, 21:305701, 2009.

\bibitem{PhysRevB.82.104502}
Tzu-Wen Huang, Ta-Kun Chen, Kuo-Wei Yeh, Chung-Ting Ke, Chi~Liang Chen, Yi-Lin
  Huang, Fong-Chi Hsu, Maw-Kuen Wu, Phillip~M. Wu, Maxim Avdeev, and Andrew~J.
  Studer.
\newblock Doping-driven structural phase transition and loss of
  superconductivity in $M_{x}$Fe$_{1-x}$Se$_{\delta}$ ($M=$ Mn, Cu).
\newblock {\em Phys. Rev. B}, 82:104502, 2010.

\bibitem{Zhang20091958}
S.~B. Zhang, H.C. Lei, X.D. Zhu, G.~Li, B.S. Wang, L.J. Li, X.B. Zhu, W.H.
  Song, Z.R. Yang, and Y.P. Sun.
\newblock Divergency of SDW and structure transition in
  Fe$_{1-x}$Ni$_x$Se$_{0.82}$ superconductors.
\newblock {\em Physica C}, 469:1958, 2009.

\bibitem{2010arXiv1010.4217G}
D.~J. {Gawryluk}, J.~{Fink-Finowicki}, A.~{Wisniewski}, R.~{Puzniak},
  V.~{Domukhovski}, R.~{Diduszko}, M.~{Kozlowski}, and M.~{Berkowski}.
\newblock {Growth conditions, structure, and superconductivity of pure and
  metal-doped FeTe$_{1-x}$Se$_x$ single crystals}.
\newblock {\em Supercond. Sci. Tech.}, 24:065011, 2011.

\bibitem{fetesenico1}
R.~Shipra, H.~Takeya, K.~Hirata, and A.~Sundaresan.
\newblock Effects of Ni and Co doping on the physical properties of tetragonal
  FeSe$_{0.5}$Te$_{0.5}$ superconductor.
\newblock {\em Physica C}, 470:528, 2010.

\bibitem{feteni1}
P.~Zajdel, M.~Zubko, J.~Kusz, and M.~A. Green.
\newblock Single crystal growth and structural properties of iron telluride
  doped with nickel.
\newblock {\em Cryst. Res. Technol.}, 45:1316, 2010.

\bibitem{2011arXiv1101.5117Z}
B.~Zeng, B.~Shen, G.~F. Chen, J.~B. He, D.~M. Wang, C.~H. Li, and H.~H. Wen.
\newblock Nodeless superconductivity of single-crystalline
  K$_{x}$Fe$_{2-y}$Se$_{2}$ revealed by the low-temperature specific heat.
\newblock {\em Phys. Rev. B}, 83:144511, 2011.

\bibitem{2011arXiv1102.1931H}
R.~{Hu}, K.~{Cho}, H.~{Kim}, H.~{Hodovanets}, W.~E. {Straszheim}, M.~A.
  {Tanatar}, R.~{Prozorov}, S.~L. {Bud'ko}, and P.~C. {Canfield}.
\newblock Anisotropic magnetism, resistivity, London penetration depth and
  magneto-optical imaging of superconducting K$_{0.86}$Fe$_{1.84}$Se$_2$ single
  crystals.
\newblock {\em Supercon. Sci. Tech.}, 24:065006, 2011.

\bibitem{2011arXiv1103.0098L}
Z.~{Li}, X.~{Ma}, H.~{Pang}, and F.~{Li}.
\newblock {Evidence of spin excitation gap in K$_{0.86}$Fe$_{1.73}$Se$_2$
  superconductor as revealed by M\"{o}ssbauer spectroscopy}.
\newblock {\em arXiv:1103.0098}, 2011.

\bibitem{2011arXiv1101.4967T}
D.~A. Torchetti, M.~Fu, D.~C. Christensen, K.~J. Nelson, T.~Imai, H.~C. Lei,
  and C.~Petrovic.
\newblock $^{77}Se$ NMR investigation of the K$_{x}$Fe$_{2-y}$Se$_{2}$
  high-$T_{c}$ superconductor ($T_{c}=33$ K).
\newblock {\em Phys. Rev. B}, 83:104508, 2011.

\bibitem{2011arXiv1102.1381Y}
R.~H. {Yuan}, T.~{Dong}, G.~F. {Chen}, J.~B. {He}, D.~M. {Wang}, and N.~L.
  {Wang}.
\newblock {Nanoscale stripe-type phase separation of antiferromagnetic order
  and superconductivity in K$_{0.75}$Fe$_{1.75}$Se$_2$}.
\newblock {\em arXiv:1102.1381}, 2011.

\bibitem{2011arXiv1103.0507M}
E.~D. Mun, M.~M. Altarawneh, C.~H. Mielke, V.~S. Zapf, R.~Hu, S.~L. Bud'ko, and
  P.~C. Canfield.
\newblock Anisotropic $H_{c2}$ of K$_{0.8}$Fe$_{1.76}$Se$_{2}$ determined up to
  60 T.
\newblock {\em Phys. Rev. B}, 83:100514, 2011.

\bibitem{2011arXiv1101.4572K}
Hisashi Kotegawa, Yudai Hara, Hiroki Nohara, Hideki Tou, Yoshikazu Mizuguchi,
  Hiroyuki Takeya, and Yoshihiko Takano.
\newblock Possible Superconducting Symmetry and Magnetic Correlations in
  K$_{0.8}$Fe$_{2}$Se$_{2}$: A $^{77}$Se-NMR Study.
\newblock {\em Phys. Rev. B}, 80:043708, 2011.

\bibitem{2011arXiv1102.3380P}
V~Yu Pomjakushin, E~V Pomjakushina, A~Krzton-Maziopa, K~Conder, and
  Z~Shermadini.
\newblock Room temperature antiferromagnetic order in superconducting
  $X_y$Fe$_{2-x}$Se$_2$, ($X=$ Rb, K): a neutron powder diffraction study.
\newblock {\em J. Phys. Conden. Matter}, 23:156003, 2011.

\bibitem{2011arXiv1101.5670L}
X~G Luo, X~F Wang, J~J Ying, Y~J Yan, Z~Y Li, M~Zhang, A~F Wang, P~Cheng, Z~J
  Xiang, G~J Ye, R~H Liu, and X~H Chen.
\newblock Crystal structure, physical properties and superconductivity in
  $A_{x}$Fe$_2$Se$_2$ single crystals.
\newblock {\em New J. Phys.}, 13:053011, 2011.

\bibitem{2011arXiv1102.2464S}
G.~Seyfarth, D.~Jaccard, P.~Pedrazzini, A.~Krzton-Maziopa, E.~Pomjakushina,
  K.~Conder, and Z.~Shermadini.
\newblock {Pressure cycle of superconducting Cs$_{0.8}$Fe$_2$Se$_2$: a
  transport study}.
\newblock {\em Solid State Communications}, 151:747, 2011.

\bibitem{2011arXiv1102.1919P}
V.~Yu. Pomjakushin, D.~V. Sheptyakov, E.~V. Pomjakushina, A.~Krzton-Maziopa,
  K.~Conder, D.~Chernyshov, V.~Svitlyk, and Z.~Shermadini.
\newblock Iron-vacancy superstructure and possible room-temperature
  antiferromagnetic order in superconducting Cs$_{y}$Fe$_{2-x}$Se$_{2}$.
\newblock {\em Phys. Rev. B}, 83:144410, 2011.

\bibitem{2011arXiv1102.2783L}
R.~H. Liu, X.~G. Luo, M.~Zhang, A.~F. Wang, J.~J. Ying, X.~F. Wang, Y.~J. Yan,
  Z.~J. Xiang, P.~Cheng, G.~J. Ye, Z.~Y. Li, and X.~H. Chen.
\newblock Coexistence of superconductivity and antiferromagnetism in single
  crystals $A_{0.8}$Fe$_{2-y}$Se$_2$ ($A=$ K, Rb, Cs, Tl/K and Tl/Rb): Evidence
  from magnetization and resistivity.
\newblock {\em Euro. Phys. Lett.}, 94:27008, 2011.

\bibitem{2011arXiv1101.1873S}
Z.~Shermadini, A.~Krzton-Maziopa, M.~Bendele, R.~Khasanov, H.~Luetkens,
  K.~Conder, E.~Pomjakushina, S.~Weyeneth, V.~Pomjakushin, O.~Bossen, and
  A.~Amato.
\newblock Coexistence of Magnetism and Superconductivity in the Iron-Based
  Compound Cs$_{0.8}$(FeSe$_{0.98}$)$_2$.
\newblock {\em Phys. Rev. Lett.}, 106:117602, 2011.

\bibitem{2011arXiv1102.2434L}
H.~{Lei}, K.~{Wang}, J.~B. {Warren}, and C.~{Petrovic}.
\newblock {Magnetic and superconducting phase diagram of
  K$_x$Fe$_{2-y}$Se$_{2-z}$S$_z$ ($0\leq z\leq2$)}.
\newblock {\em arXiv:1102.2434}, 2011.

\bibitem{2011arXiv1101.5327L}
L.~{Li}, Z.~R. {Yang}, Z.~T. {Zhang}, W.~{Tong}, C.~J. {Zhang}, S.~{Tan}, and
  Y.~H. {Zhang}.
\newblock {Coexistence of superconductivity and magnetism in
  K$_{0.8}$Fe$_2$Se$_{1.4}$S$_{0.4}$}.
\newblock {\em arXiv:1101.5327}, 2011.

\bibitem{2011arXiv1102.3506Z}
T.~{Zhou}, X.~{Chen}, J.~{Guo}, G.~{Wang}, X.~{Lai}, S.~{Wang}, S.~{Jin}, and
  K.~{Zhu}.
\newblock {Quenching of superconductivity by Co doping in
  K$_{0.8}$Fe$_2$Se$_2$}.
\newblock {\em arXiv:1102.3506}, 2011.

\bibitem{2011arXiv1102.0830B}
W.~{Bao}, Q.~{Huang}, G.~F. {Chen}, M.~A. {Green}, D.~M. {Wang}, J.~B. {He},
  X.~Q. {Wang}, and Y.~{Qiu}.
\newblock {A Novel Large Moment Antiferromagnetic Order in
  K$_{0.8}$Fe$_{1.6}$Se$_2$ Superconductor}.
\newblock {\em arXiv:1102.0830}, 2011.

\bibitem{2011arXiv1102.2217W}
Kefeng Wang, Hechang Lei, and C.~Petrovic.
\newblock Thermoelectric studies of $K_{x}$Fe$_{2-y}$Se$_{2}$ indicating a
  weakly correlated superconductor.
\newblock {\em Phys. Rev. B}, 83:174503, 2011.

\bibitem{2010arXiv1012.5637L}
Chun-Hong Li, Bing Shen, Fei Han, Xiyu Zhu, and Hai-Hu Wen.
\newblock Transport properties and anisotropy of Rb$_{1-x}$Fe$_{2-y}$Se$_2$
  single crystals.
\newblock {\em Phys. Rev. B}, 83:184521, 2011.

\bibitem{2011arXiv1103.0059R}
D.~H. Ryan, W.~N. Rowan-Weetaluktuk, J.~M. Cadogan, R.~Hu, W.~E. Straszheim,
  S.~L. Bud'ko, and P.~C. Canfield.
\newblock $^{57}$Fe M\"ossbauer study of magnetic ordering in superconducting
  K$_{0.80}$Fe$_{1.76}$Se$_{2.00}$ single crystals.
\newblock {\em Phys. Rev. B}, 83:104526, 2011.

\bibitem{2011arXiv1101.1234Y}
J~J Ying, X~F Wang, X~G Luo, Z~Y Li, Y~J Yan, M~Zhang, A~F Wang, P~Cheng, G~J
  Ye, Z~J Xiang, R~H Liu, and X~H Chen.
\newblock Pressure effect on superconductivity of $A_x$Fe$_2$Se$_2$ ($A$ = K
  and Cs).
\newblock {\em New J. Phys.}, 13:033008, 2011.

\bibitem{2011arXiv1101.4556M}
Daixiang Mou, Shanyu Liu, Xiaowen Jia, Junfeng He, Yingying Peng, Lin Zhao,
  Li~Yu, Guodong Liu, Shaolong He, Xiaoli Dong, Jun Zhang, Hangdong Wang,
  Chiheng Dong, Minghu Fang, Xiaoyang Wang, Qinjun Peng, Zhimin Wang, Shenjin
  Zhang, Feng Yang, Zuyan Xu, Chuangtian Chen, and X.~J. Zhou.
\newblock {Distinct Fermi Surface Topology and Nodeless Superconducting Gap in
  (Tl$_{0.58}$Rb$_{0.42}$)Fe$_{1.72}$Se$_2$ Superconductor}.
\newblock {\em Phys. Rev. Lett.}, 106:107001, 2011.

\bibitem{2011arXiv1102.3505G}
J.~{Guo}, X.~{Chen}, G.~{Wang}, T.~{Zhou}, X.~{Lai}, S.~{Jin}, S.~{Wang}, and
  K.~{Zhu}.
\newblock {Superconductivity on the verge of Mott localization in ternary iron
  sulfide}.
\newblock {\em arXiv:1102.3505}, 2011.

\bibitem{hongdingarpeskfese}
X.-P. Wang, T.~Qian, P.~Richard, P.~Zhang, J.~Dong, H.-D. Wang, C.-H. Dong,
  M.-H. Fang, and H.~Ding.
\newblock Strong nodeless pairing on separate electron Fermi surface sheets in
  (Tl,K)Fe$_{1.78}$Se$_2$ probed by ARPES.
\newblock {\em Euro. Phys. Lett.}, 93:57001, 2011.

\bibitem{2011arXiv1102.3888M}
Long Ma, G.~F. Ji, J.~Zhang, J.~B. He, D.~M. Wang, G.~F. Chen, Wei Bao, and
  Weiqiang Yu.
\newblock Superconducting and normal-state properties of single-crystalline
  Tl$_{0.47}$Rb$_{0.34}$Fe$_{1.63}$Se$_2$ as seen via $^{77}$Se and $^{87}$Rb
  NMR.
\newblock {\em Phys. Rev. B}, 83:174510, 2011.

\bibitem{2011arXiv1101.0896K}
Y.~{Kawasaki}, Y.~{Mizuguchi}, K.~{Deguchi}, T.~{Watanabe}, T.~{Ozaki},
  S.~{Tsuda}, T.~{Yamaguchi}, H.~{Takeya}, and Y.~{Takano}.
\newblock {Pressure study of the new iron-based superconductor
  K$_{0.8}$Fe$_2$Se$_2$}.
\newblock {\em arXiv:1101.0896}, 2011.

\bibitem{2011arXiv1101.0092G}
J.~{Guo}, L.~{Sun}, C.~{Zhang}, J.~{Guo}, X.~{Chen}, Q.~{Wu}, D.~{Gu},
  P.~{Gao}, X.~{Dai}, and Z.~{Zhao}.
\newblock {Correlation between superconductivity and resistance hump in
  K$_{0.8}$Fe$_{1.7}$Se$_2$ single crystal under high pressure}.
\newblock {\em arXiv:1101.0092}, 2011.

\bibitem{PhysRevLett.106.087005}
Xun-Wang Yan, Miao Gao, Zhong-Yi Lu, and Tao Xiang.
\newblock Electronic Structures and Magnetic Order of Ordered-Fe-Vacancy
  Ternary Iron Selenides TlFe$_{1.5}$Se$_{2}$ and $A$Fe$_{1.5}$Se$_{2}$ ($A=K$,
  Rb, or Cs).
\newblock {\em Phys. Rev. Lett.}, 106:087005, 2011.

\bibitem{2011arXiv1102.2215Y}
Xun-Wang Yan, Miao Gao, Zhong-Yi Lu, and Tao Xiang.
\newblock Ternary iron selenide K$_{0.8}$Fe$_{1.6}$Se$_2$ is an
  antiferromagnetic semiconductor.
\newblock {\em Phys. Rev. B}, 83:233205, 2011.

\bibitem{2010arXiv1012.5164S}
I.R. Shein and A.L. Ivanovskii.
\newblock Electronic structure and Fermi surface of new K intercalated iron
  selenide superconductor K$_x$Fe$_2$Se$_2$.
\newblock {\em Phys. Lett. A}, 375:1028, 2011.

\bibitem{2011arXiv1101.0051N}
I.~A. {Nekrasov} and M.~V. {Sadovskii}.
\newblock {Electronic Structure, Topological Phase Transitions and
  Superconductivity in (K,Cs)xFe$_2$Se$_2$}.
\newblock {\em Pisma ZhETF}, 93:182, 2011.

\bibitem{2010arXiv1012.5621C}
Cao Chao and Dai Jian-Hui.
\newblock {Electronic Structure of KFe$_2$Se$_2$ from First Principles
  Calculations}.
\newblock {\em Chin. Phys. Lett.}, 28:057402, 2011.

\bibitem{2011arXiv1101.0533C}
Chao Cao and Jianhui Dai.
\newblock Electronic structure and Mott localization of iron-deficient
  TlFe$_{1.5}$Se$_2$ with superstructures.
\newblock {\em Phys. Rev. B}, 83:193104, 2011.

\bibitem{kfesearpes1}
Y.~{Zhang}, L.~X. {Yang}, M.~{Xu}, Z.~R. {Ye}, F.~{Chen}, C.~{He}, J.~{Jiang},
  B.~P. {Xie}, J.~J. {Ying}, X.~F. {Wang}, X.~H. {Chen}, J.~P. {Hu}, and D.~L.
  {Feng}.
\newblock {Nodeless superconducting gap in A$_x$Fe$_2$Se$_2$ (A = K, Cs)
  revealed by angle-resolved photoemission spectroscopy}.
\newblock {\em Nature Mater.}, 10:273, 2010.

\bibitem{2010arXiv1012.6017Q}
T.~{Qian}, {X.-P.} {Wang}, {W.-C.} {Jin}, P.~{Zhang}, P.~{Richard}, G.~{Xu},
  X.~{Dai}, Z.~{Fang}, {J.-G.} {Guo}, {X.-L.} {Chen}, and H.~{Ding}.
\newblock {Absence of holelike Fermi surface in superconducting
  K$_{0.8}$Fe$_{1.7}$Se$_2$ revealed by Angle-Resolved Photoemission
  Spectroscopy}.
\newblock {\em Phys. Rev. Lett.}, 106:187001, 2011.

\bibitem{2011arXiv1102.1057Z}
Lin Zhao, Daixiang Mou, Shanyu Liu, Xiaowen Jia, Junfeng He, Yingying Peng,
  Li~Yu, Xu~Liu, Guodong Liu, Shaolong He, Xiaoli Dong, Jun Zhang, J.~B. He,
  D.~M. Wang, G.~F. Chen, J.~G. Guo, X.~L. Chen, Xiaoyang Wang, Qinjun Peng,
  Zhimin Wang, Shenjin Zhang, Feng Yang, Zuyan Xu, Chuangtian Chen, and X.~J.
  Zhou.
\newblock {Common Fermi Surface Topology and Nodeless Superconducting Gap in
  K$_{0.68}$Fe$_{1.79}$Se$_2$ and (Tl$_{0.45}$K$_{0.34}$)Fe$_{1.84}$Se$_2$
  Superconductors Revealed from Angle-Resolved Photoemission Spectroscopy}.
\newblock {\em Phys. Rev. B}, 83:140508, 2011.

\bibitem{2011arXiv1101.4988M}
T.~A. Maier, S.~Graser, P.~J. Hirschfeld, and D.~J. Scalapino.
\newblock $d$-wave pairing from spin fluctuations in the
  K$_{x}$Fe$_{2-y}$Se$_{2}$ superconductors.
\newblock {\em Phys. Rev. B}, 83:100515, 2011.

\bibitem{mazin-2009}
I.~I. Mazin and J.~Schmalian.
\newblock Pairing symmetry and pairing state in ferropnictides: Theoretical
  overview.
\newblock {\em Physica C}, 469:614, 2009.

\bibitem{epl9357003}
Fa~Wang, Fan Yang, Miao Gao, Zhong-Yi Lu, Tao Xiang, and Dung-Hai Lee.
\newblock The electron pairing of K$_x$Fe$_{2-y}$Se$_2$.
\newblock {\em Euro. Phys. Lett.}, 93:57003, 2011.

\bibitem{2011arXiv1101.4462Z}
Yi~Zhou, Dong-Hui Xu, Fu-Chun Zhang, and Wei-Qiang Chen.
\newblock Theory for superconductivity in (Tl,K)Fe$_x$Se$_2$ as a doped Mott
  insulator.
\newblock {\em Euro. Phys. Lett.}, 95:17003, 2011.

\bibitem{2011arXiv1103.1902}
Yi-Zhuang You, Fan Yang, Su-Peng Kou, and Zheng-Yu Weng.
\newblock {Phase diagram and a possible unified description of intercalated
  iron selenide superconductors}.
\newblock {\em arXiv:1103.1902}, 2011.

\bibitem{2011arXiv1102.4049S}
Tetsuro Saito, Seiichiro Onari, and Hiroshi Kontani.
\newblock Emergence of fully gapped $s_{++}$-wave and nodal $d$-wave states
  mediated by orbital and spin fluctuations in a ten-orbital model of
  KFe$_2$Se$_2$.
\newblock {\em Phys. Rev. B}, 83:140512, 2011.

\bibitem{2011arXiv1101.3687M}
Weiqiang Yu, L.~Ma, J.~B. He, D.~M. Wang, T.-L. Xia, G.~F. Chen, and Wei Bao.
\newblock $^{77}Se$ NMR Study of the Pairing Symmetry and the Spin Dynamics in
  $K_{y}Fe_{2-x}Se_{2}$.
\newblock {\em Phys. Rev. Lett.}, 106:197001, 2011.

\bibitem{2011arXiv1103.1347}
Fei Han, Bing Shen, Zhen-Yu Wang, and Hai-Hu Wen.
\newblock {Reversibly tuning the insulating and superconducting state in
  K$_x$Fe$_{2-y}$Se$_2$ crystals by post-annealing}.
\newblock {\em arXiv:1103.1347}, 2011.

\bibitem{2011arXiv1102.3674B}
W.~{Bao}, G.~N. {Li}, Q.~{Huang}, G.~F. {Chen}, J.~B. {He}, M.~A. {Green},
  Y.~{Qiu}, D.~M. {Wang}, and J.~L. {Luo}.
\newblock {Vacancy tuned magnetic high-$T_c$ superconductor
  K$_x$Fe$_{2-x/2}$Se$_2$}.
\newblock {\em arXiv:1102.3674}, 2011.

\bibitem{2011arXiv1101.3307Y}
Rong Yu, Jian-Xin Zhu, and Qimiao Si.
\newblock Mott Transition in Modulated Lattices and Parent Insulator of
  (K,Tl)$_y$Fe$_x$Se$_2$ Superconductors.
\newblock {\em Phys. Rev. Lett.}, 106:186401, 2011.

\bibitem{2011arXiv1103.4599F}
C.~{Fang}, B.~{Xu}, P.~{Dai}, T.~{Xiang}, and J.~{Hu}.
\newblock {Magnetic Frustration and Iron-Vacancy Ordering in
  Iron-Chalcogenide}.
\newblock {\em arXiv:1103.4599}, 2011.

\bibitem{subedi-2008-78}
Alaska Subedi, Lijun Zhang, David~J. Singh, and Mao-Hua Du.
\newblock Density functional study of FeS, FeSe and FeTe: Electronic structure,
  magnetism, phonons and superconductivity.
\newblock {\em Phys. Rev. B}, 78:134514, 2008.

\bibitem{xia037002}
Y.~Xia, D.~Qian, L.~Wray, D.~Hsieh, G.~F. Chen, J.~L. Luo, N.~L. Wang, and
  M.~Z. Hasan.
\newblock Fermi Surface Topology and Low-Lying Quasiparticle Dynamics of Parent
  Fe$_{1+x}$Te/Se Superconductor.
\newblock {\em Phys. Rev. Lett.}, 103:037002, 2009.

\bibitem{PhysRevLett.105.197001}
K.~Nakayama, T.~Sato, P.~Richard, T.~Kawahara, Y.~Sekiba, T.~Qian, G.~F. Chen,
  J.~L. Luo, N.~L. Wang, H.~Ding, and T.~Takahashi.
\newblock Angle-Resolved Photoemission Spectroscopy of the Iron-Chalcogenide
  Superconductor Fe$_{1.03}$Te$_{0.7}$Se$_{0.3}$: Strong Coupling Behavior and
  the Universality of Interband Scattering.
\newblock {\em Phys. Rev. Lett.}, 105:197001, 2010.

\bibitem{structure6}
D.~Fruchart, P.~Convert, P.~Wolfers, R.~Madar, J.~P. Senateur, and R.~Fruchart.
\newblock Structure antiferroma gnetique de Fe$_{1. 125}$Te accompagnee d'une
  deformation monoclinique.
\newblock {\em Mater. Res. Bull.}, 10:169, 1975.

\bibitem{Physics.2.59}
Alexander~V. Balatsky and David Parker.
\newblock Not all iron superconductors are the same.
\newblock {\em Physics}, 2:59, 2009.

\bibitem{han:067001}
Myung~Joon Han and Sergey~Y. Savrasov.
\newblock Doping Driven ($\pi$, 0) Nesting and Magnetic Properties of
  Fe$_{1+x}$Te Superconductors.
\newblock {\em Phys. Rev. Lett.}, 103:067001, 2009.

\bibitem{ma-2009}
Fengjie Ma, Wei Ji, Jiangping Hu, Zhong-Yi Lu, and Tao Xiang.
\newblock First-Principles Calculations of the Electronic Structure of
  Tetragonal $\alpha$-FeTe and $\alpha$-FeSe Crystals: Evidence for a
  Bicollinear Antiferromagnetic Order.
\newblock {\em Phys. Rev. Lett.}, 102:177003, 2009.

\bibitem{loalm1}
Ari~M. Turner, Fa~Wang, and Ashvin Vishwanath.
\newblock Kinetic magnetism and orbital order in iron telluride.
\newblock {\em Phys. Rev. B}, 80:224504, 2009.

\bibitem{fang-2009}
Chen Fang, B.~Andrei Bernevig, and Jiangping Hu.
\newblock Theory of Magnetic Order in Fe$ _{1+y}$Te$_{1-x}$Se$_x$.
\newblock {\em Euro. Phys. Lett.}, 86:67005, 2009.

\bibitem{moon057003}
Chang-Youn Moon and Hyoung~Joon Choi.
\newblock Chalcogen-Height Dependent Magnetic Interactions and Magnetic Order
  Switching in FeSe$_x$Te$_{1-x}$.
\newblock {\em Phys. Rev. Lett.}, 104:057003, 2010.

\bibitem{Zaliznyak2011}
Igor~A. Zaliznyak, Zhijun Xu, John~M. Tranquada, Genda Gu, Alexei~M. Tsvelik,
  and Matthew~B. Stone.
\newblock {Unconventional temperature enhanced magnetism in iron telluride}.
\newblock {\em arXiv:1103.5073}, 2011.

\bibitem{imai-2009-102}
T.~Imai, K.~Ahilan, F.~L. Ning, T.~M. McQueen, and R.~J. Cava.
\newblock Why Does Undoped FeSe Become A High $T_c$ Superconductor Under
  Pressure?
\newblock {\em Phys. Rev. Lett.}, 102:177005, 2009.

\bibitem{PhysRevLett.104.087003}
M.~Bendele, A.~Amato, K.~Conder, M.~Elender, H.~Keller, H.-H. Klauss,
  H.~Luetkens, E.~Pomjakushina, A.~Raselli, and R.~Khasanov.
\newblock Pressure Induced Static Magnetic Order in Superconducting
  FeSe$_{1-x}$.
\newblock {\em Phys. Rev. Lett.}, 104:087003, 2010.

\bibitem{afmafese1}
H.~Sabrowsky, M.~Rosenberg, D.~Welz, P.~Deppe, and W.~Sch\"{a}fer.
\newblock A neutron and M\"{o}sbauer study of TlFe$_x$S$_2$ compounds.
\newblock {\em J. Magn. Magn. Mater.}, 54-57:1497, 1986.

\bibitem{afmafese2}
Lennart H\"{a}gstr\"{o}m, Agneta Seidel, and Rolf Berger.
\newblock A M\"{o}sbauer study of antiferromagnetic ordering in iron deficient
  TlFe$_{2-x}$Se$_2$.
\newblock {\em J. Magn. Magn. Mater.}, 98:37, 1991.

\bibitem{C1SC00070E}
J.~Bacsa, A.~Y. Ganin, Y.~Takabayashi, K.~E. Christensen, K.~Prassides, M.~J.
  Rosseinsky, and J.~B. Claridge.
\newblock Cation vacancy order in the K$_{0.8+x}$Fe$_{1.6-y}$Se$_2$ system:
  Five-fold cell expansion accommodates 20\% tetrahedral vacancies.
\newblock {\em Chem. Sci.}, 2:1054, 2011.

\bibitem{2011arXiv1102.1344C}
C.~{Cao} and J.~{Dai}.
\newblock {Block Spin Ground State and 3-Dimensionality of
  (K,Tl)Fe$_{1.6}$Se$_2$}.
\newblock {\em arXiv:1102.1344 (to appear in Phys. Rev. Lett.)}, 2011.

\bibitem{2011arXiv1102.2882Y}
F.~{Ye}, S.~{Chi}, W.~{Bao}, X.~F. {Wang}, J.~J. {Ying}, X.~H. {Chen}, H.~D.
  {Wang}, C.~H. {Dong}, and M.~{Fang}.
\newblock {Common Structural and Magnetic Framework in the $A_2$Fe$_4$Se$_5$
  Superconductors}.
\newblock {\em arXiv:1102.2882}, 2011.

\bibitem{2011arXiv1104.2008S}
B.~{Shen}, B.~{Zeng}, G.~F. {Chen}, J.~B. {He}, D.~M. {Wang}, H.~{Yang}, and
  H.~H. {Wen}.
\newblock {Intrinsic Percolative Superconductivity in K$_x$Fe$_{2-y}$Se$_2$
  Single Crystals}.
\newblock {\em arXiv:1104.2008}, 2011.

\bibitem{2011arXiv1107.0412R}
A.~{Ricci}, N.~{Poccia}, G.~{Campi}, B.~{Joseph}, G.~{Arrighetti}, L.~{Barba},
  M.~{Reynolds}, M.~{Burghammer}, H.~{Takeya}, Y.~{Mizuguchi}, Y.~{Takano},
  M.~{Colapietro}, N.~L. {Saini}, and A.~{Bianconi}.
\newblock {Direct evidence of nanoscale phase separation in the iron
  chalcogenide superconductor K0.8Fe1.6Se2 from scanning nanofocused x-ray
  diffraction}.
\newblock {\em arXiv:1107.0412}, 2011.

\bibitem{phasesep122_1}
Alessandro Ricci, Nicola Poccia, Boby Joseph, Gianmichele Arrighetti, Luisa
  Barba, Jasper Plaisier, Gaetano Campi, Yoshikazu Mizuguchi, Hiroyuki Takeya,
  Yoshihiko Takano, Naurang~Lal Saini, and Antonio Bianconi.
\newblock Intrinsic phase separation in superconducting
  K$_{0.8}$Fe$_{1.6}$Se$_2$ ($T_c = 31.8$ K) single crystals.
\newblock {\em Superconductor Science and Technology}, 24:082002, 2011.

\bibitem{maier:020514}
T.~A. Maier and D.~J. Scalapino.
\newblock Theory of neutron scattering as a probe of the superconducting gap in
  the iron pnictides.
\newblock {\em Phys. Rev. B}, 78:020514(R), 2008.

\bibitem{maier:134520}
T.~A. Maier, S.~Graser, D.~J. Scalapino, and P.~Hirschfeld.
\newblock Neutron scattering resonance and the iron-pnictide superconducting
  gap.
\newblock {\em Phys. Rev. B}, 79:134520, 2009.

\bibitem{PhysRevLett.75.4126}
Eugene Demler and Shou-Cheng Zhang.
\newblock Theory of the Resonant Neutron Scattering of High-$T_{c}$
  Superconductors.
\newblock {\em Phys. Rev. Lett.}, 75:4126, 1995.

\bibitem{PhysRevB.64.172508}
C.~D. Batista, G.~Ortiz, and A.~V. Balatsky.
\newblock Unified description of the resonance peak and incommensuration in
  high-$T_c$ superconductors.
\newblock {\em Phys. Rev. B}, 64:172508, 2001.

\bibitem{2011arXiv1103.5073Z}
I.~A. {Zaliznyak}, Z.~{Xu}, J.~M. {Tranquada}, G.~{Gu}, A.~M. {Tsvelik}, and
  M.~B. {Stone}.
\newblock {Unconventional temperature enhanced magnetism in iron telluride}.
\newblock {\em arXiv:1103.5073}, 2011.

\bibitem{incomfetese1}
D.~N. Argyriou, A.~Hiess, A.~Akbari, I.~Eremin, M.~M. Korshunov, Jin Hu, Bin
  Qian, Zhiqiang Mao, Yiming Qiu, Collin Broholm, and W.~Bao.
\newblock Incommensurate itinerant antiferromagnetic excitations and spin
  resonance in the FeTe$_{0.6}$Se$_{0.4}$ superconductor.
\newblock {\em Phys. Rev. B}, 81:220503, 2010.

\bibitem{PhysRevLett.105.157002}
Shiliang Li, Chenglin Zhang, Meng Wang, Hui-qian Luo, Xingye Lu, Enrico
  Faulhaber, Astrid Schneidewind, Peter Link, Jiangping Hu, Tao Xiang, and
  Pengcheng Dai.
\newblock Normal-State Hourglass Dispersion of the Spin Excitations in
  FeSe$_{x}$Te$_{1-x}$.
\newblock {\em Phys. Rev. Lett.}, 105:157002, 2010.

\bibitem{fetesecomp142202}
P.~Babkevich, M.~Bendele, A.~T. Boothroyd, K.~Conder, S.~N. Gvasaliya,
  R.~Khasanov, E.~Pomjakushina, and B.~Roessli.
\newblock Magnetic excitations of Fe$_{1 + y}$Se$_x$Te$_{1-x}$ in magnetic and
  superconductive phases.
\newblock {\em J. Phys. Conden. Matter}, 22:142202, 2010.

\bibitem{lumsden-2009}
M.~D. Lumsden, A.~D. Christianson, E.~A. Goremychkin, S.~E. Nagler, H.~A. Mook,
  M.~B. Stone, D.~L. Abernathy, T.~Guidi, G.~J. MacDougall, C.~{De La Cruz},
  A.~S. Sefat, M.~A. McGuire, B.~C. Sales, and D.~Mandrus.
\newblock Evolution of spin excitations into the superconducting state in
  FeTe$_{1-x}$Se$_x$.
\newblock {\em Nature Phys.}, 6:182, 2010.

\bibitem{diallo:187206}
S.~O. Diallo, V.~P. Antropov, T.~G. Perring, C.~Broholm, J.~J. Pulikkotil,
  N.~Ni, S.~L. Bud'ko, P.~C. Canfield, A.~Kreyssig, A.~I. Goldman, and R.~J.
  McQueeney.
\newblock Itinerant Magnetic Excitations in Antiferromagnetic CaFe$_2$As$_2$.
\newblock {\em Phys. Rev. Lett.}, 102:187206, 2009.

\bibitem{zhao-2009-5}
Jun Zhao, D.~T. Adroja, Dao-Xin Yao, R.~Bewley, Shiliang Li, X.~F. Wang, G.~Wu,
  X.~H. Chen, Jiangping Hu, and Pengcheng Dai.
\newblock Spin Waves and Magnetic Exchange Interactions in CaFe$_2$As$_2$.
\newblock {\em Nature Phys.}, 5:555, 2009.

\bibitem{nature965}
P.~Dai, H.~A. Mook, G.~Aeppli, S.~M. Hayden, F.~Do\u{g}an, J.~Yu, Y.~Yanagida,
  H.~Takashima, Y.~Inaguma, M.~Itoh, and T.~Nakamura.
\newblock Resonance as a measure of pairing correlations in the high-$T_c$
  superconductor YBCO.
\newblock {\em Nature}, 406:965, 2000.

\bibitem{tranquada-2004}
J.~M. Tranquada, C.~H. Lee, K.~Yamada, Y.~S. Lee, L.~P. Regnault, and H.~M.
  R\o{}nnow.
\newblock Evidence for an incommensurate magnetic resonance in
  La$_{2-x}$Sr$_x$CuO$_4$.
\newblock {\em Phys. Rev. B}, 69:174507, 2004.

\bibitem{zhao-2010field122}
Jun Zhao, Louis-Pierre Regnault, Chenglin Zhang, Miaoying Wang, Zhengcai Li,
  Fang Zhou, Zhongxian Zhao, Chen Fang, Jiangping Hu, and Pengcheng Dai.
\newblock Neutron spin resonance as a probe of the superconducting energy gap
  of BaFe$_{1.9}$Ni$_{0.1}$As$_2$ superconductors.
\newblock {\em Phys. Rev. B}, 81:180505, 2010.

\bibitem{Bourges200545}
Ph. Bourges, B.~Keimer, S.~Pailh\`{o}es, L.P. Regnault, Y.~Sidis, and
  C.~Ulrich.
\newblock The resonant magnetic mode: A common feature of high-Tc
  superconductors.
\newblock {\em Physica C}, 424:45, 2005.

\bibitem{wbao2}
Wei Bao, A.~T. Savici, G.~E. Granroth, C.~Broholm, K.~Habicht, Y.~Qiu, Jin Hu,
  Tijiang Liu, and Z.~Q. Mao.
\newblock {A Triplet Resonance in Superconducting FeSe$_{0.4}$Te$_{0.6}$}.
\newblock {\em arXiv:1002.1617}, 2010.

\bibitem{2011arXiv1105.4923L}
S.~{Li}, X.~{Lu}, M.~{Wang}, H.-q. {Luo}, M.~{Wang}, C.~{Zhang},
  E.~{Faulhaber}, L.-P. {Regnault}, D.~{Singh}, and P.~{Dai}.
\newblock {In-plane magnetic field effect on the neutron spin resonance in
  optimally doped FeSe$_{0.4}$Te$_{0.6}$ and BaFe$_{1.9}$Ni$_{0.1}$As$_2$
  superconductors}.
\newblock {\em arXiv:1105.4923}, 2011.

\bibitem{PhysRevLett.106.057004}
O.~J. Lipscombe, G.~F. Chen, Chen Fang, T.~G. Perring, D.~L. Abernathy, A.~D.
  Christianson, Takeshi Egami, Nanlin Wang, Jiangping Hu, and Pengcheng Dai.
\newblock Spin Waves in the $(\pi,0)$ Magnetically Ordered Iron Chalcogenide
  Fe$_{1.05}$Te.
\newblock {\em Phys. Rev. Lett.}, 106:057004, 2011.

\bibitem{2011arXiv1103.1811}
C.~Stock, E.~E. Rodriguez, M.~A. Green, and J.~A. Rodriguez-Rivera.
\newblock {Gapped spin fluctuations and interstitial iron in the Fe$_{1+x}$Te
  parent compound}.
\newblock {\em arXiv:1103.1811}, 2011.

\bibitem{xulocal1}
Z.~{Xu}, J.~{Wen}, G.~{Xu}, S.~{Chi}, W.~{Ku}, G.~{Gu}, and J.~M. {Tranquada}.
\newblock {Evidence for local moment magnetism in superconducting
  FeTe$_{0.35}$Se$_{0.65}$}.
\newblock {\em arXiv:1012.2300}, 2010.

\bibitem{2011arXiv1105.4675W}
M.~{Wang}, C.~{Fang}, D.-X. {Yao}, G.~{Tan}, L.~W. {Harriger}, Y.~{Song},
  T.~{Netherton}, C.~{Zhang}, M.~{Wang}, M.~B. {Stone}, W.~{Tian}, J.~{Hu}, and
  P.~{Dai}.
\newblock {Spin Waves and magnetic exchange interactions in insulating
  Rb$_{0.89}$Fe$_{1.58}$Se$_2$}.
\newblock {\em arXiv:1105.4675}, 2011.

\end{thebibliography}
